\documentclass[prl,twocolumn,noshowpacs,preprintnumbers,amsmath,amssymb,nofootinbib,showkeys]{revtex4-2}

\usepackage{lineno}
\usepackage{bm}
\usepackage{color}
\usepackage{hyperref}
\hypersetup{hidelinks}
\usepackage{times}

\newcommand{\tb}{\textbf}

\newcommand{\dd}{{\rm d}}
\newcommand{\rmE}{{\mathrm{E}}}
\newcommand{\eps}{\epsilon}

\usepackage{tcolorbox} 

\begin{document}

 \title{Is price-impact decay statistically identifiable by propagator estimation?\\ Diffusive prices force the inferred kernel to be transient and fine-tuned}
\author{Shunta Fujiwara$^{1}$}
\author{Yuki Sato$^{2}$}
\author{Kiyoshi Kanazawa$^{1}$}
\affiliation{$^{1}$Kyoto University and $^{2}$University of Tsukuba}
\date[]{}

\keywords{econophysics, the diffusivity puzzle, price impact decay, statistical identification}
\begin{abstract}
 Despite the long memory of market-order flow, price dynamics remain diffusive---a seemingly paradoxical coexistence known in econophysics as the {\it diffusivity puzzle}. The propagator model has long been regarded as a promising solution: price impact decays in time, finely balanced against the order-flow memory, and propagators inferred from data have been reported to be consistent with this fine balance. Here we prove a no-go theorem: asymptotic impact decay is not statistically identifiable by propagator estimation, since for diffusive prices the fine-tuned decay is returned whether or not the true impact decays. It rests on universal theorems for the Wiener--Hopf deconvolution by which propagators are inferred under long-memory order flow: (i) the inferred propagator is necessarily transient, even when the true impact is permanent; (ii) under a mild condition, it decays with the fine-balance exponent, fixed by the order-flow autocorrelation alone. The fine-tuned decay is thus a property of the deconvolution, not evidence of impact decay. We demonstrate this with an exactly solvable null model that retains order splitting and the square-root law but whose impact is strictly permanent: the inferred propagator shows an artificial fine-tuned decay although the true impact never decays. Nonlinear propagator estimation, being of Wiener--Hopf type, is subject to the same theorems whenever true metaorders have long memory. 
\end{abstract}

\maketitle

\paragraph{\it Introduction.}
    Why does price remain diffusive despite the long memory of market-order flow?~\cite{BouchaudText} This long-standing question in empirical finance, known as the {\it diffusivity puzzle}, has been revisited many times and remains debated~\cite{BFL_Book2009,LilloFarmer2004,FarmerGerigLilloMike2006}. The most influential proposed solution is the {\it propagator model}~\cite{BouchaudProp1,BouchaudProp2}, where {\it transient impact} with a {\it fine-tuned decay} is arguably the key to the puzzle. Here we ask whether this impact decay can be identified from data by propagator estimation, that is, whether the inferred kernel can be read causally.

\begin{figure*}
        \includegraphics[width=170mm]{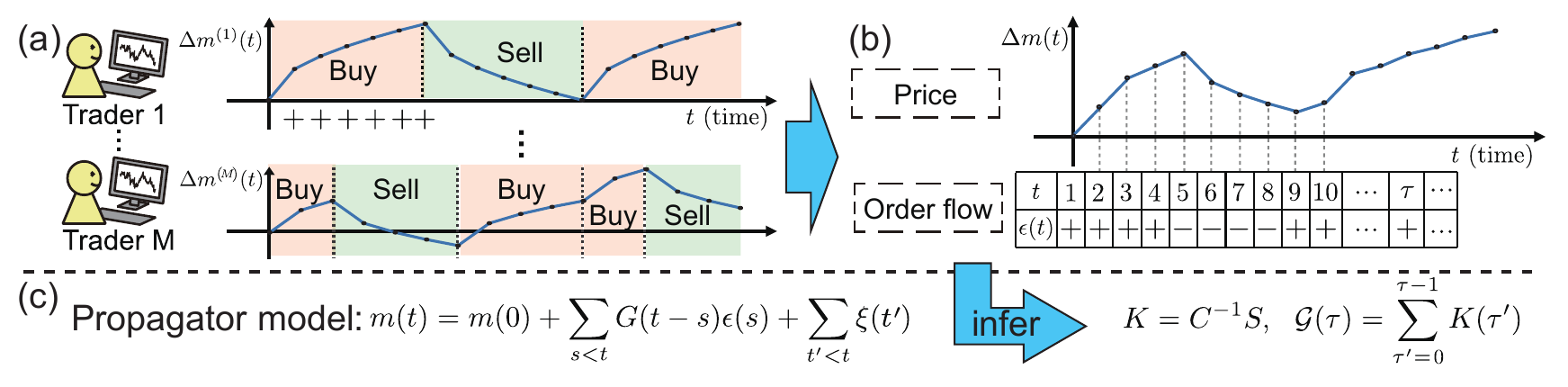}
        \caption{
            (a)~Order-splitting mechanism in the Lillo--Mike--Farmer model and the resulting square-root impact $I(Q)\propto \sqrt{Q}$ due to the metaorder of size $Q$. (b)~The price dynamics in the Sato--Fujiwara--Kanazawa (SFK) model, where the price movement is a simple sum of all traders' square-root impact at the level of metaorders. (c)~For the price and order-sign process generated by the SFK model, we apply the propagator deconvolution estimation~\eqref{eq:deconvolution} to obtain the inferred kernel $\mathcal{G}(\tau)$. Although the true impact is strictly permanent in the SFK model, the inferred kernel decays as $\mathcal{G}(\tau)\propto \tau^{-\beta_c}$. Since the propagator estimation is correct within the linear propagator class, this design provides a robustness check against model misspecification.
        }
        \label{fig:SFK}
    \end{figure*}

    It is an established empirical fact that the market-order flow has long memory and is characterized by the slow decay~\cite{LilloFarmer2004,BouchaudProp1}: 
    \begin{equation}
    C(\tau):= \rmE[\eps_0\eps_{\tau}]\propto \tau^{-\gamma},\quad \gamma \in (0,1),
    \end{equation}
    where $\eps_t$ is the market order sign, such that $\eps_t=1$ ($\eps_t=-1$) represents a buy (sell) order at time $t$. This phenomenon originates from metaorder splitting and is formulated by the Lillo--Mike--Farmer (LMF) model~\cite{LMF_PRE2005,TothPalitLilloFarmer2015,SatoPRL2023,SatoPRR2023,Sato2024_GeneralizedLMF} [see Fig.~\ref{fig:SFK}(a)]. Here, although the market-order flow has long memory, various observations confirm that the price dynamics remain diffusive at long times~\cite{BouchaudText}. Given that a market order has positive impact in general, this observation is enigmatic in empirical finance.

    This paradoxical phenomenon has been called the diffusivity puzzle, and the propagator model was proposed as a promising solution: 
    \begin{equation}
    m_t = m_0 + \sum_{s<t} G(t-s)\eps_s + \sum_{t'<t}\xi_{t'},
    \end{equation}
    where $m_t$ is the price at time $t$, $G(\tau)$ is the propagator representing the impact decay, and $\xi_t$ is the white noise term~\cite{BouchaudText,BouchaudProp1,BouchaudProp2}. Within the propagator framework, it has been argued that the diffusivity puzzle could be resolved by imposing (i) the {\it transience of impact} and (ii) the {\it fine-tuned balance} between the long memory of market-order flow and the impact decay. In other words, the propagator $G(\tau)$ is assumed to slowly decay to {\it exactly zero} at long times
    \begin{equation}
    G_{\infty}:=\lim_{\tau\to\infty}G(\tau)=0,\quad  G(\tau)\propto \tau^{-\beta_c},\quad      \beta_c=\frac{1-\gamma}{2} \label{eq:FBC}
    \end{equation}
    to ensure that the price remains diffusive. Here, $G_{\infty}$ is called the permanent impact, and $\beta_c$ is the critical exponent of the impact decay. 
 
    Propagators inferred from data, denoted by $\mathcal{G}(\tau)$ to distinguish them from the true kernel $G(\tau)$, have been reported to be consistent with Eq.~\eqref{eq:FBC}, a finding taken as evidence for the transient-impact explanation of the diffusivity puzzle~\cite{BouchaudText,BouchaudProp1,BouchaudProp2}. In this picture, the fine-tuned balance is regarded as a genuine property of the market, termed the {\it long-range resilience} of liquidity~\cite{BouchaudText,BouchaudProp1}: the empirically observed diffusivity is taken to impose the special shape of $G$, and the resulting compensation is attributed to the way liquidity providers and liquidity takers offset each other. This concept has also been generalized to a nonlinear propagator model with fine-tuned impact decay at the level of metaorders~\cite{GuillaumeBouchaud2026}. 
 
    While this picture is appealing, we examine the propagator estimation from the viewpoint of {\it statistical identification}~\cite{Lewbel2019}, a standard notion in statistics and econometrics (see End Matter). Suppose that one tries to validate an interesting hypothesis (e.g., impact decay) by applying statistical estimation (e.g., the propagator estimation) to empirical datasets. When one finds estimated results consistent with the hypothesis, one might conclude that the hypothesis is valid. However, this is insufficient: one must also rule out the possibility that the estimated result is an artifact of the estimation itself. To check this, one constructs a {\it null model}: a generative model in which the mechanism of interest---here, impact decay---is deliberately excluded, but which is otherwise plausible. One then applies the same estimation to data generated by the null model. If the estimate is again consistent with the hypothesis, the agreement is an artifact of the estimation, and we call the hypothesis {\it statistically unidentifiable} by that procedure. Throughout, we call an inferred decay artificial only in such null-model tests, where the true impact is known not to decay.
 
    In this Letter, we show that the impact decay is not statistically identifiable by propagator estimation. We first prove universal decay theorems for the deconvolution estimation: whenever the price is diffusive and the order flow has long memory, the inferred kernel is necessarily transient, $\mathcal{G}_{\infty}=0$, and, under a mild condition, its decay exponent is fixed at $\beta_c=(1-\gamma)/2$ by the order-flow autocorrelation alone, irrespective of the true impact: a property of the inversion, not of the data-generating process. Our theorems concern the population deconvolution: the inferred decay persists in the limit of infinite data and is not a finite-sample bias. These theorems apply even to a null model with strictly permanent impact and no decay. Indeed, the Sato--Fujiwara--Kanazawa (SFK) model~\cite{SatoFujiwaraKanazawa2026,SatoFujiwaraKanazawa2026Full}---an exactly solvable order-splitting model with strictly permanent, square-root impact~\cite{SatoPRL2025}---is such a null model [Fig.~\ref{fig:SFK}(a) and (b)]: it retains the empirically established ingredients, order splitting and the square-root law, and removes only the one under debate, impact decay; yet its inferred kernel still shows an artificial fine-tuned decay. Together with this null model, our theorems constitute a no-go theorem for the causal reading of the inferred kernel's asymptotics. Within the linear propagator class, the deconvolution is consistent; the non-identifiability arises once diffusive price processes outside this class, such as the SFK model, are admitted. Hence the long-range resilience of liquidity is not identified by the inferred fine-tuning.
 
     This difficulty remains even for nonlinear (more precisely, quasi-linear) propagator estimation~\cite{Brokmann2015,Bucci2020}, which reduces to the same deconvolution after a change of variables. If true metaorders have long memory, the inferred kernel is again forced to be transient by our theorems, irrespective of the true metaorder impact. Even if true metaorders have short memory, a practical problem remains: in empirical work, metaorders are typically reconstructed by the daily-segmentation algorithm, so that a metaorder executed over several days is recorded as several distinct daily metaorders. These metaorder proxies share the same sign, so that the reconstructed metaorder flow can inherit an artificial long memory, and our theorems are expected to apply to them as well. We confirm this numerically for the SFK model, using the estimation of Bucci \textit{et al.}~\cite{Bucci2020}. Thus, agreement between an inferred propagator $\mathcal{G}(\tau)$ and the fine-tuned decay law cannot, by itself, be taken as evidence for transient impact. 
 
    \paragraph{Propagator deconvolution estimation.}
         We first briefly review the standard propagator deconvolution estimation~\cite{BouchaudText}. Let us define the response function and its increment by 
         \begin{subequations}\label{eq:deconvolution}
            \begin{equation}
                 R(\tau) := \rmE[\eps_0(m_{\tau}-m_0)], \quad S(\tau) := R(\tau+1) - R(\tau).
            \end{equation}
            Also, we define the increment of the propagator kernel by 
            \begin{equation}
                K(\tau):= \mathcal{G}(\tau+1)-\mathcal{G}(\tau), \quad \mathcal{G}(0):=0,
            \end{equation}
             so that the inferred kernel is recovered as $\mathcal{G}(\tau)=\sum_{\tau'=0}^{\tau-1}K(\tau')$. Given the measured $C$ and $S$, the deconvolution estimation determines $K$ as the solution of the discrete Wiener--Hopf (WH) equation of the first kind, 
             \begin{equation}
                \sum_{\tau'=0}^{\infty}C(|\tau-\tau'|)K(\tau')=S(\tau), \quad \tau\geq 0. \label{eq:WH}
             \end{equation}
         \end{subequations}
         The estimation is thus an inverse problem: $\mathcal{G}$ is never measured directly but reconstructed from the second-order statistics $C$ and $S$ alone. In particular, within the propagator model the ``bare'' permanent impact $G_\infty$ cannot be measured directly from the ``dressed'' response $R(\tau)$; in practice, it is estimated as the deconvolved $\mathcal{G}_\infty$. In this Letter, we focus on the asymptotic decay of the inferred kernel, characterized by the estimand $T_{\rm P}:=(\mathcal{G}_\infty,\beta)$ with $\mathcal{G}_\infty:=\lim_{\tau\to\infty}\mathcal{G}(\tau)$ and $\mathcal{G}(\tau)-\mathcal{G}_\infty\propto\tau^{-\beta}$ for large $\tau$ (see End Matter).

    \paragraph{Main theorems.}
         Here we present the universal decay theorems associated with the deconvolution estimation as our first main result: Let us assume that the price dynamics $m_t$ is diffusive, the market-order correlation $C(\tau)$ is completely monotone\footnote{A function is called completely monotone if it is represented as a nonnegative superposition of exponentials. This assumption is used to apply the Tauberian theorem by excluding oscillatory non-monotonic decays. We conjecture that our theorems can be proved for more general setups. 
         } with long memory, and the response increment $S(\tau)$ is monotonic for sufficiently large timelags and converges to zero:
         \begin{equation}
             \rmE[(m_t-m_0)^2]\propto t,\quad C(\tau)\propto \tau^{-\gamma}, \quad \lim_{\tau\to \infty}S(\tau)=0 \label{eq:assumptions}
         \end{equation}
         with $\gamma \in (0,1)$. No condition is imposed at small timelags, and $S(\tau)$ may be either increasing or decreasing. Importantly, we do not assume the propagator model for the price dynamics: any diffusive-price dynamics is acceptable including nonlinear dynamics, provided that our conditions are met.

         The content of these theorems can be stated in one sentence: although the deconvolution takes two inputs $C(\tau)$ and $S(\tau)$, the decay exponent of the inferred kernel is fixed by $C(\tau)$ alone. The statements (i)--(iv) make this precise.

         (i)~If $S(\tau)$ has a power-law asymptotic $S(\tau)\propto \tau^{\eta-1}$, the exponent $\eta$ is restricted by the diffusivity of price, 
         \begin{equation}
            \eta \leq \beta_c:=\frac{1-\gamma}{2}, \label{ineq:diffusivity-CS}
         \end{equation}
         which is shown by the Cauchy--Schwarz inequality [see End Matter]. For power-law responses, the diffusivity therefore confines the exponent to the two cases treated in (ii) and (iii).
        
         (ii)~If $S(\tau)$ is absolutely summable or has a power-law asymptotic with an exponent $\eta<\beta_c$, the inferred propagator is necessarily transient with an asymptotic form: 
         \begin{equation}
             \begin{gathered}
             \sum_{\tau=0}^\infty |S(\tau)|<\infty \>\mbox{ or }\>S(\tau)\propto \tau^{\eta-1} \>\mbox{ with }\>\eta<\beta_c \\
             \Longrightarrow \quad \mathcal{G}_{\infty} = 0, \quad \mathcal{G}(\tau) \mathrel{=} A_G\tau^{-\beta_c} + o(\tau^{-\beta_c})
             \end{gathered}
         \end{equation}
         with a coefficient $A_G$. At this stage, we do not rule out the possibility that $A_G$ might be zero. 
        
         (iii)~If $S(\tau)\propto \tau^{\eta-1}$ with the boundary exponent $\eta=\beta_c$, the inferred propagator is necessarily transient with an asymptotic form with a logarithmic correction and $A_G\neq 0$: 
         \begin{equation}
             \begin{gathered}
             S(\tau)\propto \tau^{{\eta-1}} \>\mbox{ with }\> \eta=\beta_c\\
             \Longrightarrow \quad \mathcal{G}_{\infty} = 0, \quad \mathcal{G}(\tau) \mathrel{=} A_G\tau^{-\beta_c}\ln \tau + o(\tau^{-\beta_c}\ln \tau).
             \end{gathered}
         \end{equation}
        
         (iv)~In case (ii), if $R(\tau)\geq 0$ and $R(\tau)$ is not identically zero, the coefficient $A_G$ is positive: 
         \begin{equation}
             R(\tau)\geq 0 \>\mbox{ and }\> R(\tau)\not \equiv 0 \quad \Longrightarrow \quad A_G>0.
         \end{equation}
        
         The statements (i)--(iii) imply that the price diffusivity forces the inferred kernel to be transient, $\mathcal{G}_{\infty}=0$, with the fine-tuned law fixing its decay rate: $\mathcal{G}(\tau)=O(\tau^{-\beta_c})$ in the interior of the diffusive regime $\eta<\beta_c$, and $O(\tau^{-\beta_c}\ln\tau)$ at its margin $\eta=\beta_c$. The statement (iv) provides a sufficient condition for the decay to be exactly fine-tuned. See Supplementary Material (SM) for their proofs. Under the standard LMF assumption of Pareto-distributed metaorder lengths, the complete monotonicity of $C(\tau)$ can be proved (see SM).
        
         The exponent is thus set by the order-flow statistics $C(\tau)$, which carry no information about the impact, while the response function---the only input through which the true impact enters---affects only the prefactor $A_G$ and, at the margin, a logarithmic factor. This asymptotic degeneracy is the origin of the non-identifiability.

\paragraph{SFK model as a null model without true impact decay.}
 Our theorems apply even to null models with strictly permanent impact and no decay. In this Letter, we study the SFK model as a minimal null model.

 The SFK model~\cite{SatoFujiwaraKanazawa2026,SatoFujiwaraKanazawa2026Full} was recently proposed as another solution to the diffusivity puzzle [see Fig.~\ref{fig:SFK} and End Matter]. This theory attributes the diffusivity of price to the square-root law (SRL)~\cite{SatoPRL2025}. Suppose a system composed of $M$ metaorder splitters with $M\geq 1$. They split metaorders of size $Q$ into a series of child orders, where the metaorder size distribution is assumed to be a power-law form $\psi(Q)\propto Q^{-\alpha-1}$ with $\alpha=\gamma+1\in (1,2)$ under the LMF hypothesis. The price impact accumulated after the first $q$ child orders of a metaorder is given by $I(q)\propto q^{\delta}$ with $0<\delta\leq1$, without decay after the peak impact. Thus, the price dynamics is given by 
 \begin{equation}
 m_t - m_0 = \sum_{i=1}^{M}\sum_{k=1}^{N^{(i)}(t)}\eps^{(i)}_{k}I\!\left(q^{(i)}_{k}(t)\right)
 \label{eq:SFK}
 \end{equation}
 with $\eps^{(i)}_{k}$ the sign of the $k$th metaorder of splitter $i$, $N^{(i)}(t)$ the number of metaorders initiated by splitter $i$ before time $t$, and $q^{(i)}_{k}(t)\in\{0,1,\dots,Q^{(i)}_{k}\}$ the number of its child orders submitted before time $t$. At each time step, a single splitter is drawn at random and submits one child order; $q^{(i)}_{k}$ is therefore incremented by one only when splitter $i$ is drawn, and stops at the metaorder size $Q^{(i)}_{k}$ upon completion. The contribution $\eps^{(i)}_{k}I(Q^{(i)}_{k})$ then remains in the sum forever: the impact of every metaorder is strictly permanent and never decays.
 
 The exact solution of the SFK model gives $\delta\leq 1/2$ as the necessary and sufficient condition for the price diffusivity for all $\alpha \in (1,2)$. Since previous empirical studies~\cite{SatoPRL2025} report the SRL $\delta=1/2$, the SFK model ensures the diffusivity of price under empirically relevant parameters. Extensions of the SFK model are also consistent with inverse-cubic return tails and volatility clustering~\cite{SatoFujiwaraKanazawa2026Full}. Thus, this model provides a simple solution to the diffusivity puzzle without fine-tuned decay and can be a suitable null model for this Letter\footnote{The generalized SFK model with post-metaorder impact decay (with a nonzero permanent component) remains in this class, and is solved exactly in Ref.~\cite{SatoFujiwaraKanazawa2026Full}.}.

 \begin{figure*}
    \centering
    \includegraphics[width=170mm]{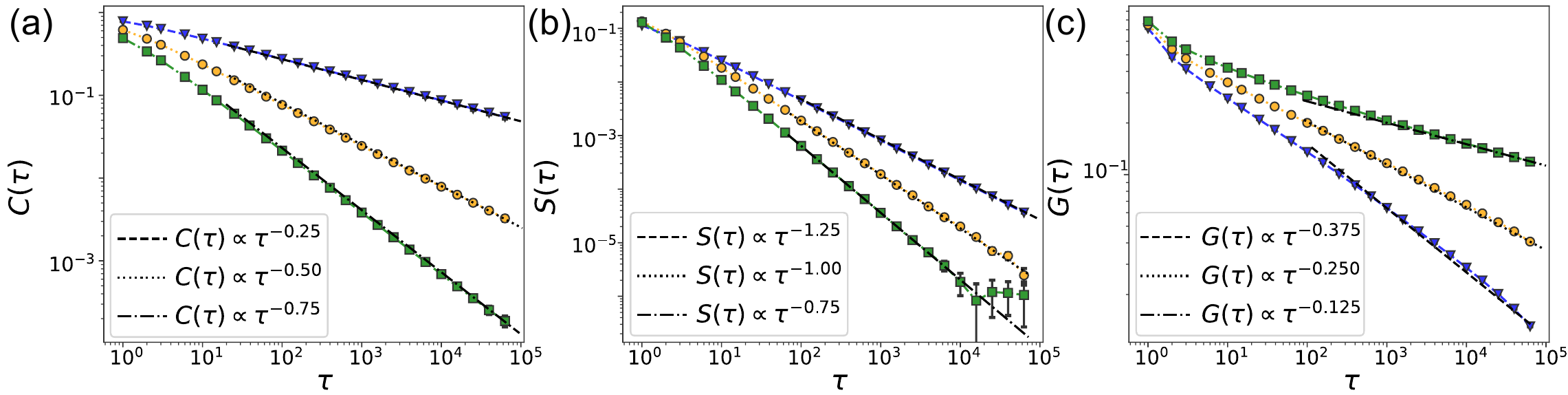}
    \caption{
        Linear propagator estimation applied to the SFK model with $M=1$, $\alpha\in \{1.25, 1.5, 1.75\}$, and $\delta=1/2$. 
        (a)~The order-sign autocorrelation, agreeing with the LMF prediction $C(\tau)\propto \tau^{-\gamma}$ with $\gamma=\alpha-1$. 
        (b)~The response increment has a power-law tail with $S(\tau)\propto\tau^{\delta-\alpha}$, agreeing with the SFK asymptotics.
        (c)~The inferred kernel $\mathcal{G}(\tau)$ exhibits transience $\lim_{\tau\to \infty}\mathcal{G}(\tau)=0$ and fine-tuned decay $\mathcal{G}(\tau)\propto \tau^{-\beta_c}$. Since the SFK model has no true decay, this apparent decay is an estimator artifact predicted by our universal decay theorems. 
    }
    \label{fig:SFK-linear-propagator}
 \end{figure*}
 While the SFK model~\eqref{eq:SFK} has no true decay, it satisfies the conditions (ii) and (iv) as shown in the SM for $\delta\leq 1/2$. As a corollary, the inferred propagator exhibits an artificial fine-tuned decay [see Fig.~\ref{fig:SFK}(c) for our analysis design]:
 \begin{equation}
 \mathcal{G}_{\infty}=0, \quad \mathcal{G}(\tau)=A_G\tau^{-\beta_c} + o(\tau^{-\beta_c}), \quad A_G > 0.
 \end{equation}
 Figure~\ref{fig:SFK-linear-propagator} shows a numerical validation of our prediction.

 The two competing pictures are separated not by the presence of impact decay but by whether the permanent component of impact vanishes: it is non-zero in the SFK class and it is zero in the fine-tuned transient picture. Our theorems force $\mathcal{G}_{\infty}=0$ in both classes, so the deconvolution cannot decide which class generated the data. Explicit construction of the null model and its numerical verification are the second main result of this Letter. 

 Within the linear propagator class alone, the deconvolution poses no identification problem: the procedure is consistent there, returning the true kernel that generated the data. The difficulty appears only once the model class is enlarged to include diffusive price processes outside the propagator class. Consider the candidate model set $\mathcal{M}=\{\mathcal{M}_{\mathrm{LP}},\mathcal{M}_{\mathrm{SFK}}\}$, composed of the transient linear propagator class $\mathcal{M}_{\mathrm{LP}}$ and the SFK class $\mathcal{M}_{\mathrm{SFK}}$. Our theorems map both to inferred kernels with $\mathcal{G}_{\infty}=0$ and the same decay exponent $\beta_c$, so the two classes are observationally equivalent with respect to the asymptotics of $\mathcal{G}$ (see End Matter). This is non-identifiability in the statistical sense. In other words, the deconvolution is fragile under misspecification of the model class: the inferred decay has not been subjected to a robustness check against generative models outside the propagator class, and our null model provides such a check. 

\paragraph{Nonlinear-propagator estimation.}
 \begin{figure*}
    \centering
    \includegraphics[width=170mm]{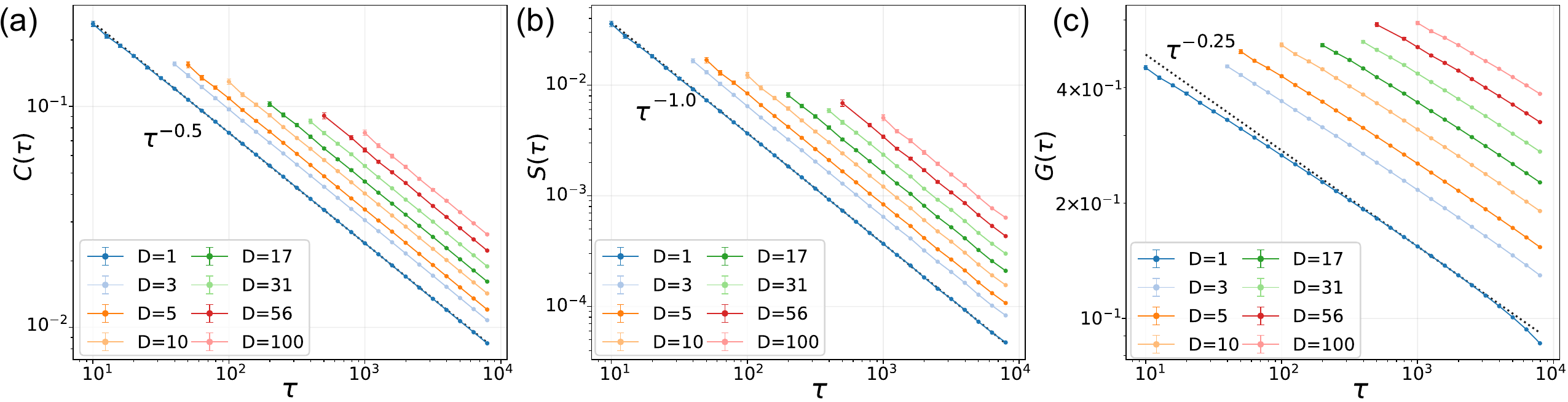}
    \caption{
        Nonlinear propagator estimation applied to the SFK model with $M=1$, $\alpha=1.5$, and $\delta=1/2$, following Ref.~\cite{Bucci2020}: the decay kernel is estimated from metaorder volumes aggregated over segments of $D$ steps, the simulation analogue of daily segmentation.
        (a)~The autocorrelation of the metaorder proxies, $C(\tau)\propto\tau^{-\gamma}$ with $\gamma\approx\alpha-1$.
        (b)~The response increment has a power-law tail $S(\tau)\propto\tau^{\delta-\alpha}$.
        (c)~The inferred kernel $\mathcal{G}(\tau)$ exhibits transient, fine-tuned decay $\mathcal{G}(\tau)\propto\tau^{-\beta_c}$ for every $D$, although the SFK model has no true decay.
    }
    \label{fig:SFK-nonlinear-propagator}
 \end{figure*}

 One might expect that enlarging the model class to nonlinear propagators would remove the misspecification and hence resolve the problem. However, nonlinear (more precisely, quasi-linear) propagator estimation~\cite{Brokmann2015,Bucci2020} does not escape our theorems, because it reduces to the same WH deconvolution after a change of variables. Thus, if true metaorders have long memory, identification fails again: since prices are empirically diffusive, the deconvolution forces the inferred kernel to be transient, even when the true dynamics lies outside the quasi-linear class and has strictly permanent impact.

 Even if true metaorders have short memory, a practical problem remains: the observational definition of metaorders is not unique, and a common choice is the daily-segmentation algorithm, in which market orders issued on different days are assigned to different metaorders. In this Letter, metaorders reconstructed by this daily-segmentation algorithm are called {\it metaorder proxies}, distinguished from true metaorders. 
 
 We find that this algorithm can undermine the identification of metaorder impact decay within the nonlinear propagator model. Indeed, this algorithm can create metaorder proxy correlation as an artifact of the algorithm: suppose a trader executes a single metaorder gradually over five days. After applying the daily-segmentation algorithm, five metaorder proxies are generated from a true single metaorder. These metaorder proxies are trivially correlated by construction, even if the true metaorder dynamics has no metaorder correlation at all. Our theorems should then apply with $C(\tau)$ now the metaorder-proxy autocorrelation.

 Let us go back to the SFK model as a concrete null model. In the SFK model, metaorder correlation is excluded by construction. Here, we apply the daily-segmentation algorithm to the metaorders generated by this null model to obtain the metaorder proxies. Even though the SFK model has no metaorder correlation, we find that the metaorder proxies have an artificial long memory: $C^{\rm SFK}_{\rm proxy}(\tau)\propto \tau^{-\gamma}$ with $\gamma \approx \alpha-1$. Then, if one applies the nonlinear propagator estimation based on metaorder proxies, the universal decay theorems should apply at the level of metaorder proxies given that price dynamics is diffusive. This is confirmed numerically in Fig.~\ref{fig:SFK-nonlinear-propagator}, based on the formulation proposed in Ref.~\cite{Bucci2020}. Daily segmentation creates correlated metaorder proxies whenever true metaorders span several days, a model-free mechanism; our argument therefore does not rely on the SFK model, and the numerical results above serve only as a demonstration.

\paragraph{Concluding discussion.}
 Deconvolution estimation has repeatedly returned transient, fine-tuned kernels, which have been read as a universal property of true impact decay. Earlier work explains this universality by invoking non-trivial mechanisms: transient impact finely balanced against order-flow memory~\cite{BFL_Book2009}, or, equivalently at the linear level, a history-dependent impact of the order-flow surprise~\cite{FarmerGerigLilloMike2006,BFL_Book2009,Taranto2018}. Our theorems show that, under diffusivity and long memory, the universality of these findings follows from the WH inversion itself, with no market mechanism required: the same exponent is returned for any diffusive price process under our conditions, including our nonlinear null model. The estimand $T_{\rm P}=(\mathcal{G}_{\infty},\beta)$ therefore identifies none of these mechanisms, nor impact decay itself.

 This suggests a diagnostic wherever a WH inversion is applied to long-memory data under a diffusivity constraint: compare the inferred exponent with the value that the input autocorrelation alone fixes. Agreement carries no information about the mechanism (see End Matter); it is then a prompt to test the estimation outside the fitted class. Under long memory, the fine-tuned decay is a property of the estimator before it is a finding about the dynamics. 
 
 Finally, our theorems concern the causal reading of the inferred kernel, not its predictive use: the kernel remains the best linear predictor of price changes from past order signs. Applications that rely on the causal reading---e.g., optimal execution, where the kernel is used to predict one's counterfactual impact---require additional identification assumptions. Methods that use the full distribution also lie outside the scope of our theorems; this encourages the use of observables beyond the second-order statistics $C(\tau)$ and $S(\tau)$ for identification.

\newpage

\subsection{Acknowledgements}

\begin{acknowledgements}
	YS was supported by JSPS KAKENHI (Grant Nos. 24KJ1328 and 26K24638). KK was supported by JSPS KAKENHI (Grant No. 25K01450).
\end{acknowledgements}

\subsection{Author contribution}
    SF and YS contributed equally to this work and share first authorship. SF proved the main theorems. YS numerically discovered the artificial decay in the propagator kernel of the SFK model. YS and SF conjectured that such decay is a generic feature of the Wiener--Hopf (WH) inversion. KK contributed to the proof of Theorem~(i) and supervised the overall project. All authors contributed to the writing and approved the final manuscript.

\newpage

\section{End Matter}

 \subsection{Statistical identification}
    We review statistical identification in statistics and econometrics~\cite{Lewbel2019}. Let us identify a property $\Delta(M)$ of a model $M$ by the estimand $T$. Let $P_M$ and $P_{M'}$ be the probability distributions for the models $M$ and $M'$, respectively. If $T(P_M)=T(P_{M'})$, the two models $M$ and $M'$ are observationally equivalent with respect to $T$. If $\Delta(M)\neq\Delta(M') \Longrightarrow T(P_M)\neq T(P_{M'})$ for all $M,M'$ in a candidate set $\mathcal{M}$, the property $\Delta$ is statistically identifiable by the estimand $T$. Taking $\Delta(M)=M$ and $T(P_M)=P_M$ recovers the standard notion of identification of the model from the full distribution. 

    In econophysics, ruling out artifacts of the estimation itself has long been a challenge, which can be naturally formulated as an identification problem. Let $P_{M_{\rm LP}}$ and $P_{M_{\rm SFK}}$ be the probability distributions for a model $M_{\rm LP}$ in the transient linear propagator class with fine-tuned decay and a model $M_{\rm SFK}$ in the SFK class, respectively, with the same $C(\tau)$. For the propagator estimation, we extract second-order statistics $C(\tau)$ and $S(\tau)$ from data. We obtain the inferred kernel $\mathcal{G}(\tau)$ by the WH inversion and focus on the propagator estimand $T_{\rm P}:=(\mathcal{G}_{\infty},\beta)$, where $\mathcal{G}_{\infty}:=\lim_{\tau\to\infty}\mathcal{G}(\tau)$ and $\mathcal{G}(\tau)-\mathcal{G}_{\infty}\propto\tau^{-\beta}$ for large $\tau$. Let $\Delta(M)=1$ if the true impact of $M$ is strictly transient and $\Delta(M)=0$ if it has a nonzero permanent impact. Since $T_{\rm P}(P_{M_{\rm LP}})= T_{\rm P}(P_{M_{\rm SFK}})$ although $\Delta(M_{\rm LP})\neq\Delta(M_{\rm SFK})$, the two models are observationally equivalent with respect to $T_{\rm P}$, and the transience of impact is not identifiable by the propagator estimand $T_{\rm P}$.  

    The constancy of $T_{\rm P}$ does not rely on the SFK model. Define $\mathcal{M}_{\rm D}$ as the set of all diffusive processes satisfying the conditions of (ii) and (iv). We prove that $T_{\rm P}(P_{M})=(0,\beta_c)$ for any $M\in\mathcal{M}_{\rm D}$, where $P_{M}$ is the corresponding probability distribution. This implies that the estimand $T_{\rm P}$ depends on $P_M$ only through $\gamma$ and carries no information beyond $\gamma$ within the diffusive class $\mathcal{M}_{\rm D}$: given $\gamma$, observing $T_{\rm P}$ does not update the posterior distribution of any property of $M$, for any prior on $\mathcal{M}_{\rm D}$. The SFK model shows that $\mathcal{M}_{\rm D}$ contains a model with strictly permanent impact. Hence, our theorems act as a no-go theorem for the causal reading of the inferred kernel through the estimand $T_{\rm P}$.

    A familiar analogy is the central limit theorem (CLT): the normal distribution is universal because the rescaled sum $X:=\sum_{i=1}^N (x_i-\mu)/\sqrt{N}$ of independent and identically distributed variables $x_i$ with mean $\mu$ and finite variance asymptotically follows the normal distribution. The asymptotic distribution of $X$ has no identifying power for the distribution of $x_i$ beyond its variance. Likewise, the asymptotic estimand $T_{\rm P}$ is fixed by $\gamma$ alone and has no identifying power for the true impact decay, because the WH inversion itself returns the fine-tuned transient kernel under our assumptions. In this sense, a universality theorem is the mirror image of a no-go theorem for identifying microscopic dynamics from the universal observables alone.

\subsection{Remark on the identification within the assumed model class}
    Suppose that the linear propagator model is independently known to be the true model. In this case, the WH deconvolution estimation returns the true kernel that generated the dataset, and the estimation is consistent. However, the WH deconvolution is unnecessary if one is interested only in the estimand $T_{\rm P}=(\mathcal{G}_{\infty},\beta)$ under the conditions of (ii) and (iv). Fine-tuned decay is mathematically necessary for price diffusivity in this class~\cite{BouchaudProp1}, so that $T_{\rm P}=(0,\beta_c)$ is known without the WH deconvolution. The WH deconvolution therefore adds no information about $T_{\rm P}$ under these conditions, even if the linear propagator class is the correct model. Its information lies instead in the prefactor $A_G$ and the kernel at finite lags. Thus, under our conditions, the WH estimation provides no additional information about the asymptotic estimand $T_{\rm P}$, whether or not the true model is known.  

 \subsection{Proof of the inequality~\eqref{ineq:diffusivity-CS}}
    Let us prove the diffusive constraint~\eqref{ineq:diffusivity-CS} by the Cauchy--Schwarz inequality.
    We define the price increment by $r_t:=m_{t+1}-m_t$ and assume that the pair $(\epsilon_t,r_t)$ is jointly second-order stationary with zero means ${\rm{E}}[\epsilon_t]={\rm{E}}[r_t]=0$. The response increment is equal to the sign--return covariance,
    \begin{equation}
        S(n)=R(n+1)-R(n)=\operatorname{Cov}(\eps_0,r_n).
    \end{equation}
    Suppose $C(n)\sim A_Cn^{-\gamma}$ with $A_C>0$ and $\gamma \in (0,1)$, $S(n)\sim A_Sn^{\eta-1}$ with $A_S\neq0$, and $\operatorname{Var}(m_T-m_0)\sim A_DT$ with $A_D \in (0,\infty)$. Let us define $X_T$ and $Y_T$ as the past order-flow imbalance and the subsequent price change,
    \begin{equation}
        X_T:=\sum_{a=1}^{T}\eps_{-a},
        \quad Y_T:=m_T-m_0=\sum_{j=0}^{T-1}r_j.
    \end{equation}
    Due to the stationarity of the process, we obtain the following asymptotics:
    \begin{equation}
        \begin{aligned}
            \frac{\operatorname{Var}(X_T)}{T}
            &=C(0)+2\sum_{k=1}^{T-1}\frac{T-k}{T}C(k)
            \sim\frac{2A_CT^{1-\gamma}}{(1-\gamma)(2-\gamma)},\\
            \operatorname{Var}(Y_T)&\sim A_DT,
        \end{aligned}
    \end{equation}
    where we used $\int_0^1(1-x)x^{-\gamma}\dd x=[(1-\gamma)(2-\gamma)]^{-1}$. We can rewrite the cross covariance as 
    \begin{gather}
            \operatorname{Cov}(X_T,Y_T)
            =\sum_{a=1}^{T}\sum_{j=0}^{T-1}S(j+a)
            =\sum_{n=1}^{2T-1}w_T(n)S(n), \notag \\ 
        w_T(n)=\begin{cases}
            n,&1\leq n\leq T,\\
            2T-n,&T<n<2T.
        \end{cases}
    \end{gather}
    If $\eta\leq-1$, the desired bound is automatically satisfied. For $\eta>-1$, the cross covariance can be evaluated as 
    \begin{equation}
        \begin{gathered}
            \operatorname{Cov}(X_T,Y_T)\sim A_SJ_\eta T^{1+\eta},\\
            J_\eta:=\int_0^1x^\eta\dd x
            +\int_1^2(2-x)x^{\eta-1}\dd x>0,
        \end{gathered}
    \end{equation}
    where the integral is finite for $\eta>-1$. Here, the Cauchy--Schwarz inequality constrains the exponent $\eta$ as follows:
    \begin{equation}
        \begin{aligned}
            |\operatorname{Cov}(X_T,Y_T)|^2
            &\leq\operatorname{Var}(X_T)\operatorname{Var}(Y_T)\\
            &\sim\frac{2A_CA_D}{(1-\gamma)(2-\gamma)}T^{3-\gamma}.
        \end{aligned}
    \end{equation}
    We obtain $2+2\eta\leq3-\gamma$, or $\eta\leq(1-\gamma)/2$, proving Eq.~\eqref{ineq:diffusivity-CS}.

 \subsection{Brief review of the Sato--Fujiwara--Kanazawa model}
    The SFK model is a recent model to solve the diffusivity puzzle without assuming the fine-tuned decay mechanism~\cite{SatoFujiwaraKanazawa2026,SatoFujiwaraKanazawa2026Full}. This model can be solved exactly within the continuous-time random-walk framework, particularly from the viewpoint of the nonlinear L\'evy-walk model. The exact solution yields $\delta < \alpha/2$ as the necessary and sufficient condition for the diffusivity. Since $\delta=1/2$ and $\alpha\in (1,2)$, this model ensures diffusivity of price under the empirically validated parameter range. Also, extensions of this model are consistent with the inverse-cubic law of returns and the volatility clustering~\cite{SatoFujiwaraKanazawa2026Full}.

    Heuristically, the SFK theory attributes the price diffusivity to the finite-variance condition of metaorder impact under nonlinear price impact. The variance of a metaorder impact follows 
    \begin{equation}
        \rmE[I^2(Q)]=\int_{1}^\infty I^2(Q)\psi(Q)\dd Q\propto \int_{1}^\infty Q^{2\delta-\alpha-1}\dd Q.
    \end{equation}
    The necessary and sufficient condition for the variance to be finite $\rmE[I^2(Q)]<\infty$ is given by $\delta < \alpha/2$. When the variance of a single metaorder impact is finite, the process is expected to converge to Brownian motion in the long run. For the finite-variance condition to be met for every admissible $\alpha$, $\delta \leq 1/2$ is the necessary and sufficient condition for the price diffusivity. The L\'evy-walk theory supports this reasoning with an exact solution. Since previous empirical studies report the SRL $\delta=1/2$~\cite{BouchaudText,SatoPRL2025}, the SFK model ensures the diffusivity of price under empirically relevant parameters. 

    Any process in which LMF metaorders add independent, strictly permanent impacts $I(Q)\propto Q^{\delta}$ is expected to remain diffusive only if $\delta<\alpha/2$, so that the impact is concave; the SFK model is the minimal representative of this class rather than a particular choice.

 \subsection{Nonlinear propagator estimation and daily segmentation}
    The nonlinear propagator estimation~\cite{Brokmann2015,Bucci2020} is a natural extension of the linear propagator estimation~\cite{BouchaudProp1,BouchaudProp2} to nonlinear price impact at the level of metaorders or aggregated volume imbalance. In particular, Ref.~\cite{Bucci2020} is based on the volume imbalance of metaorders defined within a single trading day, that is, of metaorder proxies obtained by daily segmentation, which is also used for the observation of nonlinear price impact\footnote{Our previous report~\cite{SatoPRL2025} also relies on daily segmentation. Since the SRL is measured directly rather than through a deconvolution, our theorems do not apply to it; the remaining question is whether daily segmentation biases the estimated exponent $\delta$---an ordinary estimation bias, not a failure of identification. Fig.~S16 of its SM tests this as a robustness check: the exponent changes little between the shortest and longest executions, although, under a constant execution rate~\cite{BouchaudText}, the former are rarely sliced and the latter are the most exposed.}. 
    
    The formulation based on Ref.~\cite{Bucci2020} is given as follows: $D$ is the segment length in simulation steps, the analogue of one trading day in Ref.~\cite{Bucci2020}. Let $Q_k^{(D)}$ be the metaorder volume imbalance in the $k$th segment. The return is described by the following quasi-linear propagator model\footnote{We omit the market factor and the normalizations of Ref.~\cite{Bucci2020} because the SFK model has a single asset with constant volume and volatility. Also, Ref.~\cite{Bucci2020} truncates the sum at $50$ days.}:
    \begin{equation}
        r_k^{(D)} \!=\! \sum_{l=0}^{\infty}\mathcal{K}^{(D)}_l\Psi^{(D)}_{k-l} + \xi_k^{(D)}, \hspace{0.5em}
        \Psi^{(D)}_{k}\!:=\!\eps_k^{(D)}|Q_{k}^{(D)}|^{1/2},
    \end{equation}
    where $r_k^{(D)}:=m_{(k+1)D}-m_{kD}$ is the $D$-step return, $\eps_k^{(D)}:= {\rm sgn}(Q_{k}^{(D)})$ and $\xi_k^{(D)}$ is the noise term. The inferred propagator kernel $\mathcal{G}^{(D)}_k$ is obtained from the autocorrelation $\mathcal{C}^{(D)}_k$ of the signed square-root volume imbalance $\Psi^{(D)}_{k}$ and the response increment $\mathcal{S}^{(D)}_k$ by the WH deconvolution:
    \begin{equation}
        \begin{gathered}
        \mathcal{S}^{(D)}_k = \sum_{l=0}^{\infty}\mathcal{K}^{(D)}_l\mathcal{C}^{(D)}_{|k-l|}, \quad \mathcal{G}^{(D)}_k:=\sum_ {l=0}^{k-1}\mathcal{K}^{(D)}_l,\\ 
        \mathcal{S}^{(D)}_k:=\rmE[r^{(D)}_{l}\Psi^{(D)}_{l-k}],\quad 
        \mathcal{C}^{(D)}_k:=\rmE[\Psi^{(D)}_{l}\Psi^{(D)}_{l-k}]
        \end{gathered}
    \end{equation}
    with $k\geq 0$. Thus, the quasi-linear propagator estimation reduces to the WH deconvolution.
    
    For the SFK model, we observe a power-law asymptotic behavior $\mathcal{C}^{(D)}_k \propto k^{-\gamma}$ with $\gamma \approx \alpha-1$ for various values of $D$. This apparent metaorder-proxy correlation is a statistical artifact due to the daily-segmentation algorithm, since the SFK model has no true metaorder correlation. Thus, the WH deconvolution is expected to return a transient, fine-tuned kernel $\mathcal{G}^{(D)}_k\propto k^{-\beta_c}$ for large lag $k\gg 1$. This expectation is numerically confirmed in Fig.~\ref{fig:SFK-nonlinear-propagator}, where we plot $\mathcal{C}^{(D)}_k$, $\mathcal{S}^{(D)}_k$, and $\mathcal{G}^{(D)}_k$ against $\tau=kD$, denoted by $C(\tau)$, $S(\tau)$, and $\mathcal{G}(\tau)$ in the figure\footnote{
        Ref.~\cite{Brokmann2015} performed a robustness check by two-day aggregation against apparent decay due to concave impact fitted by a quasi-linear model. In the SFK model, however, the inferred decay persists for all tested $D$ in Fig.~\ref{fig:SFK-nonlinear-propagator}, showing that such a check cannot rule out an artificial decay.
        }. 
\end{document}


\title{Supplementary Material: \\Is price-impact decay statistically identifiable by propagator estimation?\\ Diffusive prices force the inferred kernel to be transient and fine-tuned}
\author{Shunta Fujiwara$^{1}$}
\author{Yuki Sato$^{2}$}
\author{Kiyoshi Kanazawa$^{1}$}
\affiliation{$^{1}$Department of Physics, Graduate School of Science, Kyoto University, Kyoto 606-8502, Japan}
\affiliation{$^{2}$Institute of Systems and Information Engineering, University of Tsukuba, Ibaraki 305-8573, Japan}
\date{\today}

\maketitle

\tableofcontents

\clearpage

\part[Mathematical proofs for the discrete-time model]{{\Large Mathematical proofs for the discrete-time model}}

\section{Mathematical notation and related theorems} 
    \subsection{Conventions}
    \label{subsec:conventions}
        Throughout this Supplementary Material, we adopt the convention that a function written with the argument $e^{i\theta}$ or $z$ is a generating function, and that the corresponding Fourier coefficients or Taylor coefficients\footnote{
            Here, we consider the $n$th Taylor coefficient to include the factor $1/n!$.
        } are denoted with the subscript $n$. For example, the $n$th Fourier coefficient of $f(e^{i\theta})$ is denoted by $f_n$,
        \begin{equation}
            f(e^{i\theta})=\sum_{n=-\infty}^{\infty}f_n e^{in\theta},\qquad
            f_n:=\frac{1}{2\pi}\int_{-\pi}^{\pi}f(e^{i\theta})e^{-in\theta}\dd \theta,
        \end{equation}
        and the $n$th Taylor coefficient of $f(z)$ is denoted by $f_n$,
        \begin{equation}
            f(z)=\sum_{n=0}^{\infty}f_n z^n,\qquad
            f_n:=\frac{1}{n!}\frac{\dd^n f}{\dd z^n}(0).
        \end{equation}

    \subsection{Special functions}
    \label{subsec:SpecialFunctions}
        The gamma function $\Gamma(z)$ is defined for $z\in\mathbb{C}$ with $\mathrm{Re}[z]>0$ by
        \begin{equation}
            \Gamma(z):=\int_{0}^{\infty}\dd t t^{z-1}e^{-t},
        \end{equation}
        and the beta function $B(x,y)$ is defined for $x\in\mathbb{C}$ with $\mathrm{Re}[x]>0$ and $y\in\mathbb{C}$ with $\mathrm{Re}[y]>0$ by
        \begin{equation}
            B(x,y):=\int_{0}^{1}\dd t t^{x-1}(1-t)^{y-1}=\frac{\Gamma(x)\Gamma(y)}{\Gamma(x+y)}.
        \end{equation}

    \subsection{Hardy spaces and Riesz projections}
        The unit circle $\mathbb{T}$ and the open unit disk $\mathbb{D}$ are defined by 
        \begin{equation}
            \mathbb{T}:=\{z\in\mathbb{C}:|z|=1\},
            \qquad
            \mathbb{D}:=\{z\in\mathbb{C}:|z|<1\},
        \end{equation}
        respectively.
        Throughout, functions on $\mathbb{T}$ are assumed to be measurable, and functions that coincide almost everywhere on $\mathbb{T}$ are identified.
        We define the space $L^p(\mathbb{T})$ and its norm $\|\cdot\|_{p}$ for $1\leq p<\infty$ by 
        \begin{equation}
            L^p(\mathbb{T}):=\left\{f:\int_{-\pi}^{\pi}\dd \theta |f(e^{i\theta})|^p<\infty\right\},
            \qquad
            \|f\|_{p}:=\left(\frac{1}{2\pi}\int_{-\pi}^{\pi}\dd \theta |f(e^{i\theta})|^p\right)^{\frac1p},
        \end{equation}
        respectively. For $p=\infty$, we define the space $L^{\infty}(\mathbb{T})$ and its norm $\|\cdot\|_{\infty}$ by
        \begin{equation}
            L^\infty(\mathbb{T})
            := \left\{ f: \esssup_{-\pi\leq\theta<\pi} |f(e^{i\theta})|<\infty \right\},\qquad
            \|f\|_{\infty}
            :=\esssup_{-\pi\leq\theta<\pi}
            |f(e^{i\theta})|,
        \end{equation}
        respectively.

        We define the Hardy space $H^2(\mathbb{T})$ on $\mathbb{T}$ in terms of the Fourier coefficients $f_n$ by 
        \begin{align}
            H^2(\mathbb{T}):=\left\{f\in L^2(\mathbb{T}):f_n=0\quad(n<0)\right\},
        \end{align}
        where the Fourier coefficients $f_n$ are defined by
        \footnote{
            Theorem~\ref{thm:Hardy_boundary_value} below allows us to identify $f$ with a function in $H^2(\mathbb{D})$ and expand it as
            \begin{equation}
                f(z)=\sum_{n=0}^{\infty}f_n z^n,
            \end{equation}
            and this expression is consistent with the convention in Sec.~\ref{subsec:conventions}.
        }
        \begin{equation}
            f_n:=\frac{1}{2\pi}\int_{-\pi}^{\pi}\dd \theta f(e^{i\theta})e^{-in\theta}.
        \end{equation}
        The space $H^2(\mathbb{T})$ can also be written as
        \begin{equation}
            H^2(\mathbb{T})=\left\{\sum_{n=0}^{\infty}f_n e^{in\theta}: \sum_{n=0}^{\infty}\left|f_n\right|^2<\infty\right\}.
        \end{equation}
        In addition, the space $H^{2}_{-}(\mathbb{T})$ of functions having only Fourier coefficients of negative index is defined by
        \begin{align}
            H^{2}_{-}(\mathbb{T}):=\left\{f\in L^2(\mathbb{T}):f_n=0\quad(n\geq0)\right\}.
        \end{align}
        We define the Riesz projections $P^{+}$ and $P^{-}$ on $L^2(\mathbb{T})$ by 
        \begin{align}
            (P^{+}f)(e^{i\theta}):=\sum_{n=0}^{\infty}f_n e^{in\theta},\qquad
            (P^{-}f)(e^{i\theta}):=\sum_{n=-\infty}^{-1}f_n e^{in\theta},
        \end{align}
        where the series converge in $L^2(\mathbb{T})$.

        Let us define the Hardy space $H^2(\mathbb{D})$ in $\mathbb{D}$ while $H^2(\mathbb{T})$ is a space of functions on $\mathbb{T}$. We define the space $\mathrm{Hol}(\mathbb{D})$ of holomorphic functions in $\mathbb{D}$ by
        \begin{equation}  
            \mathrm{Hol}(\mathbb{D}):=\left\{f:f\text{ is defined and holomorphic in }\mathbb{D}\right\}.
        \end{equation}
        Note that the holomorphy of $f\in\mathrm{Hol}(\mathbb{D})$ allows the Taylor expansion of $f$.
        For $f\in\mathrm{Hol}(\mathbb{D})$ and $0<r<1$, we define $f_{(r)}$ by
        \begin{equation}
            f_{(r)}(z):=f(rz),\quad |z|<\frac{1}{r}.
        \end{equation}
        This definition is motivated by our aim to consider the radial limit $r\uparrow1$.
        Using $f_{(r)}$, we define the Hardy space $H^2(\mathbb{D})$ in $\mathbb{D}$ by 
        \begin{equation}
            H^2(\mathbb{D}):=\left\{f\in\mathrm{Hol}(\mathbb{D}): \sup_{0<r<1}\|f_{(r)}\|_{2}<\infty\right\}.
        \end{equation}

    \subsection{The $\ell^p$ spaces of complex sequences}
        We define the space $\ell^p$ for $1\leq p<\infty$ by 
        \begin{equation}
            \ell^p:=\left\{\left(a_n\right)_{n\geq0}:\sum_{n=0}^{\infty}|a_n|^p<\infty\right\},
        \end{equation}
        where $\left(a_n\right)_{n\geq0}$ is a complex sequence. For $p=\infty$, we define $\ell^{\infty}$ by 
        \begin{equation}
            \ell^\infty:=\left\{(a_n)_{n\geq0}:\sup_{n\geq0}|a_n|<\infty\right\}.
        \end{equation}
        Unless otherwise specified, the $\ell^{p}$ spaces consist of sequences indexed by $n\in\mathbb{Z}_{\geq0}$, and $n$ in expressions such as $n\geq0$ denotes an integer.

    \subsection{Boundary value theorem in the Hardy space}
        One motivation for introducing the Hardy spaces is that the boundary value on $\mathbb{T}$ of a function $f\in H^2(\mathbb{D})$ is naturally defined, and, conversely, a function $f\in H^2(\mathbb{D})$ is uniquely obtained from $f \in H^2(\mathbb{T})$. More precisely, the following theorem holds~(see~Theorem~2.2.2 and Convention~2.2.3 in Nikolski~\cite{Nikolski2019}).
        \begin{theorem}
        \label{thm:Hardy_boundary_value}
            For any $f\in H^2(\mathbb{D})$, the boundary value $\lim_{r\uparrow 1}f_{(r)}(e^{i\theta})$ exists in the norm of $L^2(\mathbb{T})$, and this boundary value belongs to $H^2(\mathbb{T})$:
            \begin{equation}
                bf := \lim_{r\uparrow 1}f_{(r)}\mid_{\mathbb{T}} \in H^2(\mathbb{T}).
            \end{equation}
            Moreover, the map $f\mapsto bf$ is an isometric bijection from $H^2(\mathbb{D})$ to $H^2(\mathbb{T})$.
        \end{theorem}
        Thus, we can identify $f\in H^2(\mathbb{D})$ with its boundary value $bf\in H^2(\mathbb{T})$. The two functions are expressed by the Taylor expansion in $\mathbb{D}$ and the Fourier expansion on $\mathbb{T}$, respectively, and the coefficients of these expansions are identical. Since this correspondence holds in the sense of $L^2(\mathbb{T})$ convergence, we understand the correspondence between $f\in H^2(\mathbb{D})$ and $bf\in H^2(\mathbb{T})$ in this sense unless otherwise stated. Moreover, this correspondence also holds pointwise almost everywhere, as stated in the following corollary~(see~Corollary~2.5.2 in Nikolski~\cite{Nikolski2019}).
        \begin{corollary}
        \label{cor:Hardy_boundary_value}
            For any $f\in H^2(\mathbb{D})$, its boundary value exists almost everywhere on $\mathbb{T}$ as a non-tangential limit:
            \begin{equation}
                \lim_{z\to z_0}f(z) = f(z_0) = (bf)(z_0)\quad\hbox{for a.e. } z_0 \in \mathbb{T}.
            \end{equation}
        \end{corollary}
        When we use this pointwise convergence, we state it explicitly each time.

        The following theorem shows that $P^{+}$ and $P^{-}$ are indeed orthogonal projections (see Exercise~2.8.4~(e) in Nikolski~\cite{Nikolski2019})\footnote{
            This theorem also follows from the orthogonality of the exponentials $e^{in\theta}$ and Parseval's identity.
        }.
        \begin{theorem}
        \label{thm:Riesz_projection}
            The maps $P^{+}$ and $P^{-}$ are the orthogonal projections of $L^2(\mathbb{T})$ onto $H^2(\mathbb{T})$ and $H^2_{-}(\mathbb{T})$, respectively. In particular, $P^{+}f\in H^2(\mathbb{T})$ and $P^{-}f\in H^2_{-}(\mathbb{T})$ for any $f\in L^2(\mathbb{T})$, and 
            \begin{equation}
                P^{+}L^2(\mathbb{T}) = H^2(\mathbb{T}), \quad P^{-}L^2(\mathbb{T}) = H^2_{-}(\mathbb{T}), \quad L^2(\mathbb{T}) = H^2(\mathbb{T}) \oplus H^2_{-}(\mathbb{T}).
            \end{equation}
        \end{theorem}

\section{Universal decay theorems}
\label{sec:universal-decay-theorems}
    \subsection{Assumptions}
    \label{subsec:Assumptions}
        For $0<\gamma<1$, we define the parameter $\beta_c$ by
        \begin{equation}
        \label{eq:Assumption_parameter}
            \beta_c=\frac{1-\gamma}{2}.
        \end{equation}
        For the response function $R_n$, its increment is defined by
        \begin{equation}
        \label{eq:Assumption_R}
            S_n:=R_{n+1}-R_n,
        \end{equation}
        where we set $R_0=0$.

        \begin{assumption}[Autocorrelation function]
        \label{assumption:ACF}
            For the autocorrelation function~(ACF) $C_n$, we assume that the sequence $(C_n)_{n\geq0}$ is completely monotone\footnote{
                A sequence $(a_n)_{n\geq0}$ is called completely monotone if it is represented as $a_n=\int_{[0,1]}r^n\sigma(dr)$ for $n\geq0$ with a finite nonnegative Borel measure $\sigma$ on $[0,1]$, where we take the convention that $0^0=1$. 
            } and satisfies
            \begin{equation}
                C_n\sim \mathfrak{A}_Cn^{-\gamma}, \qquad \mathfrak{A}_C>0 \qquad \text{as $n\to\infty$}.
            \end{equation}
            Note that $C_{-n}=C_n$.
        \end{assumption}

        \begin{assumption}[Response increment]
        \label{assumption:response}
            For the response increment $S_n$, we assume that $(S_n)_{n\geq0}$ is eventually monotone, and that either of the following holds:
            \begin{equation}
                (S_n)_{n\geq0}\in\ell^1,
            \end{equation}
            or, for some $\eta\leq\beta_c$ and $\mathfrak{A}_S\neq0$,
            \begin{equation}
                S_n\sim \mathfrak{A}_Sn^{\eta-1} \qquad \text{as $n\to\infty$}.
            \end{equation}
            In the proofs in this Supplementary Material, we assume $\mathfrak{A}_S>0$. For the case of $\mathfrak{A}_S<0$, by replacing $S_n$ with $-S_n$, we can reduce the proof to the case of $\mathfrak{A}_S>0$.
        \end{assumption}

        \begin{assumption}[Positivity]
        \label{assumption:positivity}
            For the response function $R_n$, we assume that 
            \begin{equation}
                R_n\geq0\quad(n\geq0) \quad\text{and}\quad R_n\not\equiv0.
            \end{equation}
        \end{assumption}

    \subsection{Main theorem}
    \label{subsec:Results}
        \begin{theorem}[Universal decay theorems]
        \label{thm:universal-decay}
            Under Assumptions~\ref{assumption:ACF} and \ref{assumption:response}, the discrete Wiener--Hopf equation
            \begin{equation}
                S_n=\sum_{k=0}^{\infty}C_{|n-k|}K_{k}, \qquad (n\geq0)
            \end{equation}
            has a unique solution $(K_n)_{n\geq0}$ in $\ell^1$.
            For this solution, we define 
            \begin{equation}
                \mathcal G_0:=0,\qquad \mathcal{G}_n:=\sum_{k=0}^{n-1} K_{k} \quad (n\geq1).
            \end{equation}
            Then, the following results hold.

            \begin{enumerate}
                \item[(ii)]
                Suppose that $(S_n)_{n\geq0}\in\ell^1$ or
                \begin{equation}
                    S_n\sim \mathfrak{A}_Sn^{\eta-1},\qquad \eta<\beta_c.
                \end{equation}
                Then, there exists $\mathfrak{A}_{\mathcal{G}}\in\mathbb{R}$ such that  
                \begin{equation}
                    \mathcal{G}_n
                    =\mathfrak{A}_{\mathcal{G}}n^{-\beta_c}+o(n^{-\beta_c}) \quad\text{as $n\to\infty$}.
                \end{equation}
                In particular, we have
                \begin{equation}
                    \mathcal{G}_\infty:=\lim_{n\to\infty}\mathcal{G}_n=0.
                \end{equation}

                \item[(iii)]
                Suppose that 
                \begin{equation}
                    S_n\sim \mathfrak{A}_Sn^{\eta-1}, \quad \eta=\beta_c.
                \end{equation}
                Then, we have
                \begin{equation}
                    \mathcal{G}_n
                    =\mathfrak{A}_{\mathcal{G}} n^{-\beta_c}\ln n +o(n^{-\beta_c}\ln n) \quad\text{as $n\to\infty$},
                \end{equation}
                where
                \begin{equation}
                    \mathfrak{A}_{\mathcal{G}}
                    =B(\beta_c,\gamma)\left(\frac{\sin(\pi\beta_c)}{\pi}\right)^2
                    \frac{\mathfrak{A}_S}{\mathfrak{A}_C}\neq0.
                \end{equation}
                In particular, we have
                \begin{equation}
                    \mathcal{G}_\infty:=\lim_{n\to\infty}\mathcal{G}_n=0.
                \end{equation}

                \item[(iv)]
                In addition to the condition of (ii), suppose that Assumption~\ref{assumption:positivity} holds.
                Then, we have 
                \begin{equation}
                    \mathfrak{A}_{\mathcal{G}}>0.
                \end{equation}
            \end{enumerate}
        \end{theorem}

\section{Proofs of the universal decay theorems~\NoCaseChange{(ii)} and \NoCaseChange{(iv)}}
\label{sec:UDT}
    The proof consists of the following three steps. First, we construct a candidate for the solution of the discrete Wiener--Hopf equation $S_n=\sum_{k=0}^{\infty}C_{|n-k|}K_{k}\,\, (n\geq0)$, and by substituting this candidate into the equation, we confirm that this candidate is indeed a solution. Second, we show that the constructed solution is unique in $\ell^1$. Finally, we conduct the asymptotic analysis of the unique solution and show the positivity of its asymptotic coefficient. Here, an explicit formula for the solution is given by    
    \begin{equation}
        K^+(z) = \frac{1}{F^+(z)}P^{+}\left[\frac{S^{+}(z)}{F^{-}(z)}\right],
    \end{equation}
    where $K^{+}(z)$ is the generating function of $K_n$ defined by $K^+(z):=\sum_{n=0}^\infty K_nz^n$, $\mathrm{PSD}(\theta)$ is the power spectral density~(PSD) of $C_n$, $F^{+}(z)$ is the outer function in the Szeg\H{o} factorization of $\mathrm{PSD}(\theta)$, and $F^{-}(z)$ is the function defined from $F^{+}(z)$ below. $P^{+}$ is the Riesz projection, and $S^+(z)$ is the generating function of $S_n$. Specifically, these functions are defined by 
    \begin{equation}
        C_n=\frac{1}{2\pi}\int_{-\pi}^{\pi}\dd \theta \mathrm{PSD}(\theta)e^{-in\theta}, \quad \mathrm{PSD}(\theta) = F^{+}(e^{i\theta})F^{-}(e^{i\theta}), \quad F^{-}(z) = \overline{F^{+}\left(\frac{1}{\overline{z}}\right)}, \quad S^+(z) = \sum_{n=0}^{\infty}S_n z^n.
    \end{equation}
    Instead of relying on a formal derivation at the theoretical-physics level, we rigorously prove properties of these functions, such as well-definedness and convergence. To this end, we use several theorems in Nikolski~\cite{Nikolski2019} and Inoue~\cite{Inoue2000}.

    \subsection{Spectral representation of the ACF}
        We can express $C_n$ in terms of $\mathrm{PSD}(\theta)$ as\footnote{
            Since $C_n$ is completely monotone and $C_n\to0$ as $n\to\infty$, a stationary process with the ACF $C_n$ has the property of reflection positivity, and thus it is purely nondeterministic~(see~p.~106 in Inoue~\cite{Inoue2000}). Therefore, there exists a positive even and integrable function $\mathrm{PSD}(\theta)$ on $(-\pi, \pi)$ such that Eq.~\eqref{eq:PSD} holds~(see~p.~68 in Inoue~\cite{Inoue2000}).
        }
        \begin{equation}
            \label{eq:PSD}
            C_n=\frac{1}{2\pi}\int_{-\pi}^{\pi}\dd \theta\, \mathrm{PSD}(\theta)e^{-in\theta},\quad 
            \int_{-\pi}^{\pi}\dd \theta\,\left| \ln\mathrm{PSD}(\theta)\right|<\infty, \quad
            \int_{-\pi}^{\pi}\dd \theta\, |\mathrm{PSD}(\theta)|<\infty, \quad \mathrm{PSD}(\theta)>0 \hbox{ for a.e. } \theta\in(-\pi,\pi),
        \end{equation}
        where $\mathrm{PSD}(\theta)$ is a positive even and integrable function on $(-\pi, \pi)$.

    \subsection{Szeg\H{o} factorization}
        We define the outer function $F^{+}(z)$ of $\mathrm{PSD}(\theta)$ in $\mathbb{D}$ by 
        \begin{equation}
            F^{+}(z):=\exp\left[\frac{1}{4\pi}\int_{-\pi}^{\pi}\dd \theta\frac{e^{i\theta}+z}{e^{i\theta}-z}\ln \mathrm{PSD}(\theta)\right]\quad(z\in\mathbb{D}).
        \end{equation}
        Since $F^{+}(z)$ is holomorphic\footnote{
                The holomorphy of $F^{+}(z)$ is shown on p.~68 in Inoue~\cite{Inoue2000}; for more details, see Theorem~2.6.1 in Nikolski~\cite{Nikolski2019}.
            }
        for $z\in\mathbb{D}$,
        $F^{+}(z)$ can be expanded in a Taylor series in $\mathbb{D}$ as
        \begin{equation}
            F^{+}(z)=\sum_{n=0}^{\infty}F_n z^n \quad(z\in\mathbb{D}),
        \end{equation}
        where $F_n$ is the $n$th Taylor coefficient of $F^{+}(z)$, consistent with the convention in Sec.~\ref{subsec:conventions}.
        The coefficient $F_n$ is real, and $\left(F_n\right)_{n\geq 0}\in \ell^2$~(see~p.~69 in Inoue~\cite{Inoue2000}). Hence, the series $\sum_{n=0}^{\infty}F_n e^{in\theta}$ converges in $L^2(\mathbb{T})$, and its sum belongs to $H^2(\mathbb{T})$. By Theorem~\ref{thm:Hardy_boundary_value}, this sum is equal to the $L^2(\mathbb{T})$ boundary value of a unique function in $H^2(\mathbb{D})$ whose Taylor coefficients are $F_n$. This function is $F^{+}(z)$, and hence $F^{+}(z)\in H^2(\mathbb{D})$.
        Moreover, $F^{+}(z)$ does not have zeros in $\mathbb{D}$~(see~p.~69 in Inoue~\cite{Inoue2000})\footnote{
            This property also follows from the definition of $F^{+}(z)$ and the properties of $\mathrm{PSD}(\theta)$.
        }.
        Therefore, we can define $J^{+}(z):=1/F^{+}(z)$, which is holomorphic in $\mathbb{D}$ and can be expanded in a Taylor series in $\mathbb{D}$ as
        \begin{equation}
            J^{+}(z)=\frac{1}{F^{+}(z)}=\sum_{n=0}^{\infty}J_n z^n \quad(z\in\mathbb{D}),
        \end{equation}
        where $J_n$ is the $n$th Taylor coefficient of $J^{+}(z)$, consistent with the convention in Sec.~\ref{subsec:conventions}. Since $F_n$ is real, $J_n$ is also real.

        Next, denoting complex conjugation by an overline, we define the holomorphic function $F^{-}(z)$ for $|z|>1$ by
        \begin{equation}
            F^{-}(z):=\overline{F^{+}\left(\frac{1}{\overline{z}}\right)}\quad(|z|>1).
        \end{equation}
        Since $F^{-}(z)$ does not have zeros in $|z|>1$, we can define $J^{-}(z):=1/F^{-}(z)$, which is holomorphic in $|z|>1$ and can be expanded as follows\footnote{
            Indeed, since $J_n$ is real, we have
            \begin{equation}
                J^{-}(z)
                =\frac{1}{F^{-}(z)}
                =\frac{1}{\overline{F^{+}\left(\frac{1}{\overline{z}}\right)}}
                =\overline{J^{+}\left(\frac{1}{\overline{z}}\right)}
                =\overline{\sum_{n=0}^{\infty}J_n\left(\frac{1}{\overline{z}}\right)^n}
                =\sum_{n=0}^{\infty}J_n z^{-n}
            \end{equation}
            for $|z|>1$.
        }:
        \begin{equation}
            J^{-}(z)=\frac{1}{F^{-}(z)}=\sum_{n=0}^{\infty}J_n z^{-n} \quad(|z|>1).
        \end{equation}

        By Theorem~\ref{thm:Hardy_boundary_value}, $F^{+}(z)\in H^2(\mathbb{D})$ has the $L^2(\mathbb{T})$ boundary value $F^{+}(e^{i\theta})\in H^2(\mathbb{T})$. Since $F^{-}(re^{i\theta})=\overline{F^{+}(r^{-1}e^{i\theta})}$ for $r>1$, the $L^2(\mathbb{T})$ boundary value of $F^{-}(z)$ is given by
        \begin{equation}  
            F^{-}(e^{i\theta})
            =\lim_{r\downarrow1}F^{-}(re^{i\theta})
            =\lim_{r\uparrow1}\overline{F^{+}(re^{i\theta})}
            =\overline{F^{+}(e^{i\theta})}.
        \end{equation}
        In what follows, $F^{\pm}(e^{i\theta})$ always denote these $L^2(\mathbb{T})$ boundary values. Moreover, by Corollary~\ref{cor:Hardy_boundary_value} and the same identity, these limits also exist pointwise for almost every $\theta$.

        Then, the Szeg\H{o} factorization of $\mathrm{PSD}(\theta)$ holds almost everywhere on $(-\pi,\pi)$~(see~Theorem~2.6.1 and Corollary~2.6.2 in Nikolski~\cite{Nikolski2019} and p.~75 in Inoue~\cite{Inoue2000}):
        \begin{equation}
        \label{eq:Szego-factorization_PSD}
            \mathrm{PSD}(\theta)= |F^{+}(e^{i\theta})|^2 = F^{+}(e^{i\theta})F^{-}(e^{i\theta}),\qquad F^{-}(e^{i\theta})=\overline{F^{+}(e^{i\theta})}.
        \end{equation}
        Also, the following relations hold~(see~Theorems~7.3 and 5.1 in Inoue~\cite{Inoue2000})\footnote{By Theorem~7.3, the conditions~(C1), (C2), and (A1) of Theorem~5.1 are satisfied.}:
        \begin{align}
            \label{eq:C_Szego-factor_F_CM}
            F_n
            &=\int_{0}^{1}r^n\nu(dr)\geq0 \quad(n\geq0),\\
            J_n
            \label{eq:C_Szego-factor_J_CM}            
            &=-\int_{0}^{1}r^{n-2}\left(1-r^2\right)\rho(dr)\leq0 \quad(n\geq2),\\
            \label{eq:C_Szego-factor_F_asymptotic}
            F_n
            &\sim \sqrt{\frac{\mathfrak{A}_C}{B(\beta_c,\gamma)}}n^{-(1-\beta_c)} \quad\text{as }n\to\infty,\\
            \label{eq:C_Szego-factor_J_asymptotic}
            J_n 
            &\sim -\sqrt{\frac{B(\beta_c,\gamma)}{\mathfrak{A}_C}}\frac{\beta_c\sin(\pi\beta_c)}{\pi} n^{-(1+\beta_c)} \quad\text{as }n\to\infty,
        \end{align}
        where $\nu$ and $\rho$ are nonnegative finite Borel measures on $[0,1)$. Note that $J_0>0$ by the definition of $F^{+}(z)$. Moreover, Appendix~\ref{app:UDT:n=1} shows that $J_1<0$.
        Therefore, we show the following lemma.

        \begin{lemma}
            \label{lemma:partial_sum}
            (i)~We have $\left(J_n\right)_{n\geq0}\in\ell^1$, that is,
            \begin{equation}
            \label{eq:partial_sum_absolute_convergence}
                \sum_{n=0}^{\infty}\left|J_n\right|<\infty.
            \end{equation}
            
            (ii)~We have $\left(J_n\right)_{n\geq0}\in\ell^2$, that is,
            \begin{equation}
            \label{eq:partial_sum_square_convergence}
                \sum_{n=0}^{\infty}\left|J_n\right|^2<\infty.
            \end{equation}
            This implies $\sum_{n=0}^{\infty}J_n e^{in\theta}\in H^2(\mathbb{T})$, and thus, by Theorem~\ref{thm:Hardy_boundary_value}, $J^{+}(z)\in H^2(\mathbb{D})$. In particular, the following equality holds in $L^2(\mathbb{T})$:
            \begin{equation}
                J^{+}(e^{i\theta})=\sum_{n=0}^{\infty}J_n e^{in\theta},
            \end{equation}
            where $J^{+}(e^{i\theta})$ denotes the $L^2(\mathbb{T})$ boundary value of $J^{+}(z)$.
            
            (iii)~The following asymptotic relation holds:
            \begin{equation}
            \label{eq:partial_sum_asymptotic}
                \sum_{j=0}^{n}J_j
                \sim \sqrt{\frac{B(\beta_c,\gamma)}{\mathfrak{A}_C}}\frac{\sin(\pi\beta_c)}{\pi}n^{-\beta_c}\quad\text{as $n\to\infty$}.
            \end{equation}

            (iv)~There exists a constant $c_0>0$ such that, for all $n\geq0$,
            \begin{equation}
            \label{eq:partial_sum_inequality}
                \left|\sum_{j=0}^{n}J_{j}\right|
                \leq c_0 (n+1)^{-\beta_c}.
            \end{equation}

            (v)~For any integer $n\geq0$, we have the following inequality:
            \begin{equation}
            \label{eq:partial_sum_positive}
                \sum_{j=0}^{n}J_{j}>0.
            \end{equation}
            
            (vi)~For any integer $n\geq1$, we have the following inequality:
            \begin{equation}
            \label{eq:reciprocal_negative}
                J_n<0.
            \end{equation}

            (vii)~We have the following equality in $L^2(\mathbb{T})$:
            \begin{equation}
            \label{eq:reciprocal_C_minus}
                J^{-}(e^{i\theta})=\sum_{n=0}^{\infty}J_n e^{-in\theta},
            \end{equation}
            where $J^{-}(e^{i\theta})$ denotes the $L^2(\mathbb{T})$ boundary value of $J^{-}(z)$.

            (viii)~The series $\sum_{n=0}^{\infty}J_n e^{-in\theta}$ converges absolutely and uniformly in $\theta$, and its sum coincides almost everywhere with $J^{-}(e^{i\theta})$. Moreover, $J^{-}(e^{i\theta})$ belongs to $L^{\infty}(\mathbb{T})$.
        \end{lemma}

        \begin{proof}
            (i)~We show Eq.~\eqref{eq:partial_sum_absolute_convergence}.
            The relations $\sum_{n=1}^{\infty}n^{-1-\beta_c}<\infty$ and 
            \begin{equation} 
                \frac{\left|J_n\right|}{n^{-1-\beta_c}}\to \sqrt{\frac{B(\beta_c,\gamma)}{\mathfrak{A}_C}}\frac{\beta_c\sin(\pi\beta_c)}{\pi}>0\quad \text{as $n\to\infty$},
            \end{equation}
            which follows from Eq.~\eqref{eq:C_Szego-factor_J_asymptotic},
            imply
            \begin{equation}
                \sum_{n=0}^{\infty}\left|J_n\right|<\infty
            \end{equation}
            by the limit comparison test.

            (ii)~We show Eq.~\eqref{eq:partial_sum_square_convergence}. Since $\left|J_n\right|\leq\sum_{h=0}^{\infty}\left|J_h\right|$ for each $n\geq0$, it follows from (i) that 
            \begin{equation}
                \left|J_n\right|^2\leq \left|J_n\right|\sum_{h=0}^{\infty}\left|J_h\right|<\infty.
            \end{equation}
            Summing this inequality over $n\geq0$, we obtain
            \begin{equation}
                \sum_{n=0}^{\infty}\left|J_n\right|^2
                \leq \left[\sum_{n=0}^{\infty}\left|J_n\right|\right]^2
                <\infty.
            \end{equation}
            Thus, the series $\sum_{n=0}^{\infty}J_n e^{in\theta}$ converges in $L^2(\mathbb{T})$, and its sum belongs to $H^2(\mathbb{T})$. By Theorem~\ref{thm:Hardy_boundary_value}, this sum is equal to the $L^2(\mathbb{T})$ boundary value of a unique function in $H^2(\mathbb{D})$ whose Taylor coefficients are $J_n$. This function is $J^{+}(z)$, and hence $J^{+}(z)\in H^2(\mathbb{D})$. In particular, the following equality holds in $L^2(\mathbb{T})$:
            \begin{equation}
                J^{+}(e^{i\theta})=\sum_{n=0}^{\infty}J_n e^{i n\theta},
            \end{equation}
            where $J^{+}(e^{i\theta})$ denotes the $L^2(\mathbb{T})$ boundary value of $J^{+}(z)$.

            (iii)~We show Eq.~\eqref{eq:partial_sum_asymptotic}.
            The relations $\sum_{n=1}^{\infty}n^{-(1-\beta_c)}=\infty$ and 
            \begin{equation} 
                \frac{F_n}{n^{-(1-\beta_c)}}\to \sqrt{\frac{\mathfrak{A}_C}{B(\beta_c,\gamma)}}>0,
            \end{equation}
            which follows from Eq.~\eqref{eq:C_Szego-factor_F_asymptotic},
            imply
            \begin{equation}
                \sum_{n=0}^{\infty}F_n=\infty
            \end{equation}
            by the limit comparison test.
            Since $F_n\geq0$ by Eq.~\eqref{eq:C_Szego-factor_F_CM}, the function $F^{+}(r)=\sum_{n=0}^{\infty}F_n r^n$ is nondecreasing in $r\in(0,1)$, and thus the limit $\lim_{r\uparrow1}F^{+}(r)$ exists in $[0,\infty]$. Moreover, for any integer $N\geq0$, we have
            \begin{equation}
                F^{+}(r)\geq\sum_{n=0}^{N}F_n r^n\quad(0<r<1).
            \end{equation}
            Letting $r\uparrow1$ in this inequality, we obtain $\lim_{r\uparrow1}F^{+}(r)\geq\sum_{n=0}^{N}F_n$. Since $N$ is arbitrary and $\sum_{n=0}^{\infty}F_n=\infty$, we conclude\footnote{
                Alternatively, this follows from the monotone convergence theorem. For any monotonically increasing sequence $\left(r_k\right)_{k\geq1}$ in $(0,1)$ such that $r_k\uparrow 1$ as $k\to\infty$, we have
                \begin{equation}
                    \lim_{k\to\infty}F^{+}(r_k)
                    =\lim_{k\to\infty}\sum_{n=0}^{\infty}F_n r_k^n
                    =\sum_{n=0}^{\infty}F_n \lim_{k\to\infty}r_k^n
                    =\sum_{n=0}^{\infty}F_n
                    =\infty,
                \end{equation}
                where we interchanged the sum and the limit by the monotone convergence theorem because $F_n\geq0$. Since the sequence $\left(r_k\right)_{k\geq1}$ is arbitrary, this implies $\lim_{r\uparrow1}F^{+}(r)=\infty$.
            }
            \begin{equation}
                \lim_{r\uparrow 1}F^{+}(r)=\infty.
            \end{equation}
            On the other hand, for $J^{+}(r)$, Eq.~\eqref{eq:partial_sum_absolute_convergence} means $\left(J_n\right)_{n\geq0}\in\ell^1$.
            Therefore, Abel's theorem allows us to interchange the sum and the limit:
            \begin{equation}
            \label{eq:partial_sum_all_0}
                0
                =\lim_{r\uparrow 1}\frac{1}{F^{+}(r)}
                =\lim_{r\uparrow 1}J^{+}(r)
                =\lim_{r\uparrow 1}\sum_{n=0}^{\infty}J_n r^n
                =\sum_{n=0}^{\infty}J_n \lim_{r\uparrow 1}r^n
                =\sum_{n=0}^{\infty}J_n.
            \end{equation}
            From this equality, the relation $\sum_{j=0}^{n}J_j+\sum_{j=n+1}^{\infty}J_j=0$ holds, and thus we can evaluate the asymptotic form as follows~(see Appendix~\ref{app:UDT:partial_sum}):
            \begin{align}
                \sum_{j=0}^{n}J_j
                =-\sum_{j=n+1}^{\infty}J_j 
                \sim \sqrt{\frac{B(\beta_c,\gamma)}{\mathfrak{A}_C}}\frac{\beta_c\sin(\pi\beta_c)}{\pi}\sum_{j=n+1}^{\infty}j^{-(1+\beta_c)} 
                \sim \sqrt{\frac{B(\beta_c,\gamma)}{\mathfrak{A}_C}}\frac{\beta_c\sin(\pi\beta_c)}{\pi}\int_{n}^{\infty}\dd x x^{-(1+\beta_c)}.
            \end{align}
            Carrying out the last integral, we obtain Eq.~\eqref{eq:partial_sum_asymptotic}.

            (iv)~We show Eq.~\eqref{eq:partial_sum_inequality}. By the definition of asymptotic equivalence, for sufficiently large $n$, we have 
            \begin{equation}
                \left|\frac{\sum_{j=0}^{n}J_{j}}{\sqrt{\frac{B(\beta_c, \gamma)}{\mathfrak{A}_C}}\frac{\sin(\pi\beta_c)}{\pi}(n+1)^{-\beta_c}}-1\right|
                <1.
            \end{equation}
            For sufficiently large $n$, we have
            \begin{equation}
            \label{eq:J_partial_sum_asymptotic_inequality}
                \left|\sum_{j=0}^{n}J_{j}\right|
                \leq 2\sqrt{\frac{B(\beta_c, \gamma)}{\mathfrak{A}_C}}\frac{\sin(\pi\beta_c)}{\pi}(n+1)^{-\beta_c},
            \end{equation}
            and, for the finitely many remaining $n$, the left-hand side is bounded by a constant. Therefore, there exists a constant $c_0>0$ such that 
            \begin{equation}
            \label{eq:J_partial_sum_inequality}
                \left|\sum_{j=0}^{n}J_{j}\right|
                \leq c_0 (n+1)^{-\beta_c}
            \end{equation}
            for all $n\geq0$\nobreak\hspace{0.15em}\footnote{
                More precisely, there exists an integer $N'\geq0$ such that, for all $n\geq N'$, Eq.~\eqref{eq:J_partial_sum_asymptotic_inequality} holds. For the finitely many remaining $n<N'$, the left-hand side is bounded by $c_0(n+1)^{-\beta_c}$ if we take $c_0$ sufficiently large. This leads to Eq.~\eqref{eq:J_partial_sum_inequality}.
            }.

            (v)~We show Eq.~\eqref{eq:partial_sum_positive}. By Eq.~\eqref{eq:partial_sum_all_0}, $\sum_{j=0}^{\infty}J_{j}=0$ holds, and thus we have
            \begin{equation}
            \label{eq:J-sum_split}
                \sum_{j=0}^{n}J_{j}
                =-\sum_{j=n+1}^{\infty}J_{j}.
            \end{equation}
            By Eq.~\eqref{eq:C_Szego-factor_J_CM}, we have 
            \begin{equation}
            \label{eq:reciprocal_C-plus_non-positive}
                J_{j}\leq 0\quad(j\geq 2),
            \end{equation}
            and Eq.~\eqref{eq:C_Szego-factor_J_asymptotic} implies $J_j<0$ for sufficiently large $j$. Hence, for $n\geq1$, the right-hand side of Eq.~\eqref{eq:J-sum_split} is positive, since every term in it has an index $j\geq n+1\geq2$; that is, 
            \begin{equation}
            \label{eq:partial_sum_positive_m_geq_1}
                \sum_{j=0}^{n}J_{j}>0,\quad (n\geq 1).
            \end{equation}
            In addition, for $n=0$, we have $\sum_{j=0}^{0}J_j=J_0>0$. Therefore, we have 
            \begin{equation}
                \sum_{j=0}^{n}J_{j}>0,\quad (n\geq0).
            \end{equation}

            (vi)~We show Eq.~\eqref{eq:reciprocal_negative}. By Eq.~\eqref{eq:C_Szego-factor_J_asymptotic}, $J_n<0$ holds for sufficiently large $n$. This implies $\rho((0,1))>0$. Indeed, if $\rho$ is supported on $\{0\}$, we have
            \begin{equation}
                J_n=-\int_{0}^{1}r^{n-2}\left(1-r^2\right)\rho(dr)=0
            \end{equation}
            for $n\geq3$, which contradicts the asymptotic form in Eq.~\eqref{eq:C_Szego-factor_J_asymptotic}. Hence, we have $\rho((0,1))>0$. On the other hand, for any $n\geq2$ and $r\in(0,1)$, we have $r^{n-2}(1-r^2)>0$. Hence, by Eq.~\eqref{eq:C_Szego-factor_J_CM}, we have
            \begin{equation}
                J_n=-\int_{0}^{1}r^{n-2}\left(1-r^2\right)\rho(dr)<0
            \end{equation}
            for any $n\geq2$.
            In addition, Appendix~\ref{app:UDT:n=1} shows that $J_1<0$. Therefore, for all $n\geq1$, we have $J_n<0$.
            
            (vii)~We show Eq.~\eqref{eq:reciprocal_C_minus}. Since $J^{-}(z)=\overline{J^{+}(1/\overline{z})}$ for $|z|>1$, the $L^2(\mathbb{T})$ boundary value of $J^{-}(z)$ is the complex conjugate of that of $J^{+}(z)$, in the same way as for $F^{-}$. Hence, by (ii) and the fact that $J_n$ is real, Eq.~\eqref{eq:reciprocal_C_minus} follows.

            (viii)~Since $(J_n)_{n\geq0}\in\ell^1$ by (i), we have 
            \begin{equation}
                \sum_{n=0}^{\infty}|J_ne^{-in\theta}|
                =\sum_{n=0}^{\infty}|J_n|
                <\infty,
            \end{equation}
            and thus the Weierstrass M-test implies that the series $\sum_{n=0}^{\infty}J_ne^{-in\theta}$ converges absolutely and uniformly in $\theta\in[-\pi,\pi)$. 
            Let us temporarily denote this sum by $J^-_{\mathrm{Fourier}}(e^{i\theta})$:
            \begin{equation}
                J^-_{\mathrm{Fourier}}(e^{i\theta}) 
                :=\sum_{n=0}^{\infty}J_ne^{-in\theta},
            \end{equation}
            which is a continuous function on $\mathbb{T}$.

            On the other hand, the holomorphic function $J^-(z)$ defined for $|z|>1$ is expanded as follows:
            \begin{equation}
                J^-(z)=\sum_{n=0}^{\infty}J_nz^{-n}\qquad(|z|>1).
            \end{equation}
            Hence, for $r>1$, we have 
            \begin{equation}
                J^-(re^{i\theta})
                =\sum_{n=0}^{\infty}J_n r^{-n}e^{-in\theta},
            \end{equation}
            which, together with the triangle inequality, gives
            \begin{equation}
                \sup_{\theta}\left|J^-(re^{i\theta})-J^-_{\mathrm{Fourier}}(e^{i\theta})\right|
                \leq\sum_{n=0}^{\infty}|J_n||r^{-n}-1|.
            \end{equation}
            The right-hand side tends to zero as $r\downarrow1$. Indeed, let $\mathfrak{d}>0$. Since $\sum_{n=0}^{\infty}|J_n|<\infty$ by (i), there exists an integer $N\geq0$ such that $\sum_{n=N+1}^{\infty}|J_n|<\mathfrak{d}$. Since $|r^{-n}-1|\leq1$ for $r>1$, we have
            \begin{equation}
                \sum_{n=0}^{\infty}|J_n||r^{-n}-1|
                <\sum_{n=0}^{N}|J_n||r^{-n}-1|+\mathfrak{d}.
            \end{equation}
            The first term on the right-hand side is a finite sum and tends to zero as $r\downarrow1$, and thus it is less than $\mathfrak{d}$ for $r>1$ sufficiently close to $1$. Since $\mathfrak{d}$ is arbitrary, we obtain\footnote{
                Alternatively, this follows from the dominated convergence theorem for series: for each $n$, $|J_n||r^{-n}-1|\to0$ as $r\downarrow1$, and $|J_n||r^{-n}-1|\leq|J_n|$ with $\sum_{n=0}^{\infty}|J_n|<\infty$.
            }
            \begin{equation}
                \lim_{r\downarrow1}\sum_{n=0}^{\infty}|J_n||r^{-n}-1|=0.
            \end{equation}
            Therefore, we have 
            \begin{equation}
                J^-(re^{i\theta})\to J^-_{\mathrm{Fourier}}(e^{i\theta})\quad\text{as }r\downarrow 1,
            \end{equation}
            uniformly in $\theta\in[-\pi, \pi)$, and, in particular, 
            \begin{equation}
                J^-(re^{i\theta})\to J^-_{\mathrm{Fourier}}(e^{i\theta})\quad\text{in }L^2(\mathbb{T}).
            \end{equation}

            Moreover, since $J^{+}(z)\in H^2(\mathbb{D})$ by (ii), Theorem~\ref{thm:Hardy_boundary_value} and 
            \begin{equation}
                J^-(re^{i\theta})=\overline{J^+(r^{-1}e^{i\theta})}
            \end{equation}
            imply that the $L^2(\mathbb{T})$ boundary value $J^-(e^{i\theta})$ of $J^-(z)$ exists and that the following relation holds:
            \begin{equation}
                J^-(re^{i\theta})\to J^-(e^{i\theta})\quad\text{in }L^2(\mathbb{T})\qquad\text{as $r\downarrow 1$}.
            \end{equation}
            
            From the above, the triangle inequality gives
            \begin{equation}
                \left\|J^-(e^{i\theta}) -J^-_{\mathrm{Fourier}}(e^{i\theta})\right\|_{2}
                \leq \left\| J^{-}(e^{i\theta})-J^-(re^{i\theta})\right\|_{2}
                +\left\| J^-(re^{i\theta})-J^-_{\mathrm{Fourier}}(e^{i\theta})\right\|_{2}
                \to 0 \quad \text{as $r\downarrow 1$}, 
            \end{equation}
            and this implies
            \begin{equation}
                J^-(e^{i\theta})
                =J^-_{\mathrm{Fourier}}(e^{i\theta})
                =\sum_{n=0}^{\infty}J_ne^{-in\theta}\quad\text{a.e.}
            \end{equation}
            This allows us to identify $J^{-}_\mathrm{Fourier}(e^{i\theta})$ with $J^-(e^{i\theta})$.

            Moreover, 
            \begin{equation}
                \operatorname*{ess\,sup}_{-\pi\leq\theta<\pi}|J^-(e^{i\theta})|
                \leq\sum_{n=0}^{\infty}|J_n|
                <\infty
            \end{equation}
            leads to
            \begin{equation}
                J^-(e^{i\theta})\in L^\infty(\mathbb{T}).
            \end{equation}

        \end{proof}

    \subsection{Additive decomposition into positive and negative modes}
    \label{sec:UDT:additive-decomposition}
        Let us define the generating function $S^{+}(z)$ of $S_n$ in $\mathbb{D}$ by 
        \begin{equation}
            S^{+}(z):=\sum_{n=0}^{\infty}S_n z^n\quad(z\in\mathbb{D}),
        \end{equation}
        so that $S_n$ is the $n$th Taylor coefficient of $S^{+}(z)$, consistent with the convention in Sec.~\ref{subsec:conventions}.
        This is well-defined and belongs to $H^2(\mathbb{D})$~(see~Appendix~\ref{app:UDT:S_plus_in_H2}).

        Since $S^{+}(e^{i\theta})\in H^2(\mathbb{T})\subset L^2(\mathbb{T})$ by Theorem~\ref{thm:Hardy_boundary_value} and $J^{-}(e^{i\theta})\in L^{\infty}(\mathbb{T})$, we have  
        \begin{equation}
            \left\|S^{+}J^{-}\right\|_{2}^2
            =\frac{1}{2\pi}\int_{-\pi}^{\pi}\dd \theta |S^{+}(e^{i\theta})J^{-}(e^{i\theta})|^2
            \leq \frac{1}{2\pi}\int_{-\pi}^{\pi}\dd \theta |S^{+}(e^{i\theta})|^2 \sup_{\theta}|J^{-}(e^{i\theta})|^2
            =\left\|S^{+}\right\|_{2}^2 \left\|J^{-}\right\|_{\infty}^2<\infty, 
        \end{equation}
        which means $S^{+}(e^{i\theta})J^{-}(e^{i\theta})\in L^2(\mathbb{T})$. Then, we define the function $\Phi(e^{i\theta})$ on $\mathbb{T}$ as follows:
        \begin{equation}
            \Phi(e^{i\theta}):=S^{+}(e^{i\theta})J^{-}(e^{i\theta})\in L^2(\mathbb{T}).
        \end{equation}
        We also define the $n$th Fourier coefficient $\Phi_n$ of $\Phi(e^{i\theta})\in L^2(\mathbb{T})$, consistent with the convention in Sec.~\ref{subsec:conventions}:
        \begin{equation}
            \Phi_n:=\frac{1}{2\pi}\int_{-\pi}^{\pi}\dd \theta \Phi(e^{i\theta})e^{-in\theta}.
        \end{equation}
        Then, we have the following lemma.
        \begin{lemma}
        \label{lemma:Phi_n_l2}
            We have $(\Phi_n)_{n\geq0}\in\ell^2$.
        \end{lemma}

        \begin{proof}
            Since $\Phi(e^{i\theta})\in L^2(\mathbb{T})$, Parseval's identity gives
            \begin{equation}
                \sum_{n=-\infty}^{\infty}|\Phi_n|^2=\frac{1}{2\pi}\int_{-\pi}^{\pi}\dd \theta |\Phi(e^{i\theta})|^2<\infty,
            \end{equation}
            which implies that $(\Phi_n)_{n\geq0}\in\ell^2$.
        \end{proof}
        
        Therefore, by Theorem~\ref{thm:Riesz_projection}, we can define $\Phi^{+}(e^{i\theta}):=P^{+}\Phi(e^{i\theta})\in H^2(\mathbb{T})$, and $\Phi^{+}(e^{i\theta})$ can be expanded in a Fourier series on $\mathbb{T}$ as 
        \begin{equation}
            \Phi^{+}(e^{i\theta})=\sum_{n=0}^{\infty}\Phi_n e^{in\theta}.
        \end{equation}
        By Theorem~\ref{thm:Hardy_boundary_value}, $\Phi^{+}(e^{i\theta})$ is equal to the $L^2(\mathbb{T})$ boundary value of a unique function in $H^2(\mathbb{D})$ whose Taylor coefficients are $\Phi_n$. This function is $\Phi^{+}(z):=\sum_{n=0}^{\infty}\Phi_n z^n$, and hence $\Phi^{+}(z)\in H^2(\mathbb{D})$.

        On the other hand, by Theorem~\ref{thm:Riesz_projection}, we can define $\Phi^{-}(e^{i\theta}):=P^{-}\Phi(e^{i\theta})\in H^2_{-}(\mathbb{T})$, and $\Phi^{-}(e^{i\theta})$ can be expanded in a Fourier series on $\mathbb{T}$ as  
        \begin{equation}
            \Phi^{-}(e^{i\theta})=\sum_{n=-\infty}^{-1}\Phi_n e^{in\theta}.
        \end{equation}
        
        By Theorem~\ref{thm:Riesz_projection}, $\Phi(e^{i\theta})$ is additively decomposed into $\Phi^{+}(e^{i\theta})$, consisting of the nonnegative modes, and $\Phi^{-}(e^{i\theta})$, consisting of the negative modes:
        \begin{equation}
            \Phi(e^{i\theta})=\Phi^{+}(e^{i\theta})+\Phi^{-}(e^{i\theta})\quad\text{a.e.}
        \end{equation}
        
        We can also express $\Phi(e^{i\theta})$ as
        \begin{equation}
            \Phi(e^{i\theta})=\left(\sum_{n=0}^{\infty}S_n e^{in\theta}\right)\left(\sum_{h=0}^{\infty}J_h e^{-i h\theta}\right).
        \end{equation}
        For $n\geq0$, the $n$th Fourier coefficient of $\Phi(e^{i\theta})$ is given as follows~(see~Appendix~\ref{app:UDT:Riesz_projection}):
        \begin{equation}
            \Phi_n=\sum_{h=0}^{\infty}J_{h}S_{n+h}.
        \end{equation}
        This is the $n$th Taylor coefficient of $\Phi^{+}(z)$ and the $n$th Fourier coefficient of $\Phi^{+}(e^{i\theta})$. The following lemma give the difference representation of $\Phi_n$.

        \begin{lemma}
        \label{lemma:Phi_n_diff}
            We can express $\Phi_n$ in terms of the differences $S_{n+h}-S_{n+h+1}$:
            \begin{equation}
            \label{eq:Phi_n_difference}
                \Phi_n=\sum_{h=0}^{\infty}J_{h}S_{n+h}=\sum_{h=0}^{\infty}\left[\sum_{j=0}^{h}J_{j}\right](S_{n+h}-S_{n+h+1}).
            \end{equation}
        \end{lemma}

        \begin{proof}
            Let $N'\geq0$ be any fixed integer. Summation by parts gives  
            \begin{align}
            \label{eq:Phi_n_diff_before}
                \sum_{h=0}^{N'}J_{h}S_{n+h} 
                &=\left[\sum_{h=0}^{N'}J_{h}\right]S_{n+N'}+\sum_{h=0}^{N'-1}\left[\sum_{j=0}^{h}J_{j}\right](S_{n+h}-S_{n+h+1}).
            \end{align}
            For the first term on the right-hand side of Eq.~\eqref{eq:Phi_n_diff_before}, Eq.~\eqref{eq:partial_sum_inequality} in Lemma~\ref{lemma:partial_sum} gives 
            \begin{equation}
                0\leq\left|\sum_{h=0}^{N'}J_{h}S_{n+N'}\right|
                \leq c_0(N'+1)^{-\beta_c}|S_{n+N'}|, 
            \end{equation}
            which vanishes as $N'\to\infty$ because $S_{n+N'}\to0$ as $N'\to\infty$. On the other hand, the left-hand side of Eq.~\eqref{eq:Phi_n_diff_before} converges as $N'\to\infty$, since the series $\sum_{h=0}^{\infty}J_h S_{n+h}$ converges absolutely:  
            \begin{equation}
                \sum_{h=0}^{\infty}\left|J_{h}S_{n+h}\right|
                \leq \sup_{k}\left|S_k\right|\sum_{h=0}^{\infty}\left|J_{h}\right|
                <\infty.
            \end{equation}
            Therefore, taking the limit $N'\to\infty$ yields 
            \begin{equation}
                \sum_{h=0}^{\infty}J_{h}S_{n+h}
                =\sum_{h=0}^{\infty}\left[\sum_{j=0}^{h}J_{j}\right](S_{n+h}-S_{n+h+1}).
            \end{equation}
            This completes the proof.
        \end{proof}

        In fact, we can show the stronger property $(\Phi_n)_{n\geq0}\in\ell^1$, which fails in case (iii) of the universal decay theorems~(see~Sec.~\ref{sec:UDT_log}).
        
        \begin{lemma}
        \label{lemma:Phi_n_l1}
            We have $(\Phi_n)_{n\geq0}\in\ell^1$.
        \end{lemma}
        
        \begin{proof}
            By Eq.~\eqref{eq:partial_sum_inequality} in Lemma~\ref{lemma:partial_sum}, there is a constant $c_0>0$ such that 
            \begin{align}
                \sum_{n=0}^{\infty}|\Phi_n|
                \leq\sum_{n=0}^{\infty}\sum_{h=0}^{\infty}\left|\sum_{j=0}^{h}J_j\right|\left|S_{n+h}-S_{n+h+1}\right|
                \leq c_0\sum_{n=0}^{\infty}\sum_{h=0}^{\infty}(h+1)^{-\beta_c}|S_{n+h}-S_{n+h+1}|,
            \end{align}
            where we used the triangle inequality.
            Since all terms are nonnegative, Tonelli's theorem allows us to interchange the order of the sums. Setting $k=n+h$ gives
            \begin{equation}
                c_0\sum_{n=0}^{\infty}\sum_{h=0}^{\infty}(h+1)^{-\beta_c}|S_{n+h}-S_{n+h+1}|
                =c_0\sum_{k=0}^{\infty}|S_{k}-S_{k+1}|\sum_{h=0}^{k}(h+1)^{-\beta_c}
                \leq \frac{c_0}{1-\beta_c}\sum_{k=0}^{\infty}(k+1)^{1-\beta_c}|S_{k}-S_{k+1}|,
            \end{equation}
            where we used an estimate for the partial sums of the $p$-series\footnote
                {
                We have $\sum_{h=0}^{k}(h+1)^{-\beta_c}=\sum_{r=1}^{k+1}r^{-\beta_c}$. Since $x^{-\beta_c}$ is decreasing in $x$, we have
                \begin{equation}
                    r^{-\beta_c}\leq\int_{r-1}^{r}\dd x x^{-\beta_c}
                \end{equation}
                for $r\geq2$, and thus
                \begin{equation}
                \label{eq:p_series_estimate}
                    \sum_{r=1}^{k+1}r^{-\beta_c}\leq 1+\int_{1}^{k+1}\dd x x^{-\beta_c}
                    =1+\frac{(k+1)^{1-\beta_c}-1}{1-\beta_c}
                    \leq \frac{1}{1-\beta_c} (k+1)^{1-\beta_c}.
                \end{equation}
                }.
            Since $\sum_{k=0}^{\infty}(k+1)^{1-\beta_c}|S_k-S_{k+1}|<\infty$~(see~Appendix~\ref{app:UDT:weighted-diff}), we have $\left(\Phi_n\right)_{n\geq0}\in\ell^1$.
        \end{proof}

    \subsection{Construction of a candidate solution}
    \label{subsec:UDT:solution-construction}
        Let us construct a candidate solution. We define $K^{+}(z)$ by 
        \begin{equation}
            K^{+}(z):=\Phi^{+}(z)J^{+}(z)\quad(z\in\mathbb{D}).
        \end{equation}
        Since $\Phi^{+}(z)$ and $J^{+}(z)$ are holomorphic in $\mathbb{D}$, the function $K^{+}(z)$ is holomorphic in $\mathbb{D}$. This holomorphy allows us to expand $K^{+}(z)$ in a Taylor series as
        \begin{equation}
            K^{+}(z)=\sum_{n=0}^{\infty}K_n z^n\quad(z\in\mathbb{D}),
        \end{equation}
        where $K_n$ is the $n$th Taylor coefficient of $K^{+}(z)$, consistent with the convention in Sec.~\ref{subsec:conventions}. This $n$th Taylor coefficient $K_n$ can be expressed in terms of $\Phi_h$ and $J_{n-h}$.
        
        \begin{lemma}
        \label{lemma:K_n_sum-formula}
            We have 
            \begin{equation}
                \label{eq:K_n_sum-formula}
                K_n=\sum_{h=0}^{n}\Phi_{h}J_{n-h}.
            \end{equation}
        \end{lemma}

        \begin{proof}
            Since
            \begin{equation}
                \sum_{h=0}^{\infty}\sum_{l=0}^{\infty}\left|\Phi_{h}\right|\left|J_{l}\right|
                =\sum_{h=0}^{\infty}\left|\Phi_{h}\right|\sum_{l=0}^{\infty}\left|J_{l}\right|
                <\infty,
            \end{equation}
            the Fubini--Tonelli theorem allows us to interchange the order of the sums, and we obtain
            \begin{equation}
                K_n=\sum_{h=0}^{n}\Phi_{h}J_{n-h}.
            \end{equation}
        \end{proof}

        Moreover, the following lemma holds.

        \begin{lemma}
        \label{lemma:UDT:K_n_l1}
            We have $(K_n)_{n\geq0}\in\ell^1$, and the series $\sum_{n=0}^{\infty}K_n z^n$ converges absolutely and uniformly for $|z|\leq1$. In particular, the radial limit $\lim_{r\uparrow1}K^{+}(re^{i\theta})$ exists for every $\theta$ and coincides with the $L^2(\mathbb{T})$ boundary value $K^{+}(e^{i\theta})=\sum_{n=0}^{\infty}K_n e^{in\theta}$. Moreover, we have
            \begin{equation}
                K^{+}(e^{i\theta})=\Phi^{+}(e^{i\theta})J^{+}(e^{i\theta})\quad\text{a.e.}
            \end{equation}
        \end{lemma}
        
        \begin{proof}
            Since $(\Phi_n)_{n\geq0}\in \ell^1$ and $\left(J_n\right)_{n\geq0}\in \ell^1$, the triangle inequality gives 
            \begin{equation}
                \sum_{n=0}^{\infty}|K_n|
                \leq \sum_{n=0}^{\infty}\sum_{h=0}^{n}|\Phi_{h}|\left|J_{n-h}\right|
                =\sum_{h=0}^{\infty}|\Phi_{h}|\sum_{n=h}^{\infty}\left|J_{n-h}\right|
                =\sum_{h=0}^{\infty}|\Phi_{h}|\sum_{k=0}^{\infty}\left|J_{k}\right|
                <\infty,
            \end{equation}
            where, since all terms are nonnegative, Tonelli's theorem allows us to interchange the order of the sums. This means $(K_n)_{n\geq0}\in\ell^1$\nobreak\hspace{0.15em}\footnote{
                This also follows from Young's convolution inequality.
            }. 
            In particular, in the same way as in the proof of Lemma~\ref{lemma:partial_sum}~(viii), we have
            \begin{equation}
                \sup_\theta\left|K^{+}(re^{i\theta})-\sum_{n=0}^{\infty}K_n e^{in\theta}\right|\leq\sum_{n=0}^{\infty}|K_n||r^n-1|\to0\quad\text{as $r\uparrow 1$}.
            \end{equation}
            Hence, $K^{+}(re^{i\theta})$ converges to $\sum_{n=0}^{\infty}K_n e^{in\theta}$ uniformly in $\theta$ as $r\uparrow1 $. Since uniform convergence implies convergence in $L^2(\mathbb{T})$, this sum coincides with the $L^2(\mathbb{T})$ boundary value $K^{+}(e^{i\theta})$. 
            
            Next, let us show that $K^{+}(e^{i\theta})$ coincides with $\Phi^{+}(e^{i\theta})J^{+}(e^{i\theta})$ almost everywhere. We have already shown that $K^+(re^{i\theta})\to K^+(e^{i\theta})$ as $r\uparrow 1$ for every $\theta$.
            On the other hand, since $J^{+}(z)\in H^2(\mathbb D)$ and $\Phi^{+}(z)\in H^2(\mathbb D)$, Corollary~\ref{cor:Hardy_boundary_value} implies  
            \begin{equation}
                J^+(re^{i\theta})
                \to J^+(e^{i\theta}),\qquad
                \Phi^+(re^{i\theta})
                \to\Phi^+(e^{i\theta})
            \end{equation}
            almost everywhere as $r\uparrow 1$.
            Therefore, for almost every $\theta$, taking the limit $r\uparrow 1$ in the equality  
            \begin{equation}
                K^+(re^{i\theta})
                =\Phi^+(re^{i\theta})J^+(re^{i\theta}),
            \end{equation}
            we obtain 
            \begin{equation}
                K^+(e^{i\theta})
                =\Phi^+(e^{i\theta})J^+(e^{i\theta}).
            \end{equation}
        \end{proof}

    \subsection{Verification of the solution}
    \label{subsec:UDT:solution-check}
        \begin{proposition}
        \label{prop:UDT:solution-check}
            The sequence $(K_n)_{n\geq0}$ constructed above satisfies the discrete Wiener--Hopf equation; that is, it is a solution in $\ell^1$.
        \end{proposition}
        We show the uniqueness of the solution in Sec.~\ref{subsec:UDT:uniqueness}.

        \begin{proof}
            We have the relations $\mathrm{PSD}(\theta)=F^{+}(e^{i\theta})F^{-}(e^{i\theta})$, $\Phi(e^{i\theta})=S^{+}(e^{i\theta})J^{-}(e^{i\theta})$, $\Phi(e^{i\theta})=\Phi^{+}(e^{i\theta})+\Phi^{-}(e^{i\theta})$, and $K^{+}(e^{i\theta})=\Phi^{+}(e^{i\theta})J^{+}(e^{i\theta})$ almost everywhere on $\mathbb{T}$. Since $F^{\pm}(e^{i\theta})J^{\pm}(e^{i\theta})=1$ almost everywhere\footnote{
                This follows from $F^{\pm}(z)J^{\pm}(z)=1$ by taking the radial limits, which exist almost everywhere by Corollary~\ref{cor:Hardy_boundary_value}.
            }, combining these gives 
            \begin{equation}
                K^{+}(e^{i\theta})F^{+}(e^{i\theta})=\Phi^{+}(e^{i\theta})=\Phi(e^{i\theta})-\Phi^{-}(e^{i\theta})
                =S^{+}(e^{i\theta})J^{-}(e^{i\theta})-\Phi^{-}(e^{i\theta}),
            \end{equation}
            and thus multiplying both sides by $F^{-}(e^{i\theta})$ gives 
            \begin{equation}
            \label{eq:verification_Fourier-before}
                K^{+}(e^{i\theta})\mathrm{PSD}(\theta)=S^{+}(e^{i\theta})-F^{-}(e^{i\theta})\Phi^{-}(e^{i\theta}).
            \end{equation}
            Since $F^{-}(e^{i\theta})$ has only nonpositive modes and $\Phi^{-}(e^{i\theta})$ has only negative modes~(see~Appendix~\ref{app:UDT:negative_modes}), the product $F^{-}(e^{i\theta})\Phi^{-}(e^{i\theta})$ has only negative modes.
            
            The relation $(K_n)_{n\geq 0}\in\ell^1$ implies $K^{+}(e^{i\theta})$ is bounded, and $\mathrm{PSD}(\theta)$ belongs to $L^1(\mathbb{T})$. Hence, $K^{+}(e^{i\theta})\mathrm{PSD}(\theta)$ belongs to $L^1(\mathbb{T})$. Moreover, since $S^{+}(e^{i\theta})$ belongs to $L^2(\mathbb{T})$ and the interval $[-\pi, \pi)$ is finite, $S^{+}(e^{i\theta})$ belongs to $L^1(\mathbb{T})$\footnote{
                By the Cauchy--Schwarz inequality, we have
                \begin{equation}
                    \left\|S^{+}(e^{i\theta})\right\|_{1}
                    = \frac{1}{2\pi}\int_{-\pi}^{\pi}\dd \theta \left|S^{+}(e^{i\theta})\right|
                    \leq \frac{1}{2\pi}\sqrt{\int_{-\pi}^{\pi}\dd \theta 1^2}\sqrt{\int_{-\pi}^{\pi}\dd \theta \left|S^{+}(e^{i\theta})\right|^2}
                    =\left\|S^{+}(e^{i\theta})\right\|_{2}<\infty.
                \end{equation}
            }.
            In addition, $F^{-}(e^{i\theta})\Phi^{-}(e^{i\theta})$ belongs to $L^1(\mathbb{T})$~(see~Appendix~\ref{app:UDT:negative_modes}). Therefore, we can take the $n$th Fourier coefficient of each term in Eq.~\eqref{eq:verification_Fourier-before}\footnote{
                Indeed, for any function $f(e^{i\theta})\in L^1(\mathbb{T})$, we have 
                \begin{equation}
                    \left|\frac{1}{2\pi}\int_{-\pi}^{\pi}\dd \theta f(e^{i\theta})e^{-in\theta}\right|
                    \leq \frac{1}{2\pi}\int_{-\pi}^{\pi}\dd \theta |f(e^{i\theta})|
                    =\left\|f(\theta)\right\|_{1}<\infty.
                \end{equation}
                This allows us to define the $n$th Fourier coefficient of $f(e^{i\theta})$.
            }.
            Then, multiplying both sides by $e^{-in\theta}$ and integrating over $\theta$, we obtain
            \begin{equation}
                \frac{1}{2\pi}\int_{-\pi}^{\pi}\dd \theta K^{+}(e^{i\theta})\mathrm{PSD}(\theta)e^{-in\theta}
                =\frac{1}{2\pi}\int_{-\pi}^{\pi}\dd \theta S^{+}(e^{i\theta})e^{-in\theta}-\frac{1}{2\pi}\int_{-\pi}^{\pi}\dd \theta F^{-}(e^{i\theta})\Phi^{-}(e^{i\theta})e^{-in\theta}.
            \end{equation}

            On the right-hand side, the second term is 0 for $n\geq0$ because it has only negative modes, and the first term is $S_n$ for $n\geq0$.
            On the other hand, on the left-hand side, we can interchange the order of the sum and the integral; that is, 
            \begin{equation}
                \frac{1}{2\pi}\int_{-\pi}^{\pi}\dd \theta \left(\sum_{h=0}^{\infty}K_h e^{ih\theta}\right)\mathrm{PSD}(\theta)e^{-in\theta}
                =\sum_{h=0}^{\infty}K_h \frac{1}{2\pi}\int_{-\pi}^{\pi}\dd \theta \mathrm{PSD}(\theta)e^{-i(n-h)\theta}.
            \end{equation}
            Indeed, by Lemma~\ref{lemma:UDT:K_n_l1}, the partial sum $\sum_{h=0}^{N'} K_h e^{ih\theta}$ converges uniformly in $\theta$ to $K^{+}(e^{i\theta})$ as $N'\to\infty$, which implies
            \begin{equation}
                \left|\frac{1}{2\pi}\int_{-\pi}^{\pi}\dd \theta \left(K^{+}(e^{i\theta})-\sum_{h=0}^{N'}K_h e^{ih\theta}\right)\mathrm{PSD}(\theta)e^{-in\theta}\right|
                \leq \left\|K^{+}(e^{i\theta})-\sum_{h=0}^{N'}K_h e^{ih\theta}\right\|_{\infty}\left\|\mathrm{PSD}(\theta)\right\|_{1}\to0\quad\text{as $N'\to\infty$}.
            \end{equation}
            Therefore, for $n\geq0$, we have  
            \begin{equation}
                \frac{1}{2\pi}\int_{-\pi}^{\pi}\dd \theta K^{+}(e^{i\theta})\mathrm{PSD}(\theta)e^{-in\theta}
                =\sum_{h=0}^{\infty}K_h \frac{1}{2\pi}\int_{-\pi}^{\pi}\dd \theta \,\mathrm{PSD}(\theta)e^{-i(n-h)\theta}
                =\sum_{h=0}^{\infty}K_h C_{n-h}.
            \end{equation}
            
            Finally, using the symmetry $C_{-n}=C_n$, we obtain 
            \begin{equation}
                \frac{1}{2\pi}\int_{-\pi}^{\pi}\dd \theta K^{+}(e^{i\theta})\mathrm{PSD}(\theta)e^{-in\theta}=\sum_{h=0}^{\infty}K_h C_{n-h}=\sum_{h=0}^{\infty}C_{|n-h|}K_h.
            \end{equation}
            Comparing this with the right-hand side, we obtain $S_n=\sum_{h=0}^{\infty}C_{|n-h|}K_h$ for $n\geq0$, which is the discrete Wiener--Hopf equation. This completes the proof.
        \end{proof}

    \subsection{Uniqueness of the solution in $\ell^1$}
    \label{subsec:UDT:uniqueness}
        \begin{proposition}
        \label{prop:UDT:uniqueness}
            The solution $(K_n)_{n\geq0}$ verified above is the unique solution of the discrete Wiener--Hopf equation in $\ell^1$.
        \end{proposition}
        
        \begin{proof}
            Suppose that there are two solutions $\left(K_n^{(\mathrm{1})}\right)_{n\geq0}$ and $\left(K_n^{(\mathrm{2})}\right)_{n\geq0}$ of the discrete Wiener--Hopf equation in $\ell^1$. Then, since both solutions satisfy the same discrete Wiener--Hopf equation, taking the difference gives 
            \begin{equation}
            \label{eq:uniqueness_before}
                0=\sum_{h=0}^{\infty}C_{|n-h|}\left(K_h^{(\mathrm{1})}-K_h^{(\mathrm{2})}\right),\quad n\geq0.
            \end{equation}
            In addition, $K_n^{(\mathrm{1})}-K_n^{(\mathrm{2})}$ belongs to $\ell^1$. Let us show that this difference is identically zero.

            Multiplying both sides of Eq.~\eqref{eq:uniqueness_before} by $\overline{K_n^{(\mathrm{1})}-K_n^{(\mathrm{2})}}$ and summing over $n\geq0$, we obtain 
            \begin{equation}
            \label{eq:uniqueness_after}
                0
                =\sum_{n=0}^{\infty}\overline{K_n^{(\mathrm{1})}-K_n^{(\mathrm{2})}}\sum_{h=0}^{\infty}C_{|n-h|}\left(K_h^{(\mathrm{1})}-K_h^{(\mathrm{2})}\right).
            \end{equation}
            Since $K_n^{(\mathrm{1})}-K_n^{(\mathrm{2})}\in\ell^1$, we have 
            \begin{align}
                &\sum_{n=0}^{\infty}\sum_{h=0}^{\infty}
                \frac1{2\pi}\int_{-\pi}^{\pi}\dd \theta\left|\mathrm{PSD}(\theta)e^{-i(n-h)\theta}
                \overline{\left(K_n^{(1)}-K_n^{(2)}\right)}\left(K_h^{(1)}-K_h^{(2)}\right)\right|\\
                &=\sum_{n=0}^{\infty}\sum_{h=0}^{\infty}\frac1{2\pi}\int_{-\pi}^{\pi}\dd \theta\,\mathrm{PSD}(\theta)\left|K_n^{(1)}-K_n^{(2)}\right|\left|K_h^{(1)}-K_h^{(2)}\right|\\
                &=C_0\left(\sum_{n=0}^{\infty}\left|K_n^{(1)}-K_n^{(2)}\right|\right)^2
                <\infty.
            \end{align}    
            Hence, the Fubini--Tonelli theorem allows us to interchange the order of the double series and the integral; that is, 
            \begin{align}
                0
                &=\sum_{n=0}^{\infty}\sum_{h=0}^{\infty}C_{|n-h|}\overline{\left(K_n^{(\mathrm{1})}-K_n^{(\mathrm{2})}\right)}\left(K_h^{(\mathrm{1})}-K_h^{(\mathrm{2})}\right) \\
                &=\sum_{n=0}^{\infty}\sum_{h=0}^{\infty}\frac{1}{2\pi}\int_{-\pi}^{\pi}\dd \theta\, \mathrm{PSD}(\theta)e^{-i(n-h)\theta}\overline{\left(K_n^{(\mathrm{1})}-K_n^{(\mathrm{2})}\right)}\left(K_h^{(\mathrm{1})}-K_h^{(\mathrm{2})}\right) \\
                &=\frac{1}{2\pi}\int_{-\pi}^{\pi}\dd \theta\, \mathrm{PSD}(\theta)\left|\sum_{n=0}^{\infty}\left(K_n^{(\mathrm{1})}-K_n^{(\mathrm{2})}\right)e^{in\theta}\right|^2.
            \end{align}
            Since $\mathrm{PSD}(\theta)>0$ almost everywhere on $\mathbb{T}$, we have 
            \begin{equation}
                \sum_{n=0}^{\infty}\left(K_n^{(\mathrm{1})}-K_n^{(\mathrm{2})}\right)e^{in\theta}=0\quad\text{a.e.}
            \end{equation}
            Renaming the summation index $n$ as $h$, multiplying both sides by $e^{-in\theta}$, and integrating over $\theta$, we obtain
            \begin{equation}
                K_n^{(\mathrm{1})}-K_n^{(\mathrm{2})}
                =\frac{1}{2\pi}\int_{-\pi}^{\pi}\dd \theta \sum_{h=0}^{\infty}\left(K_h^{(\mathrm{1})}-K_h^{(\mathrm{2})}\right)e^{ih\theta}e^{-in\theta}
                = 0,
            \end{equation}
            where we interchanged the order of the sum and the integral because $\sum_{h=0}^{\infty}\left(K_h^{(\mathrm{1})}-K_h^{(\mathrm{2})}\right)e^{ih\theta}$ converges uniformly in $\theta$ by the Weierstrass M-test.
            This means $K_n^{(\mathrm{1})}=K_n^{(\mathrm{2})}$ for all $n\geq0$.
            This completes the proof.
        \end{proof}

    \subsection{Exact formula for the inferred propagator $\mathcal{G}_n$}
    \label{subsec:UDT:propagator-exact-solution}
        Let us present the exact formula for the inferred propagator $\mathcal{G}_n$.
        \begin{proposition}
        \label{prop:UDT:propagator-exact-solution}
            The unique solution $(K_n)_{n\geq0}$ of the discrete Wiener--Hopf equation in $\ell^1$ uniquely determines the inferred propagator $\mathcal{G}_n$. It is expressed in terms of $\Phi_j$ and $J_k$:
            \begin{equation}
            \label{eq:propagator_exact_solution}
                \mathcal{G}_n=\sum_{j=0}^{n-1}\Phi_{j}\sum_{k=0}^{n-1-j}J_{k}.
            \end{equation}
        \end{proposition}
        
        \begin{proof}
            By the definitions of $\mathcal{G}_n$ and $K_n$, we have 
            \begin{equation}
                \mathcal{G}_n=\sum_{h=0}^{n-1}K_h.
            \end{equation}
            By Eq.~\eqref{eq:K_n_sum-formula}, we obtain 
            \begin{equation}
                \mathcal{G}_n=\sum_{h=0}^{n-1}\sum_{j=0}^{h}\Phi_{j}J_{h-j}
                =\sum_{j=0}^{n-1}\Phi_{j}\sum_{h=j}^{n-1}J_{h-j}
                =\sum_{j=0}^{n-1}\Phi_{j}\sum_{k=0}^{n-1-j}J_{k}.
            \end{equation}
            This completes the proof.
        \end{proof}

    \subsection{Asymptotic analysis of the inferred propagator $\mathcal{G}_n$}
        Let us conduct the asymptotic analysis of the inferred propagator $\mathcal{G}_n$. We prove the following proposition.
        \begin{proposition}
        \label{prop:UDT:propagator-asymptotic}
            The inferred propagator $\mathcal{G}_n$ has the following asymptotic behavior:
            \begin{equation}
                \lim_{n\to\infty}n^{\beta_c}\mathcal{G}_n= \mathfrak{A}_{\mathcal{G}},\quad \mathfrak{A}_{\mathcal{G}}:=\sqrt{\frac{B(\beta_c,\gamma)}{\mathfrak{A}_C}}\frac{\sin(\pi\beta_c)}{\pi}\sum_{j=0}^{\infty}\Phi_{j}.
            \end{equation}
        \end{proposition}
        For any fixed integer $N'\geq0$ and sufficiently large $n$ such that $N'<\lfloor n/2\rfloor$, we evaluate the inferred propagator $\mathcal{G}_n$ by splitting it into three parts; that is, 
        \begin{equation}
            \mathcal{G}_n=\mathcal{G}_{1}(n,N')+\mathcal{G}_{2}(n,N')+\mathcal{G}_{3}(n),
        \end{equation}
        where 
        \begin{equation}
            \mathcal{G}_{1}(n,N'):=\sum_{j=0}^{N'}\Phi_j\sum_{k=0}^{n-1-j}J_{k} \qquad
            \mathcal{G}_{2}(n,N'):=\sum_{j=N'+1}^{\lfloor n/2 \rfloor}\Phi_j\sum_{k=0}^{n-1-j}J_{k} \qquad
            \mathcal{G}_{3}(n):=\sum_{j=\lfloor n/2 \rfloor + 1}^{n-1}\Phi_j\sum_{k=0}^{n-1-j}J_{k}. 
        \end{equation}
        To establish the claim, we show that $\mathcal{G}_1(n,N')$ is the leading term and that $\mathcal{G}_2(n,N')$ and $\mathcal{G}_3(n)$ are lower-order terms in the following three lemmas.

        \begin{lemma}
        \label{lemma:UDT:G1_asymptotic}
            For any fixed integer $N'\geq0$, we have 
            \begin{equation}
                \lim_{n\to\infty}\left|n^{\beta_c}\mathcal{G}_{1}(n,N')-\sqrt{\frac{B(\beta_c,\gamma)}{\mathfrak{A}_C}}\frac{\sin(\pi\beta_c)}{\pi}\sum_{j=0}^{N'}\Phi_{j}\right|=0.
            \end{equation}
        \end{lemma}

        \begin{proof}
            For any fixed integer $N'\geq0$, by the definition of $\mathcal{G}_{1}(n,N')$, we have 
            \begin{equation}
                n^{\beta_c}\mathcal{G}_{1}(n,N')=n^{\beta_c}\sum_{j=0}^{N'}\Phi_{j}\sum_{k=0}^{n-1-j}J_{k}.
            \end{equation}
            For each $j=0,1,\dots,N'$, Eq.~\eqref{eq:partial_sum_asymptotic} gives
            \begin{equation}
                \lim_{n\to\infty}n^{\beta_c}\sum_{k=0}^{n-1-j}J_{k}
                = \sqrt{\frac{B(\beta_c,\gamma)}{\mathfrak{A}_C}}\frac{\sin(\pi\beta_c)}{\pi}.
            \end{equation}
            Since the sum over $j$ consists of finitely many terms, we can interchange the limit and the sum, which yields
            \begin{equation}
                \lim_{n\to\infty}n^{\beta_c}\sum_{j=0}^{N'}\Phi_{j}\sum_{k=0}^{n-1-j}J_{k}
                = \sum_{j=0}^{N'}\Phi_{j}\lim_{n\to\infty}n^{\beta_c}\sum_{k=0}^{n-1-j}J_{k}
                =\sqrt{\frac{B(\beta_c,\gamma)}{\mathfrak{A}_C}}\frac{\sin(\pi\beta_c)}{\pi}\sum_{j=0}^{N'}\Phi_{j}, 
            \end{equation}
            which implies, for any fixed $N'$, 
            \begin{equation}
                \lim_{n\to\infty}n^{\beta_c}\mathcal{G}_{1}(n,N')= \sqrt{\frac{B(\beta_c,\gamma)}{\mathfrak{A}_C}}\frac{\sin(\pi\beta_c)}{\pi}\sum_{j=0}^{N'}\Phi_{j}.
            \end{equation}
            This leads to
            \begin{equation}
                \lim_{n\to\infty}\left|n^{\beta_c}\mathcal{G}_{1}(n,N')-\sqrt{\frac{B(\beta_c,\gamma)}{\mathfrak{A}_C}}\frac{\sin(\pi\beta_c)}{\pi}\sum_{j=0}^{N'}\Phi_{j}\right|=0.
            \end{equation}
        \end{proof}

        \begin{lemma}
        \label{lemma:UDT:G2_asymptotic}
            The function $\mathcal{G}_{2}(n,N')$ has the following asymptotic behavior:
            \begin{equation}
                \lim_{N'\to\infty}\limsup_{n\to\infty}\left|n^{\beta_c}\mathcal{G}_{2}(n,N')\right|=0.
            \end{equation}
        \end{lemma}        

        \begin{proof}
            Let $N'\geq0$ be any fixed integer, and let $n$ be sufficiently large so that $N'<\lfloor n/2\rfloor$.

            By Eq.~\eqref{eq:partial_sum_inequality}, we have 
            \begin{equation}
                \left|\sum_{k=0}^{n-1-j}J_{k}\right|\leq c_0 (n-j)^{-\beta_c}.
            \end{equation}
            In addition, since $n-j\geq\frac{n}{2}$ for $j=N'+1, N'+2, \dots, \lfloor n/2 \rfloor$, we have  
            \begin{equation}
                \left(\frac{n}{n-j}\right)^{\beta_c}\leq 2^{\beta_c},
            \end{equation}
            and thus
            \begin{equation}
                n^{\beta_c}\left|\sum_{k=0}^{n-1-j}J_{k}\right|
                \leq c_0 2^{\beta_c}.
            \end{equation}
            Therefore, the triangle inequality give 
            \begin{align}
                n^{\beta_c}\left|\sum_{j=N'+1}^{\lfloor n/2 \rfloor}\Phi_{j}\sum_{k=0}^{n-1-j}J_{k}\right|
                \leq c_0 2^{\beta_c} \sum_{j=N'+1}^{\lfloor n/2 \rfloor}|\Phi_{j}| 
                \leq c_0 2^{\beta_c} \sum_{j=N'+1}^{\infty}|\Phi_{j}|.
            \end{align}
            The right-hand side is independent of $n$ and tends to zero as $N'\to\infty$ since $(\Phi_n)_{n\geq0}\in\ell^1$. Hence, we obtain 
            \begin{equation}
                \lim_{N'\to\infty}\limsup_{n\to\infty}\left|n^{\beta_c}\mathcal{G}_{2}(n,N')\right|=0.
            \end{equation}
        \end{proof}

        \begin{lemma}
        \label{lemma:UDT:G3_asymptotic}
            The function $\mathcal{G}_{3}(n)$ has the following asymptotic behavior:
            \begin{equation}
                \lim_{n\to\infty}\left|n^{\beta_c}\mathcal{G}_{3}(n)\right|=0.
            \end{equation}
        \end{lemma}
        
        \begin{proof}
            By Eq.~\eqref{eq:partial_sum_inequality}, we have 
            \begin{equation}
                \left|\sum_{k=0}^{n-1-j}J_{k}\right|
                \leq c_0 (n-j)^{-\beta_c},
            \end{equation}
            and thus 
            \begin{equation}
            \label{eq:G3_partial-sum}
                n^{\beta_c}\sum_{j>\frac{n}{2}}^{n-1}|\Phi_{j}|\left|\sum_{k=0}^{n-1-j}J_{k}\right|
                \leq c_0 n^{\beta_c}\sum_{j>\frac{n}{2}}^{n-1}|\Phi_{j}|(n-j)^{-\beta_c}.
            \end{equation}
            By Eqs.~\eqref{eq:Phi_n_difference} and \eqref{eq:partial_sum_inequality}, we have
            \begin{equation}
            \label{eq:Phi_n-inequality}
                \left|\Phi_j\right|\leq c_0 \sum_{h=0}^{\infty}(h+1)^{-\beta_c}\left|S_{j+h}-S_{j+h+1}\right|.
            \end{equation}
            Substituting Eq.~\eqref{eq:Phi_n-inequality} into Eq.~\eqref{eq:G3_partial-sum} gives 
            \begin{equation}
                n^{\beta_c}\sum_{j>\frac{n}{2}}^{n-1}|\Phi_{j}|\left|\sum_{k=0}^{n-1-j}J_{k}\right|
                \leq c_0^2 n^{\beta_c}\sum_{j>\frac{n}{2}}^{n-1}\sum_{h=0}^{\infty}(h+1)^{-\beta_c}\left|S_{j+h}-S_{j+h+1}\right|(n-j)^{-\beta_c}.
            \end{equation}
            Setting $k:=j+h$ (note that $k>\frac{n}{2}$), we obtain
            \begin{align}
                c_0^2 n^{\beta_c}\sum_{j>\frac{n}{2}}^{n-1}\sum_{h=0}^{\infty}(h+1)^{-\beta_c}\left|S_{j+h}-S_{j+h+1}\right|(n-j)^{-\beta_c}
                &= c_0^2 \sum_{k>\frac{n}{2}}^{\infty}\left|S_{k}-S_{k+1}\right|n^{\beta_c}\sum_{\frac{n}{2}<j<n,j\leq k}(n-j)^{-\beta_c}(k-j+1)^{-\beta_c},
            \end{align}
            where, since all terms are nonnegative, Tonelli's theorem allows us to interchange the order of the sums.
            This double series can be split into two parts
            \begin{align}
                &\quad c_0^2 \sum_{k>\frac{n}{2}}^{\infty}\left|S_{k}-S_{k+1}\right|n^{\beta_c}\sum_{\frac{n}{2}<j<n,j\leq k}(n-j)^{-\beta_c}(k-j+1)^{-\beta_c}\\
                &=c_0^2 \sum_{\frac{n}{2}<k<n}\left|S_{k}-S_{k+1}\right|n^{\beta_c}\sum_{\frac{n}{2}<j\leq k}(n-j)^{-\beta_c}(k-j+1)^{-\beta_c}
                +c_0^2 \sum_{k=n}^{\infty}\left|S_{k}-S_{k+1}\right|n^{\beta_c}\sum_{\frac{n}{2}<j<n}(n-j)^{-\beta_c}(k-j+1)^{-\beta_c}, 
            \end{align}
            and we can show the following inequalities:
            \begin{align}
                \label{eq:G3_first-term}
                n^{\beta_c}\sum_{\frac{n}{2}<j\leq k}(n-j)^{-\beta_c}(k-j+1)^{-\beta_c}
                &\leq \frac{2^{\beta_c}}{1-2\beta_c} (k+1)^{1-\beta_c},\\
                \label{eq:G3_second-term}
                n^{\beta_c}\sum_{\frac{n}{2}<j<n}(n-j)^{-\beta_c}(k-j+1)^{-\beta_c}
                &\leq \frac{1}{1-2\beta_c} (k+1)^{1-\beta_c}.
            \end{align}
            Note that each right-hand side of these inequalities is independent of $n$ and $j$.

            Let us prove Eq.~\eqref{eq:G3_first-term}. Setting $q:=k-j$, we have $(n-k+q)^{-\beta_c}\leq (q+1)^{-\beta_c}$ because $n-k+q\geq q+1$. This implies 
            \begin{align}
                n^{\beta_c}\sum_{\frac{n}{2}<j\leq k}(n-j)^{-\beta_c}(k-j+1)^{-\beta_c}
                &\leq n^{\beta_c}\sum_{q=0}^{k}(n-k+q)^{-\beta_c}(q+1)^{-\beta_c} \\
                &\leq n^{\beta_c}\sum_{q=0}^{k}(q+1)^{-2\beta_c} \\
                &\leq n^{\beta_c} \frac{1}{1-2\beta_c} (k+1)^{1-2\beta_c},
            \end{align}
            where we used Eq.~\eqref{eq:p_series_estimate} with $\beta_c$ replaced by $2\beta_c$ in the last inequality.
            Moreover, we have $n^{\beta_c}\leq 2^{\beta_c}k^{\beta_c}\leq 2^{\beta_c}(k+1)^{\beta_c}$ because $k>n/2$. This implies 
            \begin{equation}
                n^{\beta_c}\sum_{\frac{n}{2}<j\leq k}(n-j)^{-\beta_c}(k-j+1)^{-\beta_c}
                \leq \frac{2^{\beta_c}}{1-2\beta_c} (k+1)^{1-\beta_c}, 
            \end{equation}
            which is Eq.~\eqref{eq:G3_first-term}.

            Let us prove Eq.~\eqref{eq:G3_second-term}. Setting $q:=n-j$, we have $k-n+q+1\geq q+1\geq q$ because $k\geq n$. This means $(k-n+q+1)^{-\beta_c}\leq q^{-\beta_c}$, and thus
            \begin{align}
                n^{\beta_c}\sum_{\frac{n}{2}<j<n}(n-j)^{-\beta_c}(k-j+1)^{-\beta_c}
                &\leq n^{\beta_c}\sum_{q=1}^{n}q^{-\beta_c}(k-n+q+1)^{-\beta_c} \\
                &\leq n^{\beta_c}\sum_{q=1}^{n}q^{-2\beta_c} \\
                &\leq \frac{1}{1-2\beta_c} n^{\beta_c} n^{1-2\beta_c},
            \end{align}
            where we used Eq.~\eqref{eq:p_series_estimate} with $\beta_c$ replaced by $2\beta_c$ in the last inequality.
            Moreover, we have $n^{1-\beta_c}\leq k^{1-\beta_c}\leq (k+1)^{1-\beta_c}$ because $k\geq n$ and $1-\beta_c>0$. This implies
            \begin{equation}
                n^{\beta_c}\sum_{\frac{n}{2}<j<n}(n-j)^{-\beta_c}(k-j+1)^{-\beta_c}
                \leq \frac{1}{1-2\beta_c} (k+1)^{1-\beta_c},
            \end{equation}
            which is Eq.~\eqref{eq:G3_second-term}.
            
            Using Eqs.~\eqref{eq:G3_first-term} and \eqref{eq:G3_second-term}, we obtain 
            \begin{equation}
                n^{\beta_c}\sum_{j>\frac{n}{2}}^{n-1}|\Phi_{j}|\left|\sum_{k=0}^{n-1-j}J_{k}\right|
                \leq \frac{\left(2^{\beta_c}+1\right)c_0^2}{1-2\beta_c} \sum_{k>\frac{n}{2}}^{\infty}(k+1)^{1-\beta_c}\left|S_{k}-S_{k+1}\right|.
            \end{equation}
            Since $\sum_{k=0}^{\infty}(k+1)^{1-\beta_c}\left|S_{k}-S_{k+1}\right|<\infty$~(see~Appendix~\ref{app:UDT:weighted-diff}), the right-hand side tends to zero as $n\to\infty$. Hence, 
            \begin{equation}
                \lim_{n\to\infty}n^{\beta_c}\sum_{j>\frac{n}{2}}^{n-1}|\Phi_{j}|\left|\sum_{k=0}^{n-1-j}J_{k}\right|=0,
            \end{equation}
            and thus the triangle inequality yields
            \begin{equation}
                0\leq n^{\beta_c}\left|\mathcal{G}_{3}(n)\right|
                \leq n^{\beta_c}\sum_{j>\frac{n}{2}}^{n-1}|\Phi_{j}|\left|\sum_{k=0}^{n-1-j}J_{k}\right|
                \to 0\quad\text{as $n\to\infty$}.
            \end{equation}
            This completes the proof.
        \end{proof}

        By~Lemmas~\ref{lemma:UDT:G1_asymptotic}, \ref{lemma:UDT:G2_asymptotic}, and \ref{lemma:UDT:G3_asymptotic}, we can prove Proposition~\ref{prop:UDT:propagator-asymptotic}.
        \begin{proof}[Proof of Proposition~\ref{prop:UDT:propagator-asymptotic}]
            The triangle inequality gives 
            \begin{align}
                \left|n^{\beta_c}\mathcal{G}_n-\mathfrak{A}_{\mathcal{G}}\right|
                &\leq \left|n^{\beta_c}\mathcal{G}_{1}(n,N')-\sqrt{\frac{B(\beta_c,\gamma)}{\mathfrak{A}_C}}\frac{\sin(\pi\beta_c)}{\pi}\sum_{j=0}^{N'}\Phi_{j}\right|\\
                &+\sqrt{\frac{B(\beta_c,\gamma)}{\mathfrak{A}_C}}\frac{\sin(\pi\beta_c)}{\pi}\left|\sum_{j=N'+1}^{\infty}\Phi_{j}\right|\\
                &+\left|n^{\beta_c}\mathcal{G}_{2}(n,N')\right|\\
                &+\left|n^{\beta_c}\mathcal{G}_{3}(n)\right|.
            \end{align}

            Let $\mathfrak{d}>0$ be arbitrary. Since $(\Phi_j)_{j\geq0}\in\ell^1$, the second term is smaller than $\mathfrak{d}/4$ for sufficiently large $N'$. Lemma~\ref{lemma:UDT:G2_asymptotic} shows
            \begin{equation}
                \lim_{N'\to\infty}\limsup_{n\to\infty}\left|n^{\beta_c}\mathcal{G}_{2}(n,N')\right|=0,
            \end{equation}
            which implies 
            \begin{equation}
                \limsup_{n\to\infty}\left|n^{\beta_c}\mathcal{G}_{2}(n,N')\right|<\mathfrak{d}/4
            \end{equation}
            for sufficiently large $N'$. We now fix such an $N'$. By the definition of $\limsup$, the inequality 
            \begin{equation}
                \left|n^{\beta_c}\mathcal{G}_{2}(n, N')\right|<\mathfrak{d}/4
            \end{equation}
            holds for sufficiently large $n$.
            For this fixed $N'$, Lemma~\ref{lemma:UDT:G1_asymptotic} shows
            \begin{equation}
                \lim_{n\to\infty}\left|n^{\beta_c}\mathcal{G}_{1}(n,N')-\sqrt{\frac{B(\beta_c,\gamma)}{\mathfrak{A}_C}}\frac{\sin(\pi\beta_c)}{\pi}\sum_{j=0}^{N'}\Phi_{j}\right|=0.
            \end{equation}
            This means that the first term is smaller than $\mathfrak{d}/4$ for sufficiently large $n$. Moreover, Lemma~\ref{lemma:UDT:G3_asymptotic} shows
            \begin{equation}
                \lim_{n\to\infty}\left|n^{\beta_c}\mathcal{G}_{3}(n)\right|=0.
            \end{equation}
            This means that the fourth term is smaller than $\mathfrak{d}/4$ for sufficiently large $n$. 
            
            Therefore, for sufficiently large $n$, we have 
            \begin{equation}
                \left|n^{\beta_c}\mathcal{G}_n-\mathfrak{A}_{\mathcal{G}}\right|<\mathfrak{d}.
            \end{equation}
            Since $\mathfrak{d}>0$ is arbitrary, we obtain $\lim_{n\to\infty}n^{\beta_c}\mathcal{G}_n=\mathfrak{A}_{\mathcal{G}}$. 
            This completes the proof.
        \end{proof}

        \begin{remark}
        \label{rem:AG_nonzero}
            At this stage, we do not rule out the possibility that $\mathfrak{A}_{\mathcal{G}}$ might be zero. We have shown only that
            \begin{equation}
                \mathcal{G}_n = \mathfrak{A}_{\mathcal{G}} n^{-\beta_c}+o(n^{-\beta_c}).
            \end{equation}
        \end{remark}

        \begin{corollary}
        \label{cor:transient}
            The inferred propagator~$\mathcal{G}_n$ is transient; that is,
            \begin{equation} 
                \mathcal{G}_{\infty}=0.
            \end{equation}
        \end{corollary}

        \begin{proof}
            Since $\beta_c>0$, it follows from Proposition~\ref{prop:UDT:propagator-asymptotic} that
            \begin{equation}
                \mathcal{G}_{\infty}
                =\lim_{n\to\infty}\mathcal{G}_n
                =\lim_{n\to\infty}n^{-\beta_c}\left(n^{\beta_c}\mathcal{G}_n\right)
                =0\cdot \mathfrak{A}_{\mathcal{G}}
                =0.
            \end{equation}
        \end{proof}

        \subsection{Positivity of the coefficient $\mathfrak{A}_{\mathcal{G}}$}
        \begin{proposition}
        \label{prop:AG_positive}
            If, in addition, Assumption~\ref{assumption:positivity} holds, then the coefficient $\mathfrak{A}_{\mathcal{G}}$ is positive; that is, 
            \begin{equation}
                \mathfrak{A}_{\mathcal{G}}=\sqrt{\frac{B(\beta_c,\gamma)}{\mathfrak{A}_C}}\frac{\sin(\pi\beta_c)}{\pi}\sum_{j=0}^{\infty}\Phi_{j}>0.
            \end{equation}
        \end{proposition}
        
        \begin{proof}
            It suffices to show that $\sum_{j=0}^{\infty}\Phi_{j}>0$, because
            \begin{equation}
                \sqrt{\frac{B(\beta_c,\gamma)}{\mathfrak{A}_C}}\frac{\sin(\pi\beta_c)}{\pi}>0.
            \end{equation}

            By Eq.~\eqref{eq:Phi_n_difference}, summing $\Phi_n$ over $n$ gives  
            \begin{equation}
                \sum_{n=0}^{\infty}\Phi_n=\sum_{n=0}^{\infty}\sum_{h=0}^{\infty}\left[\sum_{j=0}^{h}J_{j}\right](S_{n+h}-S_{n+h+1}).
            \end{equation}
            In the proof of Lemma~\ref{lemma:Phi_n_l1}, we have verified that the double series with respect to $n$ and $h$ converges absolutely. Hence, we have 
            \begin{align}
                \sum_{n=0}^{\infty}\Phi_n
                =\sum_{n=0}^{\infty}\sum_{h=0}^{\infty}\left[\sum_{j=0}^{h}J_{j}\right](S_{n+h}-S_{n+h+1}) 
                =\sum_{h=0}^{\infty}\left[\sum_{j=0}^{h}J_{j}\right]\sum_{n=0}^{\infty}(S_{n+h}-S_{n+h+1}) 
                =\sum_{h=0}^{\infty}\left[\sum_{j=0}^{h}J_{j}\right]S_h,
            \end{align}
            where the Fubini--Tonelli theorem allows us to interchange the order of the sums, and we used $S_n\to0$ as $n\to\infty$\footnote{
                More precisely, for each $h\geq0$, considering the partial sum up to $n=N'$, we have 
                \begin{equation}
                    \sum_{n=0}^{\infty}\left(S_{n+h}-S_{n+h+1}\right)
                    =\lim_{N'\to\infty}\sum_{n=0}^{N'}\left(S_{n+h}-S_{n+h+1}\right)
                    =\lim_{N'\to\infty}\left(S_h-S_{N'+h+1}\right)
                    =S_h.
                \end{equation}
            }.
              
            Substituting $S_h=R_{h+1}-R_h$ gives
            \begin{equation}
                \sum_{n=0}^{\infty}\Phi_n
                =\sum_{h=0}^{\infty}\left[\sum_{j=0}^{h}J_{j}\right](R_{h+1}-R_h).
            \end{equation}
            Applying summation by parts to the partial sum up to $N'\geq1$, we obtain
            \begin{align}
                \sum_{h=0}^{N'}\left[\sum_{j=0}^{h}J_{j}\right](R_{h+1}-R_h)
                =\left[\sum_{j=0}^{N'}J_{j}\right]R_{N'+1}
                -\sum_{h=1}^{N'}J_{h}R_h, 
            \end{align}
            where we used $R_0=0$. In addition, the first term on the right-hand side vanishes as $N'\to\infty$~(see~Appendix~\ref{app:UDT:M-vanish}):
            \begin{equation}
            \label{eq:UDT:M-vanish}
                \lim_{N'\to\infty}\left[\sum_{j=0}^{N'}J_{j}\right]R_{N'+1}=0.
            \end{equation}
            Hence, taking the limit $N'\to\infty$ gives 
            \begin{equation}
                \sum_{n=0}^{\infty}\Phi_n
                =-\sum_{h=1}^{\infty}J_{h}R_h.
            \end{equation}
            By Eq.~\eqref{eq:reciprocal_negative} in Lemma~\ref{lemma:partial_sum}, $J_{h}<0$ for $h\geq1$. By Assumption~\ref{assumption:positivity}, $R_h\geq0$ for $h\geq1$. Hence, every term on the right-hand side is nonnegative: $-J_{h}R_h\geq0$ for $h\geq1$.
            Since $R_0=0$ and $R_n\not\equiv0$, there exists an integer $h_0\geq1$ such that $R_{h_0}>0$, and hence $-J_{h_0}R_{h_0}>0$. Therefore, we obtain  
            \begin{equation}
                \sum_{n=0}^{\infty}\Phi_n>0.
            \end{equation}
            This leads to $\mathfrak{A}_{\mathcal{G}}>0$. 
        \end{proof}

\section{Proof of the universal decay theorems~\NoCaseChange{(iii)}}
\label{sec:UDT_log}
    In this section, we give the proof of the universal decay theorems~(iii): the case of $S_n\sim \mathfrak{A}_S n^{\beta_c-1}$, in which a logarithmic correction appears. The proof follows the same strategy as that of the universal decay theorems~(ii) in Sec.~\ref{sec:UDT}. In particular, the argument is identical up to and including the derivation of the difference representation of $\Phi_n$ in Lemma~\ref{lemma:Phi_n_diff}.

    \subsection{Additive decomposition into positive and negative modes}
        Instead of $(\Phi_n)_{n\geq0}\in\ell^1$ in Sec.~\ref{sec:UDT:additive-decomposition}, we show the following asymptotic behavior.
        
        \begin{lemma}
        \label{lemma:UDT_log:Phi_n_sum_asymptotic}
            We have 
            \begin{equation}
                \label{eq:UDT_log:Phi_n_sum_asymptotic}        
                \lim_{N\to\infty}\frac{1}{\ln N}\sum_{n=0}^{N}\Phi_n
                = \sqrt{\frac{B(\beta_c,\gamma)}{\mathfrak{A}_C}}\frac{\sin(\pi\beta_c)}{\pi}\mathfrak{A}_S.
            \end{equation}
            Hence, $(\Phi_n)_{n\geq0}\not\in\ell^1$, and the series $\sum_{n=0}^{\infty}\Phi_n$ diverges logarithmically.
        \end{lemma}

        \begin{proof}
            By the eventual monotonicity of $S_n$ and the asymptotic form $S_n\sim \mathfrak{A}_S n^{\beta_c-1}$ with $\mathfrak{A}_{S}>0$, there exists an integer $N_0$ such that 
            \begin{equation}
                S_n\geq S_{n+1}\geq0
            \end{equation}
            for $n\geq N_0$.
            Eq.~\eqref{eq:partial_sum_positive}, which involves only $C_n$, holds in the present case as well; that is,  
            \begin{equation}
                \sum_{j=0}^{h}J_{j}>0
            \end{equation}
            for $h\geq0$.

            By Eq.~\eqref{eq:Phi_n_difference}, we have 
            \begin{equation}
                \sum_{n=N_0}^{N}\Phi_n
                =\sum_{n=N_0}^{N}\sum_{h=0}^{\infty}\left[\sum_{j=0}^{h}J_{j}\right]\left(S_{n+h}-S_{n+h+1}\right).
            \end{equation}
            Since all terms are nonnegative, Tonelli's theorem allows us to interchange the order of the sums. Setting $k=n+h$, we have
            \begin{align}
                \sum_{n=N_0}^{N}\Phi_n
                &=\sum_{n=N_0}^{N}\sum_{k=n}^{\infty}\left[\sum_{j=0}^{k-n}J_{j}\right]\left(S_{k}-S_{k+1}\right)\\
            \label{eq:UDT_log:Phi_n_sum_N_0_to_N_two}
                &=\sum_{k=N_0}^{N}\left(S_{k}-S_{k+1}\right)\sum_{n=N_0}^{k}\left[\sum_{j=0}^{k-n}J_{j}\right]+\sum_{k=N+1}^{\infty}\left(S_{k}-S_{k+1}\right)\sum_{n=N_0}^{N}\left[\sum_{j=0}^{k-n}J_{j}\right].
            \end{align}

            For the first term in the last line of Eq.~\eqref{eq:UDT_log:Phi_n_sum_N_0_to_N_two}, applying summation by parts to the sum over $k$ gives 
            \begin{align}
            \label{eq:UDT_log:Phi_n_sum_N_0_to_N_first_term}
                \sum_{k=N_0}^{N}\left(S_{k}-S_{k+1}\right)\sum_{n=N_0}^{k}\left[\sum_{j=0}^{k-n}J_{j}\right]
                =S_{N_0}J_0
                -S_{N+1}\sum_{n=N_0}^{N}\left[\sum_{j=0}^{N-n}J_{j}\right]
                +\sum_{k=N_0+1}^{N}S_k\sum_{j=0}^{k-N_0}J_{j}.
            \end{align}
            For the first term in Eq.~\eqref{eq:UDT_log:Phi_n_sum_N_0_to_N_first_term}, since $S_{N_0}J_0$ is a constant independent of $N$, the term $S_{N_0}J_0/\ln N$ tends to zero as $N\to\infty$. 
            For the second term in Eq.~\eqref{eq:UDT_log:Phi_n_sum_N_0_to_N_first_term}, Eqs.~\eqref{eq:partial_sum_inequality} and \eqref{eq:partial_sum_positive} give 
            \begin{equation}
                0\leq \sum_{j=0}^{N-n}J_{j}\leq c_0 (N-n+1)^{-\beta_c},
            \end{equation}
            for $N_0\leq n\leq N$, and hence  
            \begin{equation}
                0\leq S_{N+1}\sum_{n=N_0}^{N}\left[\sum_{j=0}^{N-n}J_{j}\right]
                \leq c_0 S_{N+1}\sum_{n=N_0}^{N}(N-n+1)^{-\beta_c}.
            \end{equation}
            Since $(N-x+1)^{-\beta_c}$ is an increasing function of $x$, we have 
            \begin{equation}
            \label{eq:p-estimate_N-n+1}
                \sum_{n=N_0}^{N}(N-n+1)^{-\beta_c}
                \leq 1+\int_{N_0}^{N}(N-x+1)^{-\beta_c}\dd x
                =1+\frac{(N-N_0+1)^{1-\beta_c}-1^{1-\beta_c}}{1-\beta_c}
                =O(N^{1-\beta_c}).
            \end{equation}
            In addition, since $S_n\sim \mathfrak{A}_S n^{\beta_c-1}$, we have $S_{N+1}=O(N^{\beta_c-1})$. 
            Hence, we obtain  
            \begin{equation}
                S_{N+1}\sum_{n=N_0}^{N}\left[\sum_{j=0}^{N-n}J_{j}\right]
                =O(N^{\beta_c-1})O(N^{1-\beta_c})=O(1),
            \end{equation}
            which implies 
            \begin{equation}
                \lim_{N\to\infty}\frac{S_{N+1}}{\ln N}\sum_{n=N_0}^{N}\left[\sum_{j=0}^{N-n}J_{j}\right]=0.
            \end{equation}
            For the third term in Eq.~\eqref{eq:UDT_log:Phi_n_sum_N_0_to_N_first_term}, we have the following limit~(see~Appendix~\ref{app:UDT_log:partial_sum}):
            \begin{equation}
                \lim_{N\to\infty}\frac{1}{\ln N}\sum_{k=N_0+1}^{N}S_k\sum_{j=0}^{k-N_0}J_{j}
                = \sqrt{\frac{B(\beta_c,\gamma)}{\mathfrak{A}_C}}\frac{\sin(\pi\beta_c)}{\pi}\mathfrak{A}_S.
            \end{equation}
            Therefore, for the first term in the last line of Eq.~\eqref{eq:UDT_log:Phi_n_sum_N_0_to_N_two}, we obtain 
            \begin{equation}
                \lim_{N\to\infty}\frac{1}{\ln N}\sum_{k=N_0}^{N}\left(S_{k}-S_{k+1}\right)\sum_{n=N_0}^{k}\left[\sum_{j=0}^{k-n}J_{j}\right]
                = \sqrt{\frac{B(\beta_c,\gamma)}{\mathfrak{A}_C}}\frac{\sin(\pi\beta_c)}{\pi}\mathfrak{A}_S.
            \end{equation}
            
            For the second term in the last line of Eq.~\eqref{eq:UDT_log:Phi_n_sum_N_0_to_N_two}, by Eqs.~\eqref{eq:partial_sum_inequality} and \eqref{eq:partial_sum_positive}, we have 
            \begin{equation}
                0\leq\sum_{k=N+1}^{\infty}\left(S_{k}-S_{k+1}\right)\sum_{n=N_0}^{N}\left[\sum_{j=0}^{k-n}J_{j}\right]
                \leq c_0 \sum_{k=N+1}^{\infty}\left(S_{k}-S_{k+1}\right)\sum_{n=N_0}^{N}(k-n+1)^{-\beta_c}, 
            \end{equation}
            and the right-hand side is $O(1)$.
            Indeed, since $S_k-S_{k+1}\geq0$ and $(k-n+1)^{-\beta_c}\leq (N-n+1)^{-\beta_c}$ for $k\geq N+1$ and $N_0\leq n\leq N$, we have 
            \begin{equation}
                0
                \leq c_0 \sum_{k=N+1}^{\infty}\left(S_{k}-S_{k+1}\right)\sum_{n=N_0}^{N}(k-n+1)^{-\beta_c}
                \leq c_0 S_{N+1}\sum_{n=N_0}^{N}\left(N-n+1\right)^{-\beta_c}
                =O(N^{\beta_c-1})O(N^{1-\beta_c})
                =O(1),
            \end{equation}
            where we used $S_n\to0$ as $n\to\infty$, Eq.~\eqref{eq:p-estimate_N-n+1}, and $S_{N+1}=O(N^{\beta_c-1})$. 
            Hence, we obtain 
            \begin{equation}
            \label{eq:UDT_log:partial_order}
                \lim_{N\to\infty}\frac{1}{\ln N}\sum_{k=N+1}^{\infty}\left(S_{k}-S_{k+1}\right)\sum_{n=N_0}^{N}\left[\sum_{j=0}^{k-n}J_{j}\right]
                =0.
            \end{equation}
            
            Therefore, we have 
            \begin{equation}
            \label{eq:UDT_log:Phi_n_sum_N_0_to_N}
                \lim_{N\to\infty}\frac{1}{\ln N}\sum_{n=N_0}^{N}\Phi_n
                = \sqrt{\frac{B(\beta_c,\gamma)}{\mathfrak{A}_C}}\frac{\sin(\pi\beta_c)}{\pi}\mathfrak{A}_S.
            \end{equation}
            Since $\sum_{n=0}^{N_0-1}\Phi_n$ does not depend on $N$, we also have  
            \begin{equation}
                \lim_{N\to\infty}\frac{1}{\ln N}\sum_{n=0}^{N_0-1}\Phi_n=0.
            \end{equation}
            This completes the proof.
        \end{proof}

        \begin{remark}
            We have already shown $(\Phi_n)_{n\geq0}\in\ell^2$ in Lemma~\ref{lemma:Phi_n_l2}.
        \end{remark}

    \subsection{Construction of a candidate solution}
    \label{subsec:UDT_log:solution_construction}
        In the same way as in Sec.~\ref{subsec:UDT:solution-construction}, we define $K^{+}(z)$ by 
        \begin{equation}
            K^{+}(z):=\Phi^{+}(z)J^{+}(z)\quad(z\in\mathbb{D}).
        \end{equation}
        Since $\Phi^{+}(z)$ and $J^{+}(z)$ are holomorphic in $\mathbb{D}$, the function $K^{+}(z)$ is holomorphic in $\mathbb{D}$. This holomorphy allows us to expand $K^{+}(z)$ in a Taylor series as
        \begin{equation}
            K^{+}(z)=\sum_{n=0}^{\infty}K_n z^n\quad(z\in\mathbb{D}),
        \end{equation}
        where $K_n$ is the $n$th Taylor coefficient of $K^{+}(z)$, consistent with the convention in Sec.~\ref{subsec:conventions}. This $n$th Taylor coefficient $K_n$ can be expressed in terms of $\Phi_h$ and $J_{n-h}$ as with Lemma~\ref{lemma:K_n_sum-formula}. Since $(\Phi_n)_{n\geq0}\notin\ell^1$ in the present case, we give a proof based on $(\Phi_n)_{n\geq0}\in\ell^2$ instead.

        \begin{lemma}
        \label{lemma:K_n_Taylor_coefficients}
            We have 
            \begin{equation}
            \label{eq:UDT_log:K_n_Taylor_coefficients}
                K_n=\sum_{h=0}^{n}\Phi_{h}J_{n-h}.
            \end{equation}
        \end{lemma} 

        \begin{proof}
            Fix $z\in\mathbb{D}$. Since $(\Phi_h)_{h\geq0}\in\ell^2$, the Cauchy--Schwarz inequality gives 
            \begin{equation}
                \sum_{h=0}^{\infty}|\Phi_h||z|^h
                \leq \left(\sum_{h=0}^{\infty}|\Phi_h|^2\right)^{1/2}\left(\sum_{h=0}^{\infty}|z|^{2h}\right)^{1/2}
                =\left(\sum_{h=0}^{\infty}|\Phi_h|^2\right)^{1/2}\left(\frac{1}{1-|z|^2}\right)^{1/2}
                <\infty, 
            \end{equation}
            which implies that the series $\Phi^{+}(z)=\sum_{h=0}^{\infty}\Phi_h z^h$ converges absolutely. On the other hand, since $\left(J_l\right)_{l\geq0}\in\ell^1$, we have
            \begin{equation}
                \sum_{l=0}^{\infty}\left|J_l\right||z|^l
                \leq \sum_{l=0}^{\infty}\left|J_l\right|
                <\infty,
            \end{equation}
            which implies that the series $J^{+}(z)=\sum_{l=0}^{\infty}J_l z^l$ also converges absolutely. Therefore, the double series formed by the product of these two series also converges absolutely. In fact,
            \begin{equation}
                \sum_{h=0}^{\infty}\sum_{l=0}^{\infty}|\Phi_h|\left|J_l\right||z|^{h+l}
                =\sum_{h=0}^{\infty}|\Phi_h||z|^h\sum_{l=0}^{\infty}\left|J_l\right||z|^l
                <\infty.
            \end{equation}
            Hence, the Fubini--Tonelli theorem allows us to interchange the order of the sums; that is, 
            \begin{equation}
                K^{+}(z)
                =\Phi^{+}(z)J^{+}(z)
                =\sum_{h=0}^{\infty}\Phi_h z^h\sum_{l=0}^{\infty}J_l z^l 
                =\sum_{h=0}^{\infty}\sum_{l=0}^{\infty}\Phi_{h}J_l z^{h+l} 
                =\sum_{n=0}^{\infty}\left(\sum_{h=0}^{n}\Phi_{h}J_{n-h}\right)z^n.
            \end{equation}
            Since $z\in\mathbb{D}$ is arbitrary, the uniqueness of a Taylor expansion completes the proof.
        \end{proof}

        Next, let us show that $(K_n)_{n\geq0}\in\ell^2$\nobreak\hspace{0.15em}\footnote
            {
            This also follows from Young's convolution inequality.
            }.
        
        \begin{lemma}
        \label{lemma:K_n_l2}
            We have $(K_n)_{n\geq0}\in\ell^2$.
        \end{lemma}        

        \begin{proof}
            The triangle inequality and the Cauchy--Schwarz inequality give 
            \begin{align}
                |K_n|^2
                =\left|\sum_{h=0}^{n}\Phi_{h}J_{n-h}\right|^2 
                \leq \left[\sum_{h=0}^{n}\left|\Phi_h\right|^2\left|J_{n-h}\right|\right]\left[\sum_{l=0}^{n}\left|J_{n-l}\right|\right], 
            \end{align}
            which implies 
            \begin{align}
                \sum_{n=0}^{\infty}|K_n|^2
                \leq \sum_{n=0}^{\infty}\left[\sum_{h=0}^{n}\left|\Phi_h\right|^2\left|J_{n-h}\right|\right]\left[\sum_{l=0}^{n}\left|J_{n-l}\right|\right] 
                \leq \sum_{n=0}^{\infty}\left[\sum_{h=0}^{n}\left|\Phi_h\right|^2\left|J_{n-h}\right|\right]\left[\sum_{l=0}^{\infty}\left|J_{l}\right|\right].
            \end{align}
            Since all terms are nonnegative, Tonelli's theorem allows us to interchange the order of the sums. Setting $k=n-h$, we have 
            \begin{equation}
                \sum_{n=0}^{\infty}|K_n|^2
                \leq \sum_{h=0}^{\infty}\left|\Phi_h\right|^2\sum_{k=0}^{\infty}\left|J_{k}\right|\sum_{l=0}^{\infty}\left|J_{l}\right|
                =\sum_{h=0}^{\infty}\left|\Phi_h\right|^2\left[\sum_{k=0}^{\infty}\left|J_{k}\right|\right]^2
                <\infty.
            \end{equation}
            This completes the proof.
        \end{proof}

        Lastly, let us show that $(K_n)_{n\geq0}\in\ell^1$.

        \begin{lemma}
        \label{lemma:UDT_log:K_n_l1}
            We have $(K_n)_{n\geq0}\in\ell^1$.
        \end{lemma}
        
        \begin{proof}
            By the definition of $\mathcal{G}_{n}$, we have $K_n=\mathcal{G}_{n+1}-\mathcal{G}_n$, and hence 
            \begin{equation}
                \sum_{n=0}^{\infty}|K_n|=\sum_{n=0}^{\infty}|\mathcal{G}_{n+1}-\mathcal{G}_n|.
            \end{equation}
            All we have to do is show that the sum on the right-hand side is finite.

            By the definition of $\mathcal{G}_n$  and Eq.~\eqref{eq:UDT_log:K_n_Taylor_coefficients}, we have 
            \begin{equation}
                \mathcal{G}_n
                =\sum_{h=0}^{n-1}K_h
                =\sum_{h=0}^{n-1}\sum_{j=0}^{h}\Phi_jJ_{h-j}
                =\sum_{j=0}^{n-1}\Phi_j\sum_{h=j}^{n-1}J_{h-j}
                =\sum_{j=0}^{n-1}\Phi_j\sum_{h=0}^{n-1-j}J_{h},
            \end{equation}
            where we interchanged the order of the sums.  
            Substituting Eq.~\eqref{eq:Phi_n_difference} for $\Phi_j$ gives 
            \begin{equation}
                \mathcal{G}_n
                =\sum_{j=0}^{n-1}\sum_{r=0}^{\infty}\left[\sum_{l=0}^{r}J_{l}\right](S_{j+r}-S_{j+r+1})\sum_{h=0}^{n-1-j}J_{h}.
            \end{equation}
            For fixed $n$, this double series with respect to $j$ and $r$ converges absolutely. Indeed, let $N_0$ be as in the proof of Lemma~\eqref{lemma:UDT_log:Phi_n_sum_asymptotic}, so that $S_q\geq S_{q+1}\geq0$ for $q\geq N_0$. Then we have  
            \begin{equation}
                \sum_{q=0}^{\infty}|S_q-S_{q+1}|
                =\sum_{q=0}^{N_0-1}|S_q-S_{q+1}|+\sum_{q=N_0}^{\infty}(S_q-S_{q+1})
                =\sum_{q=0}^{N_0-1}|S_q-S_{q+1}|+S_{N_0}
                <\infty,
            \end{equation}
            which means $(S_q-S_{q+1})_{q\geq0}\in\ell^1$. On the other hand, by Eq.~\eqref{eq:partial_sum_inequality}, the partial sums $\sum_{l=0}^{r}J_{l}$ and $\sum_{h=0}^{n-1-j}J_{h}$ are bounded by $c_0$. Hence we obtain 
            \begin{align}
                \sum_{j=0}^{n-1}\sum_{r=0}^{\infty}\left|\sum_{l=0}^{r}J_{l}(S_{j+r}-S_{j+r+1})\sum_{h=0}^{n-1-j}J_{h}\right|
                \leq c_0^2 \sum_{j=0}^{n-1}\sum_{r=0}^{\infty}|S_{j+r}-S_{j+r+1}|
                \leq c_0^2 n\sum_{q=0}^{\infty}|S_{q}-S_{q+1}|<\infty.
            \end{align}
            Therefore, the Fubini--Tonelli theorem allows us to interchange the order of the sums. Setting $k=j+r$ gives 
            \begin{equation}
            \label{eq:G_n_sum_equation}
                \mathcal{G}_n
                =\sum_{k=0}^{\infty}(S_{k}-S_{k+1})\sum_{j=0}^{\min\{k,n-1\}}\left[\sum_{l=0}^{k-j}J_{l}\right]\sum_{h=0}^{n-1-j}J_{h}.
            \end{equation}

            By Eqs.~\eqref{eq:partial_sum_positive} and \eqref{eq:reciprocal_negative}, we have 
            \begin{equation}
            \label{eq:partial_sum_sum_inequality}
                0
                <\sum_{l=0}^{r+1}J_{l}
                < \sum_{l=0}^{r}J_{l}
            \end{equation}
            for $r\geq 0$.
            By this inequality, we show that, for each fixed $k$, the quantity 
            \begin{equation}
            \label{eq:inner_sum_max-evaluation}
                \sum_{j=0}^{\min\{k,n-1\}}\left[\sum_{l=0}^{k-j}J_{l}\right]\sum_{h=0}^{n-1-j}J_{h},
            \end{equation}
            which is nonnegative by Eq.~\eqref{eq:partial_sum_positive}, is strictly increasing for $0\leq n\leq k+1$, attains its maximum at $n=k+1$, and is strictly decreasing for $n\geq k+1$. Note that the quantity in Eq.~\eqref{eq:inner_sum_max-evaluation} is zero for $n=0$ by the empty-sum convention.
            
            Let $k=0$. For $n\geq1$, since $\min\{k,n-1\}=0$, we have 
            \begin{equation}
                \sum_{j=0}^{\min\{k,n-1\}}\left[\sum_{l=0}^{k-j}J_{l}\right]\sum_{h=0}^{n-1-j}J_{h}
                =\sum_{j=0}^{0}\left[\sum_{l=0}^{0-j}J_{l}\right]\sum_{h=0}^{n-1-j}J_{h}
                =J_0\sum_{h=0}^{n-1}J_{h}, 
            \end{equation}
            which is $J_0^2>0$ for $n=k+1=1$. Taking the difference between the values at $n$ and $n+1$ gives 
            \begin{equation}
                J_0\sum_{h=0}^{n}J_{h}-J_0\sum_{h=0}^{n-1}J_{h}
                =J_0 J_n
                < 0.
            \end{equation}
            Therefore, for $k=0$, the quantity in Eq.~\eqref{eq:inner_sum_max-evaluation} is strictly increasing for $0\leq n\leq k+1=1$, attains its maximum at $n=k+1=1$, and is strictly decreasing for $n\geq k+1=1$.

            Next, let $k\geq1$. For $1\leq n\leq k$, since $\min\{k,n-1\}=n-1$, setting $q:=n-1-j$ gives 
            \begin{equation}
                \sum_{j=0}^{\min\{k,n-1\}}\left[\sum_{l=0}^{k-j}J_{l}\right]\sum_{h=0}^{n-1-j}J_{h}
                =\sum_{j=0}^{n-1}\left[\sum_{l=0}^{k-j}J_{l}\right]\sum_{h=0}^{n-1-j}J_{h}
                =\sum_{q=0}^{n-1}\left[\sum_{l=0}^{k-n+1+q}J_{l}\right]\sum_{h=0}^{q}J_{h},
            \end{equation}
            and replacing $n$ with $n+1$ gives 
            \begin{equation}
                \sum_{j=0}^{\min\{k,n\}}\left[\sum_{l=0}^{k-j}J_{l}\right]\sum_{h=0}^{n-j}J_{h}
                =\sum_{q=0}^{n}\left[\sum_{l=0}^{k-n+q}J_{l}\right]\sum_{h=0}^{q}J_{h}.
            \end{equation}
            By these relations and Eq.~\eqref{eq:partial_sum_sum_inequality}, the quantity in Eq.~\eqref{eq:inner_sum_max-evaluation} is strictly increasing for $1\leq n\leq k+1$. Since this quantity is zero for $n=0$ by the empty-sum convention and positive for $n=1$ by Eq.~\eqref{eq:partial_sum_positive}, it is strictly increasing for $0\leq n\leq k+1$.
            
            For $n\geq k+1$, since $\min\{k,n-1\}=k$, we have 
            \begin{equation}
                \sum_{j=0}^{\min\{k,n-1\}}\left[\sum_{l=0}^{k-j}J_{l}\right]\sum_{h=0}^{n-1-j}J_{h}
                =\sum_{j=0}^{k}\left[\sum_{l=0}^{k-j}J_{l}\right]\sum_{h=0}^{n-1-j}J_{h}, 
            \end{equation}
            and replacing $n$ with $n+1$ gives 
            \begin{equation}
                \sum_{j=0}^{\min\{k,n\}}\left[\sum_{l=0}^{k-j}J_{l}\right]\sum_{h=0}^{n-j}J_{h}
                =\sum_{j=0}^{k}\left[\sum_{l=0}^{k-j}J_{l}\right]\sum_{h=0}^{n-j}J_{h}.
            \end{equation}
            By these relations and Eq.~\eqref{eq:partial_sum_sum_inequality}, the quantity in Eq.~\eqref{eq:inner_sum_max-evaluation} is strictly decreasing for $n\geq k+1$.
            
            Therefore, the quantity in Eq.~\eqref{eq:inner_sum_max-evaluation} is strictly increasing for $0\leq n \leq k+1$ and strictly decreasing for $n\geq k+1$. This means the quantity in Eq.~\eqref{eq:inner_sum_max-evaluation} has its maximum at $n=k+1$, and the value is 
            \begin{equation}
            \label{eq:inner_sum_maximum}
                \sum_{j=0}^{k}\left[\sum_{l=0}^{k-j}J_{l}\right]\sum_{h=0}^{k-j}J_{h}
                =\sum_{j=0}^{k}\left[\sum_{l=0}^{j}J_{l}\right]\sum_{h=0}^{j}J_{h}
                =\sum_{j=0}^{k}\left[\sum_{l=0}^{j}J_{l}\right]^2.
            \end{equation}

            By this strict unimodality of the quantity in Eq.~\eqref{eq:inner_sum_max-evaluation}, the inequality 
            \begin{equation}
            \label{eq:G_difference_inequality}
                \sum_{n=0}^{\infty}|K_n|
                =\sum_{n=0}^{\infty}\left|\mathcal{G}_{n+1}-\mathcal{G}_n\right|
                \leq 2\sum_{k=0}^{\infty}\left|S_k-S_{k+1}\right|\sum_{j=0}^{k}\left[\sum_{l=0}^{j}J_{l}\right]^2
            \end{equation}
            holds (see Appendix~\ref{app:UDT_log:G-difference_inequality}).

            In addition, by Eq.~\eqref{eq:partial_sum_inequality} and the estimate in Eq.~\eqref{eq:p_series_estimate} with $\beta_c$ replaced by $2\beta_c\in(0, 1)$, we have 
            \begin{equation}
                \sum_{j=0}^{k}\left[\sum_{l=0}^{j}J_{l}\right]^2
                \leq c_0^2\sum_{j=0}^{k}(j+1)^{-2\beta_c}
                =c_0^2\sum_{j=1}^{k+1}j^{-2\beta_c}
                \leq \frac{c_0^2}{1-2\beta_c}(k+1)^{1-2\beta_c}.
            \end{equation}

            Therefore, we have 
            \begin{equation}
                \sum_{n=0}^{\infty}|K_n|
                \leq \frac{2c_0^2}{1-2\beta_c}\sum_{k=0}^{\infty}(k+1)^{1-2\beta_c}\left|S_k-S_{k+1}\right|,
            \end{equation}
            whose right-hand side is finite~(see~Appendix~\ref{app:UDT_log:weighted-diff}). This leads to $(K_n)_{n\geq0}\in\ell^1$.
        \end{proof}

        Since we have shown $(K_n)_{n\geq0}\in\ell^1$, the following propositions follow in the same way as in Sec.~\ref{sec:UDT}.
        \begin{proposition}
        \label{prop:UDT_log:solution-check}
            The sequence $(K_n)_{n\geq0}$ constructed above satisfies the discrete Wiener--Hopf equation; that is, it is a solution in $\ell^1$.
        \end{proposition}
        \begin{proposition}
        \label{prop:UDT_log:solution-uniqueness}
            The solution $(K_n)_{n\geq0}$ verified above is the unique solution of the discrete Wiener--Hopf equation in $\ell^1$.
        \end{proposition}
        \begin{proposition}
        \label{prop:UDT_log:propagator_exact_solution}
            The unique solution $(K_n)_{n\geq0}$ of the discrete Wiener--Hopf equation in $\ell^1$ uniquely determines the inferred propagator $\mathcal{G}_n$. It is expressed in terms of $\Phi_j$ and $J_k$:
            \begin{equation}
                \mathcal{G}_n
                =\sum_{j=0}^{n-1}\Phi_j\sum_{k=0}^{n-1-j}J_{k}.
            \end{equation}
        \end{proposition}

    \subsection{Asymptotic analysis of the inferred propagator $\mathcal{G}_n$}
        Let us conduct the asymptotic analysis of the inferred propagator $\mathcal{G}_n$. We prove the following proposition.
        \begin{proposition}
            \label{prop:UDT_log:propagator-asymptotic}
            The inferred propagator $\mathcal{G}_n$ has the following asymptotic behavior:
            
            \begin{equation}
                \lim_{n\to\infty}\frac{n^{\beta_c}}{\ln n} \mathcal{G}_n
                =\mathfrak{A}_{\mathcal{G}},\qquad \mathfrak{A}_{\mathcal{G}}:=B(\beta_c,\gamma)\left(\frac{\sin(\pi\beta_c)}{\pi}\right)^2 \frac{\mathfrak{A}_S}{\mathfrak{A}_C}.
            \end{equation}
            
        \end{proposition}

        In order to prove this proposition, for sufficiently large $n$, we evaluate the inferred propagator $\mathcal{G}_n$ by splitting it into three parts; that is, 
        \begin{equation}
                \mathcal{G}_n=\mathcal{G}_{1}(n)+\mathcal{G}_{2}(n)+\mathcal{G}_{3}(n), 
        \end{equation}
        where 
        \begin{equation}
            \mathcal{G}_{1}(n):=\sum_{j=0}^{\lfloor n/\ln n\rfloor}\Phi_j\sum_{k=0}^{n-1-j}J_{k}\qquad
            \mathcal{G}_{2}(n):=\sum_{j=\lfloor n/\ln n\rfloor+1}^{\lfloor n/2\rfloor}\Phi_j\sum_{k=0}^{n-1-j}J_{k}\qquad
            \mathcal{G}_{3}(n):=\sum_{j=\lfloor n/2\rfloor+1}^{n-1}\Phi_j\sum_{k=0}^{n-1-j}J_{k}.
        \end{equation}
        To establish the claim, we show that $\mathcal{G}_1(n)$ is the leading term and that $\mathcal{G}_2(n)$ and $\mathcal{G}_3(n)$ are lower-order terms in the following three lemmas.
        
        \begin{lemma}
            \label{lemma:UDT_log:G1_asymptotic}
            The function $\mathcal{G}_{1}(n)$ has the following asymptotic behavior:
            \begin{equation}
                \lim_{n\to\infty}\frac{n^{\beta_c}}{\ln n}\mathcal{G}_{1}(n)
                =\mathfrak{A}_{\mathcal{G}}
            \end{equation}
        \end{lemma}

        \begin{proof}
            By Eq.~\eqref{eq:partial_sum_asymptotic}, we have 
            \begin{equation}
                \sum_{k=0}^{r}J_{k}
                \sim\sqrt{\frac{B(\beta_c,\gamma)}{\mathfrak{A}_C}}\frac{\sin(\pi\beta_c)}{\pi}r^{-\beta_c} \quad\text{as $r\to\infty$}.
            \end{equation}
            For $0\leq j\leq\lfloor n/\ln n\rfloor$, we have 
            \begin{equation}
                n-1-\left\lfloor \frac{n}{\ln n}\right\rfloor
                \leq n-1-j
                \leq n-1,
            \end{equation}
            and $n-1-\left\lfloor \frac{n}{\ln n}\right\rfloor\to\infty$ and $n-1\to\infty$ as $n\to\infty$. Hence, we have the uniform estimate 
            \begin{equation}
                \sup_{0\leq j\leq\lfloor n/\ln n\rfloor}\left|\frac{\sum_{k=0}^{n-1-j}J_{k}}{\sqrt{\frac{B(\beta_c,\gamma)}{\mathfrak{A}_C}}\frac{\sin(\pi\beta_c)}{\pi}(n-1-j)^{-\beta_c}}-1\right|
                \to 0\quad\text{as $n\to\infty$}.
            \end{equation}
            In addition, the relation 
            \begin{equation}
                \sup_{0\leq j\leq\lfloor n/\ln n\rfloor}\left|\frac{(n-1-j)^{-\beta_c}}{n^{-\beta_c}}-1\right|
                =\sup_{0\leq j\leq\lfloor n/\ln n\rfloor}\left|\left(\frac{n-1-j}{n}\right)^{-\beta_c}-1\right|
                \to 0\quad\text{as $n\to\infty$}
            \end{equation}
            holds\footnote{
                Indeed, we have 
                \begin{equation}
                    \sup_{0\leq j \leq \lfloor n/\ln n\rfloor}\left|\frac{n-1-j}{n}-1\right|\leq \frac{1}{n}+\frac{1}{\ln n}\to0,
                \end{equation}
                which implies 
                \begin{equation}
                    \sup_{0\leq j \leq \lfloor n/\ln n\rfloor}\left|\left(\frac{n-1-j}{n}\right)^{-\beta_c}-1\right|\to0.
                \end{equation}
                }.
            Therefore, we obtain 
            \begin{equation}
                \sup_{0\leq j\leq\lfloor n/\ln n\rfloor}\left|\frac{\sum_{k=0}^{n-1-j}J_{k}}{\sqrt{\frac{B(\beta_c,\gamma)}{\mathfrak{A}_C}}\frac{\sin(\pi\beta_c)}{\pi}n^{-\beta_c}}-1\right|
                \to 0\quad\text{as $n\to\infty$}.
            \end{equation}
            
            Let $\mathfrak{d}>0$ be arbitrary. By the triangle inequality, the relation above implies 
            \begin{align}
            \label{eq:UDT_log:G1_triangle_inequality}
                \left|\sum_{j=0}^{\lfloor n/\ln n\rfloor}\Phi_j\sum_{k=0}^{n-1-j}J_{k}-\sqrt{\frac{B(\beta_c,\gamma)}{\mathfrak{A}_C}}\frac{\sin(\pi\beta_c)}{\pi}n^{-\beta_c}\sum_{j=0}^{\lfloor n/\ln n\rfloor}\Phi_j\right|
                \leq \mathfrak{d}\sqrt{\frac{B(\beta_c,\gamma)}{\mathfrak{A}_C}}\frac{\sin(\pi\beta_c)}{\pi}n^{-\beta_c}\sum_{j=0}^{\lfloor n/\ln n\rfloor}\left|\Phi_j\right|
            \end{align}
            for sufficiently large $n$. Let $N_0$ be as in the proof of Lemma~\ref{lemma:UDT_log:Phi_n_sum_asymptotic}. Then, by Eqs.~\eqref{eq:partial_sum_positive} and \eqref{eq:Phi_n_difference}, we have $\Phi_j\geq0$ for all $j\geq N_0$. This leads to 
            \begin{equation}
                \sum_{j=0}^{\lfloor n/\ln n\rfloor}\left|\Phi_j\right|
                =\sum_{j=0}^{N_0-1}\left|\Phi_j\right|+\sum_{j=N_0}^{\lfloor n/\ln n\rfloor}\Phi_j.
            \end{equation}
            The first term  on the right-hand side is a constant independent of $n$. For the second term on the right-hand side, the relation $\ln \lfloor n/\ln n \rfloor/\ln n \to 1$ and Eq.~\eqref{eq:UDT_log:Phi_n_sum_N_0_to_N} give 
            \begin{equation}
                \lim_{n\to\infty}\frac{1}{\ln n}\sum_{j=N_0}^{\lfloor n/\ln n\rfloor}\Phi_j
                = \sqrt{\frac{B(\beta_c,\gamma)}{\mathfrak{A}_C}}\frac{\sin(\pi\beta_c)}{\pi}\mathfrak{A}_S.
            \end{equation}
            This means 
            \begin{equation}
                \sum_{j=0}^{\lfloor n/\ln n\rfloor}\left|\Phi_j\right|
                =O(\ln n).
            \end{equation}
            Since $\mathfrak{d}>0$ is arbitrary, this and Eq.~\eqref{eq:UDT_log:G1_triangle_inequality} gives 
            \begin{equation}
                \sum_{j=0}^{\lfloor n/\ln n\rfloor}\Phi_j\sum_{k=0}^{n-1-j}J_{k}
                =\sqrt{\frac{B(\beta_c,\gamma)}{\mathfrak{A}_C}}\frac{\sin(\pi\beta_c)}{\pi}n^{-\beta_c}\sum_{j=0}^{\lfloor n/\ln n\rfloor}\Phi_j+o(n^{-\beta_c}\ln n).
            \end{equation}
            By $\lfloor n /\ln n\rfloor \to\infty$ and $\ln\lfloor n/\ln n\rfloor/\ln n\to1$ as $n\to\infty$, Eq.~\eqref{eq:UDT_log:Phi_n_sum_asymptotic} gives 
            \begin{equation}
                \lim_{n\to\infty}\frac{1}{\ln n}\sum_{j=0}^{\lfloor n/\ln n\rfloor}\Phi_j
                =\lim_{n\to\infty}\frac{1}{\ln \lfloor n/\ln n \rfloor}\sum_{j=0}^{\lfloor n/\ln n \rfloor}\Phi_j
                = \sqrt{\frac{B(\beta_c,\gamma)}{\mathfrak{A}_C}}\frac{\sin(\pi\beta_c)}{\pi}\mathfrak{A}_S,
            \end{equation}
            which implies 
            \begin{align}
                \sum_{j=0}^{\lfloor n/\ln n\rfloor}\Phi_j\sum_{k=0}^{n-1-j}J_{k}
                &=\sqrt{\frac{B(\beta_c,\gamma)}{\mathfrak{A}_C}}\frac{\sin(\pi\beta_c)}{\pi}n^{-\beta_c}\sqrt{\frac{B(\beta_c,\gamma)}{\mathfrak{A}_C}}\frac{\sin(\pi\beta_c)}{\pi}\mathfrak{A}_S\ln n + o(n^{-\beta_c}\ln n)\\
                &=\mathfrak{A}_{\mathcal{G}}n^{-\beta_c}\ln n+o(n^{-\beta_c}\ln n).
            \end{align}
            This completes the proof.
        \end{proof}

        \begin{lemma}
            \label{lemma:UDT_log:G2_asymptotic}
            The function $\mathcal{G}_{2}(n)$ has the following asymptotic behavior:
            \begin{equation}
                \lim_{n\to\infty}\frac{n^{\beta_c}}{\ln n}\mathcal{G}_{2}(n)
                =0.
            \end{equation}
        \end{lemma}
            
        \begin{proof}
            Let $N_0$ be as in the proof of Lemma~\ref{lemma:UDT_log:G1_asymptotic}, so that $\Phi_j\geq0$ for all $j\geq N_0$. Since $\lfloor n/\ln n\rfloor+1\geq N_0$ for sufficiently large $n$, we have $\Phi_j\geq0$ for $\lfloor n/\ln n\rfloor+1\leq j\leq\lfloor n/2\rfloor$. In addition, Eqs.~\eqref{eq:partial_sum_inequality} and \eqref{eq:partial_sum_positive} give 
            \begin{equation}
                0
                <\sum_{k=0}^{n-1-j}J_{k}
                \leq c_0 (n-j)^{-\beta_c}.
            \end{equation}
            Since $j\leq \lfloor n/2\rfloor$, the relation $n-j\geq n/2$ holds, and thus $(n-j)^{-\beta_c}\leq 2^{\beta_c}n^{-\beta_c}$. Hence, we have 
            \begin{equation}
                0
                \leq \mathcal{G}_{2}(n)
                =\sum_{j=\lfloor n/\ln n\rfloor+1}^{\lfloor n/2\rfloor}\Phi_j\sum_{k=0}^{n-1-j}J_{k}
                \leq c_0 2^{\beta_c}n^{-\beta_c}\sum_{j=\lfloor n/\ln n\rfloor+1}^{\lfloor n/2\rfloor}\Phi_j.
            \end{equation}
            Moreover, by Lemma~\ref{lemma:UDT_log:Phi_n_sum_asymptotic} and $\ln\lfloor n/2\rfloor/\ln n\to1$, we have 
            \begin{equation}
                \frac{1}{\ln n}\sum_{j=0}^{\lfloor n/2\rfloor}\Phi_j\to\sqrt{\frac{B(\beta_c,\gamma)}{\mathfrak{A}_C}}\frac{\sin(\pi\beta_c)}{\pi}\mathfrak{A}_S.
            \end{equation} 
            The same limit holds for the sum up to $\lfloor n/\ln n\rfloor$, as shown in the proof of Lemma~\ref{lemma:UDT_log:G1_asymptotic}. Taking the difference gives 
            \begin{equation}
            \label{eq:UDT_log:Phi_n_sum_floor-to-floor}
                \frac{1}{\ln n}\sum_{j=\lfloor n/\ln n\rfloor+1}^{\lfloor n/2\rfloor}\Phi_j
                \to 0\quad\text{as $n\to\infty$}.
            \end{equation}
            Combining these estimates, we obtain 
            \begin{equation}
                0\leq \frac{n^{\beta_c}}{\ln n}\mathcal{G}_2(n)\leq \frac{c_02^{\beta_c}}{\ln n}\sum_{j=\lfloor n/\ln n\rfloor+1}^{\lfloor n/2\rfloor}\Phi_j\to0\quad\text{as $n\to\infty$}.
            \end{equation}
            This completes the proof.
        \end{proof}

        \begin{lemma}
            \label{lemma:UDT_log:G3_asymptotic}
            The function $\mathcal{G}_{3}(n)$ has the following asymptotic behavior:
            \begin{equation}
                \lim_{n\to\infty}\frac{n^{\beta_c}}{\ln n}\mathcal{G}_{3}(n)
                =0. 
            \end{equation}
        \end{lemma}

        \begin{proof}
            By Eq.~\eqref{eq:Phi_n_difference}, we have 
            \begin{equation}
                \mathcal{G}_{3}(n)
                =\sum_{j=\lfloor n/2\rfloor+1}^{n-1}\sum_{r=0}^{\infty}\left[\sum_{l=0}^{r}J_{l}\right](S_{j+r}-S_{j+r+1})\sum_{h=0}^{n-1-j}J_{h}.
            \end{equation}
            Let $N_0$ be as in the proof of Lemma~\ref{lemma:UDT_log:G1_asymptotic}, so that $(S_{n})_{n\geq N_0}$ is nonincreasing. Since $j\geq\lfloor n/2\rfloor+1\geq N_0$ for sufficiently large $n$, we have $S_{j+r}-S_{j+r+1}\geq0$. By Eq.~\eqref{eq:partial_sum_positive}, the sums $\sum_{l=0}^{r}J_{l}$ and $\sum_{h=0}^{n-1-j}J_{h}$ are positive.
            Hence, Tonelli's theorem allows us to interchange the order of the sums. Setting $k=j+r$ gives  
            \begin{equation}
                \mathcal{G}_{3}(n)
                =\sum_{k=\lfloor n/2\rfloor+1}^{\infty}(S_{k}-S_{k+1})\sum_{j=\lfloor n/2\rfloor+1}^{\min\{k,n-1\}}\left[\sum_{l=0}^{k-j}J_{l}\right]\sum_{h=0}^{n-1-j}J_{h}.
            \end{equation}
            Since each term over $j$ is nonnegative, we have  
            \begin{equation}
                \sum_{j=\lfloor n/2\rfloor+1}^{\min\{k,n-1\}}\left[\sum_{l=0}^{k-j}J_{l}\right]\sum_{h=0}^{n-1-j}J_{h}
                \leq \sum_{j=0}^{\min\{k,n-1\}}\left[\sum_{l=0}^{k-j}J_{l}\right]\sum_{h=0}^{n-1-j}J_{h}.
            \end{equation}
            For fixed $k$, as shown in the proof of Lemma~\ref{lemma:UDT_log:K_n_l1}, we have 
            \begin{equation}
                \sum_{j=0}^{\min\{k,n-1\}}\left[\sum_{l=0}^{k-j}J_{l}\right]\sum_{h=0}^{n-1-j}J_{h}
                \leq \sum_{j=0}^{k}\left[\sum_{l=0}^{j}J_{l}\right]^2
                \leq \frac{c_0^2}{1-2\beta_c}(k+1)^{1-2\beta_c}.
            \end{equation}
            Hence, we obtain 
            \begin{equation}
                0
                \leq \mathcal{G}_{3}(n)
                \leq \frac{c_0^2}{1-2\beta_c}\sum_{k=\lfloor n/2\rfloor+1}^{\infty}(S_{k}-S_{k+1})(k+1)^{1-2\beta_c}.
            \end{equation}
            In addition, the relation 
            \begin{equation}
            \label{eq:UDT_log:G3_asymptotic}
                \sum_{k=\lfloor n/2\rfloor+1}^{\infty}(S_{k}-S_{k+1})(k+1)^{1-2\beta_c}
                =O(n^{-\beta_c})
            \end{equation}
            holds~(see~Appendix~\ref{app:UDT_log:G3}).
            Therefore, we obtain 
            \begin{equation}
                0\leq \frac{n^{\beta_c}}{\ln n}\mathcal{G}_{3}(n)
                \leq \frac{c_0^2}{1-2\beta_c}\frac{n^{\beta_c}}{\ln n}\sum_{k=\lfloor n/2\rfloor+1}^{\infty}(S_{k}-S_{k+1})(k+1)^{1-2\beta_c}
                =O\left(\frac{1}{\ln n}\right)
                \to 0\quad\text{as $n\to\infty$}.
            \end{equation}  
            This completes the proof.
        \end{proof}

        By~Lemmas~\ref{lemma:UDT_log:G1_asymptotic}, \ref{lemma:UDT_log:G2_asymptotic}, and \ref{lemma:UDT_log:G3_asymptotic}, we can prove Proposition~\ref{prop:UDT_log:propagator-asymptotic}.
        \begin{proof}[Proof of Proposition~\ref{prop:UDT_log:propagator-asymptotic}]
            For sufficiently large $n$, we have split $\mathcal{G}_n$ into three parts; that is, 
            \begin{equation}
                \mathcal{G}_n=\mathcal{G}_{1}(n)+\mathcal{G}_{2}(n)+\mathcal{G}_{3}(n).
            \end{equation}
            By Lemmas~\ref{lemma:UDT_log:G1_asymptotic}, \ref{lemma:UDT_log:G2_asymptotic}, and \ref{lemma:UDT_log:G3_asymptotic}, we have 
            \begin{equation}
                \lim_{n\to\infty}\frac{n^{\beta_c}}{\ln n}\mathcal{G}_{1}(n)
                =\mathfrak{A}_{\mathcal{G}},\qquad
                \lim_{n\to\infty}\frac{n^{\beta_c}}{\ln n}\mathcal{G}_{2}(n)
                =0,\qquad
                \lim_{n\to\infty}\frac{n^{\beta_c}}{\ln n}\mathcal{G}_{3}(n)
                =0.
            \end{equation}
            This completes the proof.
        \end{proof}

        \begin{corollary}
            The inferred propagator~$\mathcal{G}_n$ is transient; that is, 
            \begin{equation}
                \mathcal{G}_{\infty}=0.
            \end{equation}
        \end{corollary}

        \begin{proof}
            Since $\beta_c>0$, it follows from Proposition~\ref{prop:UDT_log:propagator-asymptotic} that 
            \begin{equation}
                    \lim_{n\to\infty}\mathcal{G}_n
                    =\lim_{n\to\infty}\frac{\ln n}{n^{\beta_c}}\frac{n^{\beta_c}}{\ln n}\mathcal{G}_n
                    =0\cdot \mathfrak{A}_{\mathcal{G}}
                    =0.
            \end{equation}
        \end{proof}
        
        \begin{corollary}
            The coefficient in Proposition~\ref{prop:UDT_log:propagator-asymptotic} is nonzero; that is, 
            \begin{equation}
                \mathfrak{A}_{\mathcal{G}}\neq0.
            \end{equation}
        \end{corollary}
        
        \begin{proof}
            Since $B(\beta_c,\gamma)>0$, $\mathfrak{A}_C>0$, $\sin(\pi\beta_c)>0$, and $\mathfrak{A}_S\neq0$, it follows from Proposition~\ref{prop:UDT_log:propagator-asymptotic} that 
            \begin{equation}
                \mathfrak{A}_{\mathcal{G}}=B(\beta_c,\gamma)\left(\frac{\sin(\pi\beta_c)}{\pi}\right)^2 \frac{\mathfrak{A}_S}{\mathfrak{A}_C}
                \neq0.
            \end{equation}
        \end{proof}

\begin{appendices}
\renewcommand{\thesubsection}{\thesection.\arabic{subsection}}
\makeatletter
\renewcommand{\p@subsection}{}
\makeatother
\section{Auxiliary results for the proofs of the universal decay theorems~\NoCaseChange{(ii)} and \NoCaseChange{(iv)}}
\label{app:UDT_no_log}
    In this appendix, we give the proofs of the auxiliary results used in Sec.~\ref{sec:UDT}.
    
    \subsection{Proof of $J_1<0$}
    \label{app:UDT:n=1}
        We show $J_1<0$.
        \begin{proof}
            Fix $z\in\mathbb{D}$. Since $\left(F_h\right)_{h\geq0}\in\ell^2$, the Cauchy--Schwarz inequality gives 
            \begin{equation}
                \sum_{h=0}^{\infty}\left|F_h\right||z|^h
                \leq \sqrt{\sum_{h=0}^{\infty}\left|F_h\right|^2}\sqrt{\sum_{h=0}^{\infty}|z|^{2h}}
                =\sqrt{\sum_{h=0}^{\infty}\left|F_h\right|^2}\frac{1}{\sqrt{1-|z|^2}}
                <\infty.
            \end{equation}
            In addition, since $\left(J_l\right)_{l\geq0}\in\ell^1$, we have 
            \begin{equation}
                \sum_{l=0}^{\infty}\left|J_l\right| |z|^l
                \leq \sum_{l=0}^{\infty}\left|J_l\right|
                <\infty.
            \end{equation}
            These imply 
            \begin{equation}
                \sum_{h=0}^{\infty}\sum_{l=0}^{\infty}\left|F_h\right|\left|J_l\right||z|^{h+l}
                =\left(\sum_{h=0}^{\infty}\left|F_h\right| |z|^h\right)\sum_{l=0}^{\infty}\left|J_{l}\right||z|^l
                <\infty.
            \end{equation}
            The Fubini--Tonelli theorem allows us to interchange the order of the sums. Setting $n=h+l$ gives 
            \begin{equation}
                1=F^{+}(z)J^{+}(z)
                =\sum_{n=0}^{\infty}\left[\sum_{h=0}^{n}F_{h}J_{n-h}\right]z^n.
            \end{equation}
            Since $z\in\mathbb{D}$ is arbitrary, by the uniqueness of the Taylor expansion, we find 
            \begin{equation}
                1=F_0J_0
            \end{equation}
            for $n=0$, and 
            \begin{equation}
                0=F_0J_1+F_1J_0
            \end{equation}
            for $n=1$.
            By the definition of $F^{+}(z)$, we have $F_0>0$. This leads to 
            \begin{equation}
                J_1=-\frac{F_1}{\left(F_0\right)^2}.
            \end{equation}
            Moreover, we have $\nu((0,1))>0$. Indeed, if $\nu$ is supported on $\{0\}$, we have $F_n=\int_0^1 r^n\nu(dr)=0$ for $n\geq1$, which contradicts the asymptotic form in Eq.~\eqref{eq:C_Szego-factor_F_asymptotic}. Therefore, by Eq.~\eqref{eq:C_Szego-factor_F_CM}, we obtain 
            \begin{equation}
                F_1
                =\int_0^{1}r\nu(dr)
                >0.
            \end{equation}
            This completes the proof.
        \end{proof}

    \subsection{Evaluation of the partial sum}
    \label{app:UDT:partial_sum}
        We establish the following asymptotic equivalences as $n\to\infty$:
        \begin{align}
            -\sum_{j=n+1}^{\infty}J_j 
            \sim \sqrt{\frac{B(\beta_c,\gamma)}{\mathfrak{A}_C}}\frac{\beta_c\sin(\pi\beta_c)}{\pi}\sum_{j=n+1}^{\infty}j^{-(1+\beta_c)} 
            \sim \sqrt{\frac{B(\beta_c,\gamma)}{\mathfrak{A}_C}}\frac{\beta_c\sin(\pi\beta_c)}{\pi}\int_{n}^{\infty}\dd x x^{-(1+\beta_c)}.
        \end{align}
        First, we show 
        \begin{equation}
            -\sum_{j=n+1}^{\infty}J_j
            \sim \sqrt{\frac{B(\beta_c,\gamma)}{\mathfrak{A}_C}}\frac{\beta_c\sin(\pi\beta_c)}{\pi}\sum_{j=n+1}^{\infty}j^{-(1+\beta_c)}\quad\text{as $n\to\infty$}.
        \end{equation}
        
        \begin{proof}
            Let $\mathfrak{d}>0$ be arbitrary. By Eq.~\eqref{eq:C_Szego-factor_J_asymptotic} and the definition of the asymptotic equivalence, for sufficiently large $N_1$, we have 
            \begin{equation}
                (1-\mathfrak{d})\sqrt{\frac{B(\beta_c,\gamma)}{\mathfrak{A}_C}}\frac{\beta_c\sin(\pi\beta_c)}{\pi}j^{-(1+\beta_c)}
                \leq-J_j
                \leq(1+\mathfrak{d})\sqrt{\frac{B(\beta_c,\gamma)}{\mathfrak{A}_C}}\frac{\beta_c\sin(\pi\beta_c)}{\pi}j^{-(1+\beta_c)}
            \end{equation}
            for all $j\geq N_1$.
            For $n\geq N_1$, summing this over $j\geq n+1$ gives 
            \begin{equation}
                (1-\mathfrak{d})\sqrt{\frac{B(\beta_c,\gamma)}{\mathfrak{A}_C}}\frac{\beta_c\sin(\pi\beta_c)}{\pi}\sum_{j=n+1}^{\infty}j^{-(1+\beta_c)}
                \leq-\sum_{j=n+1}^{\infty}J_j
                \leq(1+\mathfrak{d})\sqrt{\frac{B(\beta_c,\gamma)}{\mathfrak{A}_C}}\frac{\beta_c\sin(\pi\beta_c)}{\pi}\sum_{j=n+1}^{\infty}j^{-(1+\beta_c)}.
            \end{equation}
            This leads to 
            \begin{equation}
                (1-\mathfrak{d})
                \leq
                \frac{-\sum_{j=n+1}^{\infty}J_j}{\sqrt{\frac{B(\beta_c,\gamma)}{\mathfrak{A}_C}}\frac{\beta_c\sin(\pi\beta_c)}{\pi}\sum_{j=n+1}^{\infty}j^{-(1+\beta_c)}}
                \leq(1+\mathfrak{d}).
            \end{equation}
            This implies 
            \begin{equation}
                \left|\frac{-\sum_{j=n+1}^{\infty}J_j}{\sqrt{\frac{B(\beta_c,\gamma)}{\mathfrak{A}_C}}\frac{\beta_c\sin(\pi\beta_c)}{\pi}\sum_{j=n+1}^{\infty}j^{-(1+\beta_c)}}-1\right|
                \leq\mathfrak{d}.
            \end{equation}
            for all $n\geq N_1$. Since $\mathfrak{d}>0$ is arbitrary, this completes the proof.
        \end{proof}

        Second, we show 
        \begin{equation}
            \sum_{j=n+1}^{\infty}j^{-(1+\beta_c)}
            \sim \int_{n}^{\infty}\dd x x^{-(1+\beta_c)} \quad\text{as $n\to\infty$}.
        \end{equation}
        \begin{proof}
            Since $x^{-1-\beta_c}$ is decreasing in $x$, we have 
            \begin{equation}
                \frac{1}{\beta_c}(n+1)^{-\beta_c}
                = \int_{n+1}^{\infty}\dd x x^{-(1+\beta_c)}
                \leq \sum_{j=n+1}^{\infty}j^{-(1+\beta_c)}
                \leq \int_{n}^{\infty}\dd x x^{-(1+\beta_c)}
                = \frac{1}{\beta_c}n^{-\beta_c}.
            \end{equation}
            This implies 
            \begin{equation}
                \left(\frac{n}{n+1}\right)^{\beta_c}
                \leq \frac{\sum_{j=n+1}^{\infty}j^{-(1+\beta_c)}}{\frac{1}{\beta_c}n^{-\beta_c}}
                \leq 1.
            \end{equation}
            Taking the limit $n\to\infty$ gives 
            \begin{equation}
                \lim_{n\to\infty}\frac{\sum_{j=n+1}^{\infty}j^{-(1+\beta_c)}}{\frac{1}{\beta_c}n^{-\beta_c}}=1.
            \end{equation}
            This completes the proof.
        \end{proof}

    \subsection{Proof of $S^{+}(z)\in H^2(\mathbb{D})$}
    \label{app:UDT:S_plus_in_H2}
        We show that $S^{+}(z)\in H^2(\mathbb{D})$.
        \begin{proof}
            For the case of $\left(S_n\right)_{n\geq0}\in\ell^1$, we have  
            \begin{equation}
                \sum_{n=0}^{\infty}|S_n|^2
                \leq\left[\sum_{n=0}^{\infty}|S_n|\right]^2
                <\infty,
            \end{equation}
            which means $\left(S_n\right)_{n\geq0}\in\ell^2$. 
            For the case of $S_n\sim \mathfrak{A}_S n^{\eta-1}$ with $\eta\leq\beta_c$, since $2\eta-2\leq2\beta_c-2<-1$, the relations $\sum_{n=1}^{\infty}n^{2\eta-2}<\infty$ and 
            \begin{equation} 
                \frac{|S_n|^2}{n^{2\eta-2}}\to \mathfrak{A}_S^2>0 \quad \text{as $n\to\infty$}
            \end{equation}
            imply $(S_n)_{n\geq0}\in\ell^2$ by the limit comparison test.
            Therefore, in both cases, we obtain $\left(S_n\right)_{n\geq0}\in\ell^2$. Then, the series $\sum_{n=0}^{\infty}S_ne^{in\theta}$ converges in $L^2(\mathbb{T})$, and its sum belongs to $H^2(\mathbb{T})$. By Theorem~\ref{thm:Hardy_boundary_value}, this sum is equal to the $L^2(\mathbb{T})$ boundary value of a unique function in $H^2(\mathbb{D})$ whose Taylor coefficients are $S_n$. This function is $S^{+}(z)$, and hence $S^{+}(z)\in H^2(\mathbb{D})$. In particular, the following equality holds in $L^2(\mathbb{T})$:
            \begin{equation}
                S^{+}(e^{i\theta})=\sum_{n=0}^{\infty}S_ne^{in\theta}.
            \end{equation}
        \end{proof}

    \subsection{Fourier coefficients of $\Phi(e^{i\theta})$ for $n\geq0$}
    \label{app:UDT:Riesz_projection}
        We show that, for $n\geq0$, the $n$th Fourier coefficient of $\Phi(e^{i\theta})$ is given by
        \begin{equation}
            \Phi_n=\sum_{h=0}^{\infty}J_h S_{n+h}.
        \end{equation}

        \begin{proof}
            Since $\left(J_h\right)_{h\geq0}\in\ell^1$, the triangle inequality gives 
            \begin{equation}
                \left\|S^{+}(e^{i\theta})\left[\sum_{h=0}^{N'}J_h e^{-ih\theta}-J^{-}(e^{i\theta})\right]\right\|_{2}
                \leq \left\|S^{+}(e^{i\theta})\right\|_{2}\sup_{\theta}\left|\sum_{h=N'+1}^{\infty}J_h e^{-ih\theta}\right|
                \leq \left\|S^{+}(e^{i\theta})\right\|_{2}\sum_{h=N'+1}^{\infty}\left|J_h\right|
                \to 0\quad\text{as $N'\to\infty$},
            \end{equation}
            which means 
            \begin{equation}
                S^{+}(e^{i\theta})\sum_{h=0}^{N'}J_h e^{-ih\theta}
                \to S^{+}(e^{i\theta})J^{-}(e^{i\theta})\quad \text{as $N'\to\infty$}\text{ in }L^2(\mathbb{T}).
            \end{equation}
            Let us show that we can interchange the order of the limit and the integral; that is,  
            \begin{equation}
                \lim_{N'\to\infty}\frac{1}{2\pi}\int_{-\pi}^{\pi}\dd \theta S^{+}(e^{i\theta})\left[\sum_{h=0}^{N'}J_h e^{-ih\theta}\right]e^{-in\theta}
                =\frac{1}{2\pi}\int_{-\pi}^{\pi}\dd \theta S^{+}(e^{i\theta})J^{-}(e^{i\theta}) e^{-in\theta}.
            \end{equation}
            Indeed, by the Cauchy--Schwarz inequality, we have  
            \begin{align}
                \left|\frac{1}{2\pi}\int_{-\pi}^{\pi}\dd \theta S^{+}(e^{i\theta})\left[\sum_{h=0}^{N'}J_h e^{-ih\theta}-J^{-}(e^{i\theta})\right]e^{-in\theta}\right|
                \leq \left\|S^{+}(e^{i\theta})\left[\sum_{h=0}^{N'}J_h e^{-ih\theta}-J^{-}(e^{i\theta})\right]\right\|_{2}
                \to0 \quad \text{as $N'\to\infty$},
            \end{align}
            which implies 
            \begin{equation}
                \lim_{N'\to\infty}\frac{1}{2\pi}\int_{-\pi}^{\pi}\dd \theta S^{+}(e^{i\theta})\left[\sum_{h=0}^{N'}J_h e^{-ih\theta}\right]e^{-in\theta}
                = \frac{1}{2\pi}\int_{-\pi}^{\pi}\dd \theta S^{+}(e^{i\theta})J^{-}(e^{i\theta}) e^{-in\theta}
                =\Phi_n.
            \end{equation}
            For the left-hand side, interchanging the order of the finite sum and the integral yields  
            \begin{equation}
                \lim_{N'\to\infty}\frac{1}{2\pi}\int_{-\pi}^{\pi}\dd \theta S^{+}(e^{i\theta})\left[\sum_{h=0}^{N'}J_h e^{-ih\theta}\right]e^{-in\theta}
                =\lim_{N'\to\infty}\sum_{h=0}^{N'}J_h S_{n+h}
                =\sum_{h=0}^{\infty}J_h S_{n+h}
            \end{equation}
            for $n\geq0$.
            This completes the proof.
        \end{proof}

    \subsection{Finiteness of a weighted sum of $|S_n-S_{n+1}|$}
    \label{app:UDT:weighted-diff}
        We show that 
        \begin{equation}
        \label{eq:weighted_diff_finite}
            \sum_{n=0}^{\infty}(n+1)^{1-\beta_c}\left|S_n-S_{n+1}\right|<\infty.
        \end{equation}

        \begin{proof}
           By the eventual monotonicity of $S_n$ and $S_n\to0$ as $n\to\infty$, there exists an integer $N_0$ such that 
            \begin{equation}
                \left|S_n-S_{n+1}\right|=|S_n|-|S_{n+1}|
            \end{equation}
            for $n\geq N_0$.
            Since the sum up to $N_0-1$ is a constant, it suffices to prove that 
            \begin{equation}
                \sum_{n=N_0}^{\infty}(n+1)^{1-\beta_c}\left(\left|S_n\right|-\left|S_{n+1}\right|\right)<\infty.
            \end{equation}
            Let $N'\geq N_0$ be any fixed integer. Summation by parts gives
            \begin{align}
                0&\leq\sum_{n=N_0}^{N'}(n+1)^{1-\beta_c}\left|S_n-S_{n+1}\right|\\
                &=\sum_{n=N_0}^{N'}(n+1)^{1-\beta_c}\left(|S_n|-|S_{n+1}|\right)\\
                &=(N_0+1)^{1-\beta_c}|S_{N_0}|-(N'+1)^{1-\beta_c}|S_{N'+1}|+\sum_{n=N_0+1}^{N'}\left((n+1)^{1-\beta_c}-n^{1-\beta_c}\right)|S_{n}|\\
                &\leq(N_0+1)^{1-\beta_c}|S_{N_0}| 
                +\sum_{n=N_0+1}^{N'}\left((n+1)^{1-\beta_c}-n^{1-\beta_c}\right)|S_{n}|.
            \end{align}
            Since the first term in the last line is a constant, it suffices to bound the second term in the last line uniformly in $N'$. By the mean value theorem for $x^{1-\beta_c}$, there exists $\xi\in(n,n+1)$ such that 
            \begin{equation}
                (n+1)^{1-\beta_c}-n^{1-\beta_c}=(1-\beta_c)\xi^{-\beta_c}.
            \end{equation}
            Since $(1-\beta_c)x^{-\beta_c}$ is nonincreasing and $\xi>n$, we have 
            \begin{equation}
                (n+1)^{1-\beta_c}-n^{1-\beta_c}\leq(1-\beta_c)n^{-\beta_c},
            \end{equation}
            and thus we obtain 
            \begin{equation}
                \sum_{n=N_0+1}^{N'}\left((n+1)^{1-\beta_c}-n^{1-\beta_c}\right)|S_{n}|
                \leq(1-\beta_c)\sum_{n=N_0+1}^{N'}n^{-\beta_c}|S_n|.
            \end{equation}
            For the case of $\left(S_n\right)_{n\geq0}\in\ell^1$, since $n^{-\beta_c}\leq1$, we have 
            \begin{equation}
                \sum_{n=N_0+1}^{N'}n^{-\beta_c}|S_n|\leq\sum_{n=N_0+1}^{N'}|S_n|\leq\sum_{n=0}^{\infty}|S_n|<\infty.
            \end{equation}
            For the case of $S_n\sim \mathfrak{A}_S n^{\eta-1}$ with $\eta<\beta_c$, by taking $N_0$ larger if necessary, we may assume that
            \begin{equation}
                \left|S_n\right|\leq 2\mathfrak{A}_S n^{\eta-1}
            \end{equation}
            for all $n\geq N_0$. This leads to 
            \begin{equation}
                n^{-\beta_c}|S_n|\leq 2\mathfrak{A}_S n^{-\beta_c}n^{\eta-1}
                =2\mathfrak{A}_S n^{-1-(\beta_c-\eta)},
            \end{equation}
            and since $\beta_c-\eta>0$, we have 
            \begin{equation}
                \sum_{n=N_0+1}^{N'}n^{-\beta_c}|S_n|
                \leq 2\mathfrak{A}_S\sum_{n=N_0+1}^{N'} n^{-1-(\beta_c-\eta)}
                \leq 2\mathfrak{A}_S\sum_{n=1}^{\infty} n^{-1-(\beta_c-\eta)}
                <\infty.
            \end{equation}
            Therefore, in both cases, the partial sums are uniformly bounded from above. Since all terms are nonnegative, it follows that 
            \begin{equation}
                \sum_{n=N_0}^{\infty}(n+1)^{1-\beta_c}\left(\left|S_n\right|-\left|S_{n+1}\right|\right)<\infty.
            \end{equation}
            This completes the proof.
        \end{proof}

    \subsection{The product $F^{-}(e^{i\theta})\Phi^{-}(e^{i\theta})$ has only negative modes}
    \label{app:UDT:negative_modes}
        We show that $F^{-}(e^{i\theta})\Phi^{-}(e^{i\theta})$ has only negative modes; that is, 
        \begin{equation}
            \frac{1}{2\pi}\int_{-\pi}^{\pi}\dd \theta F^{-}(e^{i\theta})\Phi^{-}(e^{i\theta}) e^{-in\theta}=0
        \end{equation}
        for $n\geq 0$.

        \begin{proof}
            Let us show that we can interchange the order of the limit and the integral; that is,
            \begin{equation}
            \label{eq:negative_modes_limit_exchange}
                \lim_{N'\to\infty}\frac{1}{2\pi}\int_{-\pi}^{\pi}\dd \theta \sum_{j=0}^{N'}F_je^{-ij\theta}\sum_{k=1}^{N'}\Phi_{-k}e^{-ik\theta} e^{-in\theta}
                =\frac{1}{2\pi}\int_{-\pi}^{\pi}\dd \theta F^{-}(e^{i\theta})\Phi^{-}(e^{i\theta}) e^{-in\theta}.
            \end{equation}
            Let $N'$ be any fixed integer. We consider the product of the partial sums 
            \begin{equation}
                \sum_{j=0}^{N'}F_je^{-ij\theta}\sum_{k=1}^{N'}\Phi_{-k}e^{-ik\theta}
                =\sum_{j=0}^{N'}\sum_{k=1}^{N'}F_j\Phi_{-k}e^{-i(j+k)\theta}.
            \end{equation}
            Since $\left(F_j\right)_{j\geq0}\in\ell^2$ and $\left(\Phi_{-k}\right)_{k\geq1}\in\ell^2$, Parseval's identity gives 
            \begin{equation}
                \sum_{j=0}^{N'}F_je^{-ij\theta}
                \to F^{-}(e^{i\theta}) \quad\text{as $N'\to\infty$ in }L^2(\mathbb{T}),\qquad
                \sum_{k=1}^{N'}\Phi_{-k}e^{-ik\theta}
                \to \Phi^{-}(e^{i\theta})\quad\text{as $N'\to\infty$ in }L^2(\mathbb{T}).
            \end{equation}
            The triangle inequality and the Cauchy--Schwarz inequality yield 
            \begin{align}
                &\quad\left\|F^{-}(e^{i\theta})\Phi^{-}(e^{i\theta})-\sum_{j=0}^{N'}F_je^{-ij\theta}\sum_{k=1}^{N'}\Phi_{-k}e^{-ik\theta}\right\|_{1} \\
                &\leq \left\|F^{-}(e^{i\theta})\Phi^{-}(e^{i\theta})-F^{-}(e^{i\theta})\sum_{k=1}^{N'}\Phi_{-k}e^{-ik\theta}\right\|_{1}
                +\left\|F^{-}(e^{i\theta})\sum_{k=1}^{N'}\Phi_{-k}e^{-ik\theta}-\sum_{j=0}^{N'}F_je^{-ij\theta}\sum_{k=1}^{N'}\Phi_{-k}e^{-ik\theta}\right\|_{1} \\
                &\leq \left\|F^{-}(e^{i\theta})\right\|_{2}\left\|\Phi^{-}(e^{i\theta})-\sum_{k=1}^{N'}\Phi_{-k}e^{-ik\theta}\right\|_{2}
                +\left\|\sum_{k=1}^{N'}\Phi_{-k}e^{-ik\theta}\right\|_{2}\left\|F^{-}(e^{i\theta})-\sum_{j=0}^{N'}F_je^{-ij\theta}\right\|_{2} \\
                &\to 0\quad \text{as $N'\to\infty$}. 
            \end{align}
            This means 
            \begin{equation}
                \sum_{j=0}^{N'}F_je^{-ij\theta}\sum_{k=1}^{N'}\Phi_{-k}e^{-ik\theta}
                \to F^{-}(e^{i\theta})\Phi^{-}(e^{i\theta}) 
                \quad\text{as $N'\to\infty$ in $L^1(\mathbb{T})$}.
            \end{equation}
            Therefore, for any fixed $n\geq0$, we obtain 
            \begin{align}
                &\quad\left|\frac{1}{2\pi}\int_{-\pi}^{\pi}\dd \theta F^{-}(e^{i\theta})\Phi^{-}(e^{i\theta}) e^{-in\theta}-\frac{1}{2\pi}\int_{-\pi}^{\pi}\dd \theta \sum_{j=0}^{N'}F_je^{-ij\theta}\sum_{k=1}^{N'}\Phi_{-k}e^{-ik\theta} e^{-in\theta}\right| \\
                &\leq \frac{1}{2\pi}\int_{-\pi}^{\pi}\dd \theta \left|F^{-}(e^{i\theta})\Phi^{-}(e^{i\theta})-\sum_{j=0}^{N'}F_je^{-ij\theta}\sum_{k=1}^{N'}\Phi_{-k}e^{-ik\theta}\right|\\
                &\to 0\quad \text{as $N'\to\infty$},
            \end{align}
            which leads to Eq.~\eqref{eq:negative_modes_limit_exchange}.
            
            For any fixed $N'$ and $n\geq0$, each term of the double sum contains $e^{-i(j+k+n)\theta}$ with $j+k+n\geq1$, and hence  
            \begin{equation}
                \frac{1}{2\pi}\int_{-\pi}^{\pi}\dd \theta \sum_{j=0}^{N'}F_je^{-ij\theta}\sum_{k=1}^{N'}\Phi_{-k}e^{-ik\theta} e^{-in\theta}=0
            \end{equation}
            Taking the limit $N'\to\infty$, we obtain 
            \begin{equation}
                \frac{1}{2\pi}\int_{-\pi}^{\pi}\dd \theta F^{-}(e^{i\theta})\Phi^{-}(e^{i\theta}) e^{-in\theta}
                =\lim_{N'\to\infty}\frac{1}{2\pi}\int_{-\pi}^{\pi}\dd \theta \sum_{j=0}^{N'}F_je^{-ij\theta}\sum_{k=1}^{N'}\Phi_{-k}e^{-ik\theta} e^{-in\theta}
                =0.
            \end{equation}
        \end{proof}

    \subsection{Vanishing of the boundary term in the proof of the positivity of $\mathfrak{A}_{\mathcal{G}}$}
    \label{app:UDT:M-vanish}
        We show Eq.~\eqref{eq:UDT:M-vanish}, that is,  
        \begin{equation}
            \lim_{N'\to\infty}\left[\sum_{j=0}^{N'}J_{j}\right]R_{N'+1}=0.
        \end{equation}

        \begin{proof}
            By Eq.~\eqref{eq:partial_sum_inequality}, we have 
            \begin{equation}
                0
                \leq\left|\sum_{j=0}^{N'}J_{j}\right|\leq c_0 (N'+1)^{-\beta_c}.
            \end{equation}
            Moreover, since $R_0=0$ and $S_h=R_{h+1}-R_h$, we have $R_{N'+1}=\sum_{h=0}^{N'}S_h$.
            
            For the case of $(S_h)_{h\geq0}\in\ell^1$, we have 
            \begin{equation}
                0
                \leq\left|R_{N'+1}\right|
                =\left|\sum_{h=0}^{N'}S_h\right|
                \leq \sum_{h=0}^{N'}|S_h|\leq \sum_{h=0}^{\infty}|S_h|
                <\infty,
            \end{equation}
            which implies $R_{N'+1}=O(1)$. This leads to 
            \begin{equation}
                \left|\left[\sum_{j=0}^{N'}J_{j}\right]R_{N'+1}\right|
                \leq c_0 (N'+1)^{-\beta_c} O(1)
                =O((N')^{-\beta_c})\to0\quad \text{as $N'\to\infty$}.
            \end{equation}
            For the case of $S_h\sim \mathfrak{A}_S h^{\eta-1}$ with $\eta<\beta_c$, there exists an integer $N_0$ such that $\left|S_h\right|\leq 2\mathfrak{A}_S h^{\eta-1}$ for $h\geq N_0$. This implies 
            \begin{equation}
                0
                \leq \left|R_{N'+1}\right|\leq \sum_{h=0}^{N'}|S_h|
                \leq \sum_{h=0}^{N_0-1}|S_h|+2\mathfrak{A}_S\sum_{h=N_0}^{N'} h^{\eta-1},
            \end{equation}
            and thus we have 
            \begin{align}
                \left|R_{N'+1}\right|=
                \begin{cases}
                    O(N'^{\eta}) & (\eta>0),\\
                    O(\ln N') & (\eta=0),\\
                    O(1) & (\eta<0).
                \end{cases}
            \end{align}
            Since $\eta<\beta_c$, we obtain 
            \begin{align}
                \left|\left[\sum_{j=0}^{N'}J_{j}\right]R_{N'+1}\right|=
                \begin{cases}
                    O(N'^{\eta-\beta_c}) & (\eta>0),\\
                    O(N'^{-\beta_c}\ln N') & (\eta=0),\\
                    O(N'^{-\beta_c}) & (\eta<0)
                \end{cases}
                \to0\quad \text{as $N'\to\infty$}.
            \end{align}
            This completes the proof.
        \end{proof}

\section{Auxiliary results for the proof of the universal decay theorems~\NoCaseChange{(iii)}}
\label{app:UDT_log}
    In this appendix, we give the proofs of the auxiliary results used in Sec.~\ref{sec:UDT_log}.

    \subsection{Asymptotic behavior of a weighted sum of $S_k$}
    \label{app:UDT_log:partial_sum}
        Let $N_0$ be as in the proof of Lemma~\ref{lemma:UDT_log:Phi_n_sum_asymptotic}. We establish the following asymptotic equivalence as $N\to\infty$:
        \begin{equation}
            \sum_{k=N_0+1}^{N}S_k\sum_{j=0}^{k-N_0}J_{j}
            \sim \sqrt{\frac{B(\beta_c,\gamma)}{\mathfrak{A}_C}}\frac{\sin(\pi\beta_c)}{\pi}\mathfrak{A}_S\ln N.
        \end{equation}

        \begin{proof}
            By the asymptotic form $S_n\sim \mathfrak{A}_S n^{\beta_c-1}$ and Eq.~\eqref{eq:partial_sum_asymptotic}, we find 
            \begin{equation}
                \lim_{k\to\infty}\frac{S_k\left[\sum_{j=0}^{k-N_0}J_{j}\right]}{\sqrt{\frac{B(\beta_c,\gamma)}{\mathfrak{A}_C}}\frac{\sin(\pi\beta_c)}{\pi}\mathfrak{A}_S k^{-1}}
                =\lim_{k\to\infty}\frac{S_k}{\mathfrak{A}_S k^{\beta_c-1}}\cdot\lim_{k\to\infty}\frac{\sum_{j=0}^{k-N_0}J_{j}}{\sqrt{\frac{B(\beta_c,\gamma)}{\mathfrak{A}_C}}\frac{\sin(\pi\beta_c)}{\pi}(k-N_0)^{-\beta_c}}\cdot
                \lim_{k\to\infty}\frac{(k-N_0)^{-\beta_c}}{k^{-\beta_c}}
                =1.
            \end{equation}
            Let $0<\mathfrak{d}<1$ be arbitrary. By this limit, there exists an integer $N_{\mathfrak{d}}\geq \max\{N_0+1, 2\}$ such that
            \begin{equation}
            \label{eq:mathfrak-d_inequality}
                (1-\mathfrak{d})\sqrt{\frac{B(\beta_c,\gamma)}{\mathfrak{A}_C}}\frac{\sin(\pi\beta_c)}{\pi}\mathfrak{A}_S k^{-1}
                \leq S_k\left[\sum_{j=0}^{k-N_0}J_{j}\right]
                \leq (1+\mathfrak{d})\sqrt{\frac{B(\beta_c,\gamma)}{\mathfrak{A}_C}}\frac{\sin(\pi\beta_c)}{\pi}\mathfrak{A}_S k^{-1}
            \end{equation}
            for all $k\geq N_{\mathfrak{d}}$.
            Let us split the sum over $k$ into two parts; that is, 
            \begin{equation}
                \sum_{k=N_0+1}^{N}S_k\sum_{j=0}^{k-N_0}J_{j}
                =\sum_{k=N_0+1}^{N_{\mathfrak{d}}-1}S_k\sum_{j=0}^{k-N_0}J_{j}+\sum_{k=N_{\mathfrak{d}}}^{N}S_k\sum_{j=0}^{k-N_0}J_{j}.
            \end{equation}
            The first term on the right-hand side does not depend on $N$. Let us focus on the second term on the right-hand side. Summing Eq.~\eqref{eq:mathfrak-d_inequality} over $k$ from $k=N_{\mathfrak{d}}$ to $k=N$ gives 
            \begin{align}
            \label{eq:mathfrak-d_sum-inequality}
                (1-\mathfrak{d})\sqrt{\frac{B(\beta_c,\gamma)}{\mathfrak{A}_C}}\frac{\sin(\pi\beta_c)}{\pi}\mathfrak{A}_S\sum_{k=N_{\mathfrak{d}}}^{N}k^{-1}
                \leq \sum_{k=N_{\mathfrak{d}}}^{N}S_k\sum_{j=0}^{k-N_0}J_{j}
                \leq (1+\mathfrak{d})\sqrt{\frac{B(\beta_c,\gamma)}{\mathfrak{A}_C}}\frac{\sin(\pi\beta_c)}{\pi}\mathfrak{A}_S\sum_{k=N_{\mathfrak{d}}}^{N}k^{-1}.
            \end{align}
            Since $x^{-1}$ is deceasing, we have 
            \begin{equation}
                \ln\frac{N+1}{N_{\mathfrak{d}}}
                =\int_{N_{\mathfrak{d}}}^{N+1}x^{-1}\dd x
                \leq \sum_{k=N_{\mathfrak{d}}}^{N}k^{-1}
                \leq \int_{N_{\mathfrak{d}}-1}^{N}x^{-1}\dd x
                =\ln\frac{N}{N_{\mathfrak{d}}-1},
            \end{equation}
            which, together with Eq.~\eqref{eq:mathfrak-d_sum-inequality}, yields 
            \begin{equation}
                (1-\mathfrak{d})\sqrt{\frac{B(\beta_c,\gamma)}{\mathfrak{A}_C}}\frac{\sin(\pi\beta_c)}{\pi}\mathfrak{A}_S\ln\frac{N+1}{N_{\mathfrak{d}}}
                \leq \sum_{k=N_{\mathfrak{d}}}^{N}S_k\sum_{j=0}^{k-N_0}J_{j}
                \leq (1+\mathfrak{d})\sqrt{\frac{B(\beta_c,\gamma)}{\mathfrak{A}_C}}\frac{\sin(\pi\beta_c)}{\pi}\mathfrak{A}_S\ln\frac{N}{N_{\mathfrak{d}}-1}.
            \end{equation}
            Adding the partial sum $\sum_{k=N_0+1}^{N_{\mathfrak{d}}-1}S_k\sum_{j=0}^{k-N_0}J_{j}$ gives 
            \begin{align}
                &\quad \sum_{k=N_0+1}^{N_{\mathfrak{d}}-1}S_k\sum_{j=0}^{k-N_0}J_{j}
                + (1-\mathfrak{d})\sqrt{\frac{B(\beta_c,\gamma)}{\mathfrak{A}_C}}\frac{\sin(\pi\beta_c)}{\pi}\mathfrak{A}_S\ln\frac{N+1}{N_{\mathfrak{d}}} \\
                &\leq \sum_{k=N_0+1}^{N}S_k\sum_{j=0}^{k-N_0}J_{j} \\
                &\leq \sum_{k=N_0+1}^{N_{\mathfrak{d}}-1}S_k\sum_{j=0}^{k-N_0}J_{j}
                + (1+\mathfrak{d})\sqrt{\frac{B(\beta_c,\gamma)}{\mathfrak{A}_C}}\frac{\sin(\pi\beta_c)}{\pi}\mathfrak{A}_S\ln\frac{N}{N_{\mathfrak{d}}-1}.
            \end{align}
            Dividing by $\sqrt{B(\beta_c,\gamma)/\mathfrak{A}_C}\,\sin(\pi\beta_c)\mathfrak{A}_S\ln N/\pi$, which is positive, and taking the limit $N\to\infty$, we obtain  
            \begin{equation}
                1-\mathfrak{d}
                \leq \liminf_{N\to\infty}\frac{\sum_{k=N_0+1}^{N}S_k\sum_{j=0}^{k-N_0}J_{j}}{\sqrt{\frac{B(\beta_c,\gamma)}{\mathfrak{A}_C}}\frac{\sin(\pi\beta_c)}{\pi}\mathfrak{A}_S\ln N}
                \leq \limsup_{N\to\infty}\frac{\sum_{k=N_0+1}^{N}S_k\sum_{j=0}^{k-N_0}J_{j}}{\sqrt{\frac{B(\beta_c,\gamma)}{\mathfrak{A}_C}}\frac{\sin(\pi\beta_c)}{\pi}\mathfrak{A}_S\ln N}
                \leq 1+\mathfrak{d}.
            \end{equation}
            Since $\mathfrak{d}$ is arbitrary, we obtain 
            \begin{equation}
                \lim_{N\to\infty}\frac{\sum_{k=N_0+1}^{N}S_k\sum_{j=0}^{k-N_0}J_{j}}{\sqrt{\frac{B(\beta_c,\gamma)}{\mathfrak{A}_C}}\frac{\sin(\pi\beta_c)}{\pi}\mathfrak{A}_S\ln N}
                =1.
            \end{equation}
            This completes the proof.
        \end{proof}

    \subsection{Upper bound for the sum of $\left|\mathcal{G}_{n+1}-\mathcal{G}_n\right|$}
    \label{app:UDT_log:G-difference_inequality}
        We show Eq.~\eqref{eq:G_difference_inequality}, that is, 
        \begin{equation}
            \sum_{n=0}^{\infty}|\mathcal{G}_{n+1}-\mathcal{G}_n|
            \leq 2\sum_{k=0}^{\infty}\left|S_k-S_{k+1}\right|\sum_{j=0}^{k}\left[\sum_{l=0}^{j}J_{l}\right]^2.
        \end{equation}

        \begin{proof}
            As shown in the proof of Lemma~\ref{lemma:UDT_log:K_n_l1}, we have 
            \begin{equation}
                \mathcal{G}_n
                =\sum_{k=0}^{\infty}(S_{k}-S_{k+1})\sum_{j=0}^{\min\{k,n-1\}}\sum_{l=0}^{k-j}J_{l}\sum_{h=0}^{n-1-j}J_{h}.
            \end{equation}
            Since the series representations of $\mathcal{G}_{n+1}$ and $\mathcal{G}_{n}$ are absolutely convergent for each fixed $n$, we can take their difference term by term, and we obtain 
            \begin{align}
                \mathcal{G}_{n+1}-\mathcal{G}_n 
                &=\sum_{k=0}^{\infty}(S_{k}-S_{k+1})\sum_{j=0}^{\min\{k,n\}}\sum_{l=0}^{k-j}J_{l}\sum_{h=0}^{n-j}J_{h}
                -\sum_{k=0}^{\infty}(S_{k}-S_{k+1})\sum_{j=0}^{\min\{k,n-1\}}\sum_{l=0}^{k-j}J_{l}\sum_{h=0}^{n-1-j}J_{h} \\
                &=\sum_{k=0}^{\infty}(S_{k}-S_{k+1})\left[\sum_{j=0}^{\min\{k,n\}}\sum_{l=0}^{k-j}J_{l}\sum_{h=0}^{n-j}J_{h}-\sum_{j=0}^{\min\{k,n-1\}}\sum_{l=0}^{k-j}J_{l}\sum_{h=0}^{n-1-j}J_{h}\right].
            \end{align}
            The triangle inequality gives 
            \begin{align}
                |\mathcal{G}_{n+1}-\mathcal{G}_n| 
                &\leq \sum_{k=0}^{\infty}|S_{k}-S_{k+1}|\left|\sum_{j=0}^{\min\{k,n\}}\sum_{l=0}^{k-j}J_{l}\sum_{h=0}^{n-j}J_{h}-\sum_{j=0}^{\min\{k,n-1\}}\sum_{l=0}^{k-j}J_{l}\sum_{h=0}^{n-1-j}J_{h}\right|, 
            \end{align}
            and summing over $n\geq0$ gives 
            \begin{align}
                \sum_{n=0}^{\infty}|\mathcal{G}_{n+1}-\mathcal{G}_n| 
                &\leq \sum_{n=0}^{\infty}\sum_{k=0}^{\infty}|S_{k}-S_{k+1}|\left|\sum_{j=0}^{\min\{k,n\}}\sum_{l=0}^{k-j}J_{l}\sum_{h=0}^{n-j}J_{h}-\sum_{j=0}^{\min\{k,n-1\}}\sum_{l=0}^{k-j}J_{l}\sum_{h=0}^{n-1-j}J_{h}\right|. 
            \end{align}
            Since all terms of the sums with respect to $n$ and $k$ are nonnegative, Tonelli's theorem allows us to interchange the order of the sums. This leads to 
            \begin{align}
                \sum_{n=0}^{\infty}|\mathcal{G}_{n+1}-\mathcal{G}_n| 
                &\leq \sum_{k=0}^{\infty}|S_{k}-S_{k+1}|\sum_{n=0}^{\infty}\left|\sum_{j=0}^{\min\{k,n\}}\sum_{l=0}^{k-j}J_{l}\sum_{h=0}^{n-j}J_{h}-\sum_{j=0}^{\min\{k,n-1\}}\sum_{l=0}^{k-j}J_{l}\sum_{h=0}^{n-1-j}J_{h}\right|.
            \end{align}
            For any fixed $k$, the expression in the absolute value is the difference between the values at $n+1$ and $n$ of the quantity in Eq.~\eqref{eq:inner_sum_max-evaluation}. By the nonnegativity and the strict unimodality of this quantity, the sum over $n$ is bounded by twice the maximum value in Eq.~\eqref{eq:inner_sum_maximum}. This fact leads to 
            \begin{align}
                \sum_{n=0}^{\infty}|\mathcal{G}_{n+1}-\mathcal{G}_n| 
                &\leq 2\sum_{k=0}^{\infty}|S_{k}-S_{k+1}|\sum_{j=0}^{k}\left[\sum_{l=0}^{j}J_{l}\right]^2.
            \end{align}
        \end{proof}

    \subsection{Finiteness of a weighted sum of $|S_k-S_{k+1}|$}
    \label{app:UDT_log:weighted-diff}
        We show that 
        \begin{equation}
            \sum_{k=0}^{\infty}(k+1)^{1-2\beta_c}\left|S_k-S_{k+1}\right|<\infty.
        \end{equation}

        \begin{proof}
            By the eventual monotonicity of $S_k$ and the asymptotic form $S_k\sim \mathfrak{A}_{S}k^{\beta_c-1}$, there exists an integer $N_0$ such that $0\leq S_{k+1}\leq S_k\leq 2\mathfrak{A}_S k^{\beta_c-1}$ for all $k\geq N_0$. Since $|S_k-S_{k+1}|=S_k-S_{k+1}$ for $k\geq N_0$ and the sum up to $N_0-1$ is a constant, it suffices to prove that 
            \begin{equation}
                \sum_{k=N_0}^{\infty}(k+1)^{1-2\beta_c}(S_k-S_{k+1})<\infty.
            \end{equation}

            Let $N'\geq N_0$ be any fixed integer. Summation by parts gives 
            \begin{align}
                0
                &\leq\sum_{k=N_0}^{N'}(k+1)^{1-2\beta_c}(S_k-S_{k+1}) \\
                &=(N_0+1)^{1-2\beta_c}S_{N_0}-(N'+1)^{1-2\beta_c}S_{N'+1}+\sum_{k=N_0+1}^{N'}\left(\left(k+1\right)^{1-2\beta_c}-k^{1-2\beta_c}\right)S_k\\
                &\leq (N_0+1)^{1-2\beta_c}S_{N_0}+\sum_{k=N_0+1}^{N'}\left(\left(k+1\right)^{1-2\beta_c}-k^{1-2\beta_c}\right)S_k.
            \end{align}
            Since the first term in the last line is a constant, it suffices to bound the second term in the last line uniformly in $N'$. By the mean value theorem, in the same way as in Appendix~\ref{app:UDT:weighted-diff}, we have 
            \begin{equation}
                \left(k+1\right)^{1-2\beta_c}-k^{1-2\beta_c}\leq (1-2\beta_c)k^{-2\beta_c}, 
            \end{equation}
            which leads to 
            \begin{equation}
                \sum_{k=N_0+1}^{N'}\left(\left(k+1\right)^{1-2\beta_c}-k^{1-2\beta_c}\right)S_k
                \leq (1-2\beta_c)\sum_{k=N_0+1}^{N'}k^{-2\beta_c}S_k.
            \end{equation}
            Since $S_k\leq 2\mathfrak{A}_S k^{\beta_c-1}$ for all $k\geq N_0$, we find 
            \begin{equation}
                (1-2\beta_c)\sum_{k=N_0+1}^{N'}k^{-2\beta_c}S_k
                \leq 2(1-2\beta_c)\mathfrak{A}_{S}\sum_{k=N_0+1}^{N'}k^{-1-\beta_c}
                \leq 2(1-2\beta_c)\mathfrak{A}_{S}\sum_{k=N_0+1}^{\infty}k^{-1-\beta_c}
                <\infty,
            \end{equation}
            where we used $\beta_c>0$.
            Therefore, the partial sums are uniformly bounded from above. Since all terms are nonnegative, it follows that
            \begin{equation}
                \sum_{k=N_0}^{\infty}(k+1)^{1-2\beta_c}(S_k-S_{k+1})<\infty.
            \end{equation}
            This completes the proof.
        \end{proof}

    \subsection{Order estimate of a weighted sum of $S_k-S_{k+1}$ over $k\geq\lfloor n/2\rfloor+1$}
    \label{app:UDT_log:G3}
        We show Eq.~\eqref{eq:UDT_log:G3_asymptotic}, that is, 
        \begin{equation}
            \sum_{k=\lfloor n/2\rfloor+1}^{\infty}(S_{k}-S_{k+1})(k+1)^{1-2\beta_c}
            =O(n^{-\beta_c}).
        \end{equation}

        \begin{proof}
            Let $N_0$ be as in the proof of Lemma~\ref{lemma:UDT_log:Phi_n_sum_asymptotic}, so that $S_k\geq S_{k+1}\geq0$ for all $k\geq N_0$. Take $n$ so large that $\lfloor n/2\rfloor+1\geq N_0$ and $S_k\leq 2\mathfrak{A}_S k^{\beta_c-1}$ for all $k\geq\lfloor n/2\rfloor+2$. In particular, $S_k\geq S_{k+1}\geq0$ for all $k\geq\lfloor n/2\rfloor+1$. Let $N'\geq\lfloor n/2\rfloor+1$ be any fixed integer. Summation by parts gives 
            \begin{align}
                &\quad\sum_{k=\lfloor n/2\rfloor+1}^{N'}(S_{k}-S_{k+1})(k+1)^{1-2\beta_c} \\
                &=S_{\lfloor n/2\rfloor+1}\left(\left\lfloor \frac{n}{2}\right\rfloor+2\right)^{1-2\beta_c}-S_{N'+1}(N'+1)^{1-2\beta_c}+\sum_{k=\lfloor n/2\rfloor+2}^{N'}S_k\left[(k+1)^{1-2\beta_c}-k^{1-2\beta_c}\right].
            \end{align}
            Since $S_k\sim \mathfrak{A}_S k^{\beta_c-1}$, we have $S_k=O(k^{\beta_c-1})$. Hence, the first term in the last line is $O(n^{-\beta_c})$, and the second term in the last line is $O(N'^{-\beta_c})\to0$ as $N'\to\infty$. Let us evaluate the third term in the last line. 
            By the mean value theorem, in the same way as in Appendix~\ref{app:UDT:weighted-diff}, we have 
            \begin{equation}
                (k+1)^{1-2\beta_c}-k^{1-2\beta_c} \leq (1-2\beta_c)k^{-2\beta_c}, 
            \end{equation}
            which implies 
            \begin{align}
                \sum_{k=\lfloor n/2\rfloor+2}^{N'}S_k\left[(k+1)^{1-2\beta_c}-k^{1-2\beta_c}\right]
                \leq (1-2\beta_c)\sum_{k=\lfloor n/2\rfloor+2}^{N'}S_k k^{-2\beta_c} 
                \leq 2(1-2\beta_c)\mathfrak{A}_S\sum_{k=\lfloor n/2\rfloor+2}^{N'}k^{-1-\beta_c}, 
            \end{align}
            where we used $S_k\leq 2\mathfrak{A}_S k^{\beta_c-1}$.
            In the same way as in Eq.~\eqref{eq:p_series_estimate}, we have 
            \begin{align}
                \sum_{k=\lfloor n/2\rfloor+2}^{N'}k^{-1-\beta_c}
                &\leq \left(\left\lfloor \frac{n}{2}\right\rfloor+2\right)^{-1-\beta_c}+\int_{\lfloor n/2\rfloor+2}^{N'}\dd x x^{-1-\beta_c}\\
                &\leq \left(\left\lfloor \frac{n}{2}\right\rfloor+2\right)^{-1-\beta_c}+\int_{\lfloor n/2\rfloor+2}^{\infty}\dd x x^{-1-\beta_c}\\
                &= \left(\left\lfloor \frac{n}{2}\right\rfloor+2\right)^{-1-\beta_c}+\frac{1}{\beta_c}\left(\left\lfloor \frac{n}{2}\right\rfloor+2\right)^{-\beta_c}
                <\infty, 
            \end{align}
            which means that this sum is $O(n^{-\beta_c})$ uniformly in $N'$. Therefore, since the third term is nondecreasing in $N'$, it converges as $N'\to\infty$, and its limit is $O(n^{-\beta_c})$. Taking the limit $N'\to\infty$ yields 
            \begin{equation}
                \sum_{k=\lfloor n/2\rfloor+1}^{\infty}(S_{k}-S_{k+1})(k+1)^{1-2\beta_c}
                =O(n^{-\beta_c}).
            \end{equation}
        \end{proof}
\end{appendices}

\clearpage
\part[Demonstration with the SFK model]{{\Large Demonstration with the SFK model}}
\section{Model description}
    In this part, we consider the Sato--Fujiwara--Kanazawa~(SFK) model introduced in \cite{SatoFujiwaraKanazawa2026, SatoFujiwaraKanazawa2026Full}, a price-formation model that extends the Lillo--Mike--Farmer~(LMF) model by incorporating nonlinear permanent price impact. We assume $M\in\mathbb{Z}_{\geq1}$ order-splitting traders characterized by the same microscopic parameters: the metaorder-size distribution $\psi(\mathfrak{L})$ and the price impact function $I(\mathfrak{Q})$. The metaorder signs and sizes are statistically independent. Each trader executes their metaorders by splitting them into a sequence of unit-volume market orders (called child orders). The metaorder of trader $i\in\{1,\dots,M\}$ at time $n$ is characterized by three state variables: the metaorder sign $\epsilon^{(i)}_{n}\in\{-1,+1\}$, the remaining volume $\mathfrak{R}^{(i)}_n$, and the executed volume $\mathfrak{Q}^{(i)}_n$. 

    We next briefly review the model dynamics. At each time step, a randomly selected trader $i=\mathfrak{i}_{n}$ submits a market order; that is, all traders have the same order-submission intensity $\lambda:=1/M$ at each time step. After trader $i$ submits an order, its remaining volume is decremented by one, and the executed volume is incremented by one. At the macroscopic level, the submitted market order changes the price $m_{n}$ according to the incremental price impact, such that 
    \begin{align}
        \Delta m_n
        :=m_{n+1}-m_n 
        = \epsilon^{(i)}_n \Delta I\left(\mathfrak{Q}^{(i)}_n\right),
        \quad  \Delta I(\mathfrak{Q}):=I(\mathfrak{Q}+1)-I(\mathfrak{Q}).
    \end{align}    
    When trader $i$ is selected at time $n+1$ ($i=\mathfrak{i}_{n+1}$) and its remaining volume is one ($\mathfrak{R}^{(i)}_n=1$), that is, all the child orders of its current metaorder have been submitted, the state variables of trader $i$ are reinitialized: a new metaorder sign $\epsilon^{(i)}_{n+1}$ is randomly assigned, a new metaorder size $\mathfrak{L}$ is randomly drawn from $\psi(\mathfrak{L})$ and assigned as $\mathfrak{R}^{(i)}_{n+1}=\mathfrak{L}$, and the executed volume is reset to zero, $\mathfrak{Q}^{(i)}_{n+1}=0$. Formally, the model dynamics is written as 
    \begin{subequations}
    \label{eq:review-LMF-SDE}
        \begin{align}
            \mathfrak{R}^{(i)}_{n+1} &= 
            \begin{cases}
                \mathfrak{R}^{(i)}_n & \mbox{if $i \neq \mathfrak{i}_{n+1}$} \\
                \mathfrak{R}^{(i)}_n - 1& \mbox{if $i = \mathfrak{i}_{n+1}$ and $\mathfrak{R}^{(i)}_n>1$} \\
                \mathfrak{L} & \mbox{if $i= \mathfrak{i}_{n+1}$ and $\mathfrak{R}^{(i)}_n=1$, $\mathfrak{L}$ obeys $\psi(\mathfrak{L})$}
            \end{cases}, \\	
            \mathfrak{Q}^{(i)}_{n+1} &= 
            \begin{cases}
                \mathfrak{Q}^{(i)}_n & \mbox{if $i \neq \mathfrak{i}_{n+1}$} \\
                \mathfrak{Q}^{(i)}_n + 1 & \mbox{if $i = \mathfrak{i}_{n+1}$ and $\mathfrak{R}^{(i)}_n>1$} \\
                0 & \mbox{if $i= \mathfrak{i}_{n+1}$ and $\mathfrak{R}^{(i)}_n=1$}
            \end{cases}, \\
            \epsilon^{(i)}_{n+1} &= 
                \begin{cases}
                    \epsilon^{(i)}_n & \mbox{if $i \neq \mathfrak{i}_{n+1}$ or $\mathfrak{R}^{(i)}_n>1$} \\
                        +1 & \mbox{with prob. $1/2$, if $i = \mathfrak{i}_{n+1}$ and $\mathfrak{R}^{(i)}_n=1$} \\
                        -1 & \mbox{with prob. $1/2$, if $i = \mathfrak{i}_{n+1}$ and $\mathfrak{R}^{(i)}_n=1$}
                \end{cases}, \\
                \epsilon_{n+1} &= \epsilon^{(\mathfrak{i}_{n+1})}_{n+1},\\
                \Delta m_n &= \epsilon^{(\mathfrak{i}_n)}_n \Delta I\left(\mathfrak{Q}^{(\mathfrak{i}_n)}_n\right).
            \end{align}
        \end{subequations}
    
    In this part, we assume a power-law price impact $I(\mathfrak{Q}) = \mathfrak{Q}^\delta$ with $0<\delta<1$ and a power-law metaorder-size distribution 
    \begin{align}
        \psi(\mathfrak{L})=\frac{\mathfrak{L}^{-\alpha-1}}{\zeta(\alpha+1)},       \qquad \mathfrak{L}\in\mathbb{Z}_{\geq1},\quad 1<\alpha<2,
    \end{align}
    where $\zeta(\alpha):=\sum_{k=1}^{\infty}k^{-\alpha}$ is the Riemann zeta function.

\section{Exact autocorrelation function and response function}
    In this section, we exactly derive the sign-autocorrelation function, the response function, and the response increment.
    
    \subsection{Exact autocorrelation function}
    \label{subsec:SFK:acf}
        Using the exact autocorrelation function of the market order flow for $n\geq1$ derived in \cite{Sato2024_GeneralizedLMF} and the fact that $C_0 = 1$, for $n\geq0$, we obtain  
        \begin{equation}
        \label{eq:SFK:exact-acf}
            C_n
            =\left(1-\frac{\lambda}{\zeta(\alpha)\Gamma(\alpha+1)}\int_0^{\infty}\frac{\dd x  x^{\alpha}\left(1-\lambda+\lambda e^{-x}\right)^{-1}}{(e^{x}-1)^2}\right)0^n
            +\frac{\lambda}{\zeta(\alpha)\Gamma(\alpha+1)}\int_0^{\infty} \frac{\dd x  x^{\alpha}(1-\lambda+\lambda e^{-x})^{n-1}}{(e^{x}-1)^2},  
        \end{equation}
        where we take the convention that $0^0=1$ and $0^n=0$ for $n\geq1$.
        This expression shows that $(C_n)_{n\geq0}$ is completely monotone. Indeed, for the base of the power, we have 
        \begin{equation}
            0\leq 1-\lambda +\lambda e^{-x}\leq 1 \quad \text{for $x\geq0$ and $0<\lambda\leq1$},
        \end{equation}
        and both weights are nonnegative; that is, 
        \begin{equation}
            \frac{\lambda}{\zeta(\alpha)\Gamma(\alpha+1)}\frac{x^{\alpha}(1-\lambda+\lambda e^{-x})^{-1}}{(e^{x}-1)^2}\geq0
        \end{equation}
        for the second term, and 
        \begin{align}
            1-\frac{\lambda}{\zeta(\alpha)\Gamma(\alpha+1)}\int_0^{\infty}\frac{\dd x  x^{\alpha}\left(1-\lambda+\lambda e^{-x}\right)^{-1}}{(e^{x}-1)^2}
            \geq 1- \frac{1}{\zeta(\alpha)\Gamma(\alpha+1)}\int_0^{\infty}\frac{\dd x  x^{\alpha} e^{x}}{(e^{x}-1)^2} 
            =0
        \end{align}
        for the first term.

        The asymptotic expansion of Eq.~\eqref{eq:SFK:exact-acf} yields
        \begin{align}
        \label{eq:SFK:asymptotic-acf}
            C_n \sim \frac{\lambda}{\alpha(\alpha-1)\zeta(\alpha)}(\lambda n)^{-\gamma} \quad \text{as $n\to\infty$},
        \end{align}
        where $\gamma=\alpha-1$. For the detailed derivation, see Appendix~\ref{app:acf}.

    \subsection{Exact response function}
    \label{subsec:SFK:response}
        Next, we derive the exact response function $R_n$ defined by 
        \begin{align}
            R_n:=E[\epsilon_0(m_{n}-m_0)].
        \end{align}
        Since order signs belonging to different metaorders are independent, using the exact joint-stationary distribution $P_{\mathrm{st}}(\mathfrak{R},\mathfrak{Q})=(\mathfrak{R}+\mathfrak{Q})^{-\alpha-1}/\zeta(\alpha)$ derived in \cite{SatoFujiwaraKanazawa2026Full}, for $n\geq0$, we can write the exact response function as 
        \begin{align}
        \label{eq:SFK:exact-response}
            R_n
            &= \sum_{i=1}^{M}\lambda\sum_{N=1}^{n}\mathcal{B}_{n-1,\lambda}(N-1)\sum_{\mathfrak{Q}=0}^{\infty}\left[\sum_{\mathfrak{R}=N}^{\infty}P_{\mathrm{st}}(\mathfrak{R},\mathfrak{Q})\left\{(\mathfrak{Q}+N)^{\delta}-\mathfrak{Q}^{\delta}\right\}+\sum_{\mathfrak{R}=1}^{N-1}P_{\mathrm{st}}(\mathfrak{R},\mathfrak{Q})\left\{(\mathfrak{Q}+\mathfrak{R})^{\delta}-\mathfrak{Q}^{\delta}\right\}\right]\\
            &=\sum_{N=1}^{n}\mathcal{B}_{n-1,\lambda}(N-1)R_{N}^{(1)},\quad 
            R_{n}^{(1)}=\sum_{\mathfrak{Q}=0}^{\infty}\sum_{\mathfrak{R}=1}^{\infty}P_{\mathrm{st}}(\mathfrak{R},\mathfrak{Q})\left[(\mathfrak{Q}+\min\{n,\mathfrak{R}\})^{\delta}-\mathfrak{Q}^{\delta}\right],
        \end{align}
        where $\mathcal{B}_{n,\lambda}(k):=\binom{n}{k}\lambda^k (1-\lambda)^{n-k}$ is the binomial distribution, and $R_{n}^{(1)}$ denotes the exact response function in the single-body system. Obviously, we have $R_{n}\geq0$ for all $n\geq0$ and $R_n\not\equiv0$.

    \subsection{Exact response increment}
    \label{subsec:SFK:response-diff}
        By the exact response function $R_n$ derived in Sec.~\ref{subsec:SFK:response}, we derive the response increment $S_n := R_{n+1} - R_n$. Taking the first difference of $R_n$ with respect to $n$, for $n\geq0$, we obtain 
        \begin{align}
        \label{eq:SFK:exact-increment}
            S_n 
            &=\left(\frac{\zeta(\alpha-\delta+1)}{\zeta(\alpha)}-\frac{\lambda \delta}{\zeta(\alpha)\Gamma(\alpha+1)\Gamma(1-\delta)}\int_0^{\infty}\int_0^{\infty}\frac{\dd x \dd y  x^{\alpha} y^{-1-\delta}(1-e^{-y})\left(1-\lambda+\lambda e^{-(x+y)}\right)^{-1}}{(e^{x}-1)(e^{x+y}-1)}\right)0^{n} \\
            &\quad + \frac{\lambda \delta}{\zeta(\alpha)\Gamma(\alpha+1)\Gamma(1-\delta)}\int_0^{\infty}\int_0^{\infty}\frac{\dd x \dd y  x^{\alpha} y^{-1-\delta}(1-e^{-y})\left(1-\lambda+\lambda e^{-(x+y)}\right)^{n-1}}{(e^{x}-1)(e^{x+y}-1)}, 
        \end{align}
        where we take a convention that $0^0=1$ and $0^n=0$ for $n\geq1$.
        This expression shows that $(S_n)_{n\geq0}$ is completely monotone. Indeed, for the base of the power, we have 
        \begin{equation}
            0\leq 1-\lambda +\lambda e^{-(x+y)}\leq 1 \quad \text{for $x,y\geq0$ and $0<\lambda\leq1$},
        \end{equation}
        and both weights are nonnegative; that is, 
        \begin{equation}
            \frac{\lambda \delta}{\zeta(\alpha)\Gamma(\alpha+1)\Gamma(1-\delta)}\frac{x^{\alpha} y^{-1-\delta}(1-e^{-y})\left(1-\lambda+\lambda e^{-(x+y)}\right)^{-1}}{(e^{x}-1)(e^{x+y}-1)}\geq0
        \end{equation}
        for $n\geq1$, and 
        \begin{align}
            &\quad \frac{\zeta(\alpha-\delta+1)}{\zeta(\alpha)}-\frac{\lambda \delta}{\zeta(\alpha)\Gamma(\alpha+1)\Gamma(1-\delta)}\int_0^{\infty}\int_0^{\infty}\frac{\dd x \dd y  x^{\alpha} y^{-1-\delta}(1-e^{-y})\left(1-\lambda+\lambda e^{-(x+y)}\right)^{-1}}{(e^{x}-1)(e^{x+y}-1)}\\
            &\geq \frac{\zeta(\alpha-\delta+1)}{\zeta(\alpha)}-\frac{\delta}{\zeta(\alpha)\Gamma(\alpha+1)\Gamma(1-\delta)}\int_0^{\infty}\int_0^{\infty}\frac{\dd x \dd y  x^{\alpha} y^{-1-\delta}(1-e^{-y})e^{x+y}}{(e^{x}-1)(e^{x+y}-1)}\\
            &=0
        \end{align}
        for $n=0$.

        The asymptotic expansion of Eq.~\eqref{eq:SFK:exact-increment} yields
        \begin{align}
        \label{eq:SFK:asymptotic-response}
            S_n \sim \frac{\lambda\delta}{\alpha(\alpha-\delta)\zeta(\alpha)}(\lambda n)^{(\delta-\alpha+1)-1}\quad \text{as $n\to\infty$}.
        \end{align}
        For the detailed derivation, see Appendix~\ref{app:increment}.

\section{Application of the universal decay theorems to the SFK model}
     For the SFK model, the autocorrelation function $C_n$ satisfies Assumption~\ref{assumption:ACF}, the response increment $S_n$ Assumption~\ref{assumption:response} with $\eta=\delta-\alpha+1$ for $2\delta\leq\alpha$, and the response function $R_n$ Assumption~\ref{assumption:positivity}. Therefore, by the universal decay theorems~(ii), (iii), and (iv), we obtain the following asymptotic forms of the inferred propagator $\mathcal{G}_n$: 
    \begin{align}
        \mathcal{G}_n
        &= \mathfrak{A}_{\mathcal{G}} n^{-\beta_c}+o(n^{-\beta_c}),\quad \mathfrak{A}_{\mathcal{G}}>0,\quad \mathcal{G}_{\infty}=0\quad (2\delta<\alpha), \\
        \mathcal{G}_n
        &= \mathfrak{A}_{\mathcal{G}} n^{-\beta_c}\ln n + o(n^{-\beta_c}\ln n),\quad \mathfrak{A}_{\mathcal{G}}>0,\quad \mathcal{G}_{\infty}=0\quad (2\delta=\alpha).
    \end{align}

\begin{appendices}
\renewcommand{\thesubsection}{\thesection.\arabic{subsection}}
\makeatletter
\renewcommand{\p@subsection}{}
\makeatother
\section{Derivations for the SFK model with a power-law metaorder-size distribution}
\label{app:SFK}
In this appendix, we give the derivations for the exact and asymptotic forms of the sign-autocorrelation function and the response increment.

    \subsection{Exact sign-autocorrelation function}
    \label{app:acf}
        Let us derive the exact sign-autocorrelation function. Using the exact solution derived in \cite{Sato2024_GeneralizedLMF}, we can write the autocorrelation function in the present model as 
        \begin{align}
            C_n
            &= \frac{\lambda}{\zeta(\alpha)} \sum_{N=1}^{n}\mathcal{B}_{n-1,\lambda}(N-1)\sum_{L=1}^{\infty}L(L+N)^{-\alpha-1}
        \end{align}
        for $n\geq1$.
        Using the integral representation of the power-law function, $(L+N)^{-\alpha-1}=(\Gamma(\alpha+1))^{-1}\int_0^{\infty}\dd x x^{\alpha}e^{-x(L+N)}$, we obtain
        \begin{align}
            C_n
            &=\frac{\lambda}{\zeta(\alpha)\Gamma(\alpha+1)}\sum_{N=1}^{n}\mathcal{B}_{n-1,\lambda}(N-1)\sum_{L=1}^{\infty}L\int_{0}^{\infty}\dd x x^{\alpha} e^{-x(L+N)}\\
            &=\frac{\lambda}{\zeta(\alpha)\Gamma(\alpha+1)}\int_{0}^{\infty} \dd x \frac{x^{\alpha}\left(1-\lambda+\lambda e^{-x}\right)^{n-1}}{(e^{x}-1)^2}
        \end{align}
        for $n\geq1$. Together with $C_0=1$, this gives Eq.~\eqref{eq:SFK:exact-acf}.
        By Laplace's method, we obtain the asymptotic form as follows:
        \begin{equation}
            C_n
            \sim \frac{\lambda}{\zeta(\alpha)\Gamma(\alpha+1)}\int_{0}^{\infty}\dd x x^{\alpha-2}e^{-\lambda n x}
            =\frac{\lambda}{\alpha(\alpha-1)\zeta(\alpha)}\left(\lambda n\right)^{-(\alpha-1)}.
        \end{equation}

    \subsection{Exact response increment}
    \label{app:increment}
        Let us derive the exact response increment. Taking the first difference of Eq.~\eqref{eq:SFK:exact-response} gives 
         \begin{align}
            S_n
            =\lambda\sum_{N=1}^{n}\mathcal{B}_{n-1,\lambda}(N-1) \left[R_{N+1}^{(1)} - R_{N}^{(1)}\right]
        \end{align}
        for $n\geq1$.
        By the exact joint-stationary distribution $P_{\mathrm{st}}(\mathfrak{R}, \mathfrak{Q})=(\mathfrak{R}+\mathfrak{Q})^{-\alpha-1}/\zeta(\alpha)$, we have 
        \begin{equation}
            R^{(1)}_{N+1}-R^{(1)}_N
            =\frac{1}{\zeta(\alpha)}\sum_{\mathfrak{Q}=0}^{\infty}\sum_{\mathfrak{R}=N+1}^{\infty}\left(\mathfrak{R}+\mathfrak{Q}\right)^{-\alpha-1}\left[(\mathfrak{Q}+N+1)^{\delta}-(\mathfrak{Q}+N)^{\delta}\right].
        \end{equation}
        Using the integral representation of the power-law function, 
        \begin{align}
            (\mathfrak{R}+\mathfrak{Q})^{-\alpha-1}
            &=\frac{1}{\Gamma(\alpha+1)}\int_{0}^{\infty}\dd x x^{\alpha}e^{-x(\mathfrak{R}+\mathfrak{Q})},\\
            (\mathfrak{Q}+N+1)^{\delta}-(\mathfrak{Q}+N)^{\delta}
            &=\frac{\delta}{\Gamma(1-\delta)}\int_{0}^{\infty}\dd y y^{-1-\delta}\left(1-e^{-y}\right)e^{-y(\mathfrak{Q}+N)},
        \end{align}
        we obtain 
        \begin{align}
        S_n
        &=\frac{\lambda\delta}{\zeta(\alpha)\Gamma(\alpha+1)\Gamma(1-\delta)} \int_0^{\infty}\dd x \int_0^{\infty}\dd y \frac{x^{\alpha}y^{-1-\delta}(1-e^{-y})\left(1-\lambda+\lambda e^{-(x+y)}\right)^{n-1}}{(e^{x}-1)(e^{x+y}-1)}
        \end{align}
        for $n\geq1$. Tigether with $S_0=R_1-R_0=R_1$, this gives to Eq.~\eqref{eq:SFK:exact-increment} because 
        \begin{equation}
            S_0
            =R_1
            =R_1^{(1)}
            =\frac{1}{\zeta(\alpha)}\sum_{\mathfrak{Q}=0}^{\infty}\sum_{\mathfrak{R}=1}^{\infty}\left(\mathfrak{R}+\mathfrak{Q}\right)^{-\alpha-1}\left[(\mathfrak{Q}+1)^{\delta}-\mathfrak{Q}^{\delta}\right]
            =\frac{\zeta(\alpha-\delta+1)}{\zeta(\alpha)}.
        \end{equation}
        By Laplace's method, we obtain the asymptotic form as follows:
        \begin{align}
            S_n
            \sim \frac{\lambda\delta}{\zeta(\alpha)\Gamma(\alpha+1)\Gamma(1-\delta)}\int_0^{\infty}\dd x \int_0^{\infty}\dd y \frac{x^{\alpha-1}y^{-\delta}}{x+y}e^{-\lambda n(x+y)}
            =\frac{\lambda\delta}{\alpha(\alpha-\delta)\zeta(\alpha)}\left(\lambda n\right)^{(\delta-\alpha+1)-1}.
        \end{align}
\end{appendices}